\documentclass[
  aip,
  jcp,
  reprint,
  citeautoscript,
  superscriptaddress,
  longbibliography,
  floatfix,
  amsmath,
  amssymb
]{revtex4-2}

\usepackage[utf8]{inputenc}
\usepackage[T1]{fontenc}
\usepackage{mathptmx}
\usepackage{graphicx}
\usepackage{dcolumn}
\usepackage{bm}
\usepackage{xcolor}
\usepackage{url}
\usepackage{hyperref}
\hypersetup{colorlinks=true,linkcolor=blue,citecolor=blue,urlcolor=blue}

\definecolor{forestGreen}{RGB}{34,139,34}

\newcommand\beq{\begin{equation}}
\newcommand\eeq{\end{equation}}
\newcommand\beqa{\begin{eqnarray}}
\newcommand\eeqa{\end{eqnarray}}
\newcommand{\dd}{\mathrm{d}}

\newcommand{\al}{\alpha}

\newcommand{\bcal}[1]{\pmb{\mathcal{#1}}}

\newcommand{\fullwidthgraphic}[1]{%
  \IfFileExists{#1}{\includegraphics[width=\textwidth]{#1}}{%
    \fbox{\parbox[c][4.8cm][c]{0.94\textwidth}{\centering\texttt{\detokenize{#1}}\\[1ex]Figure file to be supplied}}%
  }%
}

\makeatletter
\DeclareFontFamily{OMX}{MnSymbolE}{}
\DeclareSymbolFont{MnLargeSymbols}{OMX}{MnSymbolE}{m}{n}
\SetSymbolFont{MnLargeSymbols}{bold}{OMX}{MnSymbolE}{b}{n}
\DeclareFontShape{OMX}{MnSymbolE}{m}{n}{
    <-6>  MnSymbolE5
   <6-7>  MnSymbolE6
   <7-8>  MnSymbolE7
   <8-9>  MnSymbolE8
   <9-10> MnSymbolE9
  <10-12> MnSymbolE10
  <12->   MnSymbolE12
}{}
\DeclareFontShape{OMX}{MnSymbolE}{b}{n}{
    <-6>  MnSymbolE-Bold5
   <6-7>  MnSymbolE-Bold6
   <7-8>  MnSymbolE-Bold7
   <8-9>  MnSymbolE-Bold8
   <9-10> MnSymbolE-Bold9
  <10-12> MnSymbolE-Bold10
  <12->   MnSymbolE-Bold12
}{}
\let\llangle\@undefined
\let\rrangle\@undefined
\DeclareMathDelimiter{\llangle}{\mathopen}{MnLargeSymbols}{'164}{MnLargeSymbols}{'164}
\DeclareMathDelimiter{\rrangle}{\mathclose}{MnLargeSymbols}{'171}{MnLargeSymbols}{'171}
\makeatother

\begin{document}

\title{Synchronization-controlled transport near an absorbing transition: Effective Enskog kinetic theory of a driven granular monolayer}

\author{David Gonz\'alez M\'endez}
\email{dgonzalezm@unex.es}
\affiliation{Departamento de F\'{\i}sica, Universidad de Extremadura, E-06071 Badajoz, Spain}

\author{Rub\'en G\'omez Gonz\'alez}
\email{ruben@unex.es}
\affiliation{Departamento de F\'{i}sica Aplicada, Universidad de Extremadura, E-06800 M\'{e}rida, Spain}

\author{Rapha\"{e}l Maire}
\email{maire@ub.edu}
\affiliation{Departament de F\'{i}sica de la Mat\`{e}ria Condensada, Universitat de Barcelona, E-08028 Barcelona, Spain}

\author{Vicente Garz\'{o}}
\email{vicenteg@unex.es}
\thanks{URL: \url{https://fisteor.cms.unex.es/investigadores/vicente-garzo-puertos/}}
\affiliation{Departamento de F\'{\i}sica, Universidad de Extremadura, E-06071 Badajoz, Spain}
\affiliation{Instituto de Computaci\'on Cient\'{\i}fica Avanzada (ICCAEx), E-06071 Badajoz, Spain}

\author{Ignacio Pagonabarraga}
\email{ipagonabarraga@ub.edu}
\affiliation{Departament de F\'{i}sica de la Mat\`{e}ria Condensada, Universitat de Barcelona, E-08028 Barcelona, Spain}

\begin{abstract}
Synchronization with vibrating boundaries can qualitatively alter the dynamics of confined granular fluids and control the collision-induced transfer of energy to horizontal motion. We develop an Enskog kinetic theory for the active phase of a driven granular monolayer within the framework of the $\Delta$-model, which accounts for the transfer of energy from the vertical to the horizontal degrees of freedom of grains. Synchronization effects are incorporated through a state-dependent effective collision rule governed by the interplay between the local collision frequency and a characteristic synchronization time. A viscous drag force is also included to account for energy dissipation arising from interactions with the confining plates. Within this framework, we characterize the homogeneous active state and use it as the reference state for a Chapman--Enskog expansion. The solution of the Enskog equation to Navier--Stokes order then yields explicit expressions for the shear and bulk viscosities, the thermal conductivity, the diffusive heat coefficient, and the first-order contribution to the cooling rate, consistently accounting for inelasticity, damping, finite-density effects, and synchronization. Event-driven molecular-dynamics simulations show that the theory accurately describes the steady temperature and shear and bulk viscosities over a broad region of the active phase. Close to the discontinuous synchronization-controlled absorbing transition, deviations are mainly attributed to the mean-field coarse-graining of the microscopic synchronization dynamics: fluctuations of the collision-dependent injection amplitude and correlations between synchronization and pre-collisional velocities become increasingly important. These results provide a kinetic description of synchronization-controlled transport and delineate its range of validity near the absorbing transition.
\end{abstract}

\date{\today}
\maketitle

\section{Introduction}
Like equilibrium matter, granular materials exhibit solid-, liquid-, and gas-like states~\cite{RevModPhys.68.1259}. Unlike equilibrium matter, however, they continually dissipate energy through interparticle interactions and require external driving to sustain their motion, keeping them out of equilibrium~\cite{RevModPhys.71.435,RevModPhys.71.S374}. Vertically vibrated grains confined between closely spaced horizontal plates provide an experimentally accessible setup for investigating the physics of such granular systems~\cite{olafsen1998clustering,prevost2004nonequilibrium,melby2005dynamics}. Individual trajectories can be tracked while density, confinement, and driving parameters are varied. Experiments on such vibrated granular systems find many collective effects such as crystallization, melting, and solid--fluid coexistence~\cite{olafsen2005twodimensional,reis2006crystallization,gotzendorfer2005sublimation,vegareyes2008effect,komatsu2015roles,luu2013capillarylike,plati2024quasicrystalline,plati2025selfassembly}, with transitions controlled by non-equilibrium driving~\cite{clerc2008liquidsolidlike,castillo2012fluctuations,castillo2015universality,guzman2018critical}. Clustering and liquid--gas phase separation have also been observed experimentally and numerically in the absence of attractive interactions~\cite{kudrolli1997cluster,falcon1999cluster,roeller2011liquidgas,clewett2012emergent,risso2018effective,maynar2019understanding}. Segregation in related systems raises the industrially relevant questions of controlling mixing and separation of grains~\cite{huerta2004vibrationinduced,breu2003reversing,mullin2000coarsening,reis2002granular,aumaitre2003granular,schnautz2005horizontal,BRM05,brito2008segregation,G08a,G11}. Standing waves and localized excitations in such systems further illustrate how diverse the collective responses are~\cite{melo1995hexagons,umbanhowar1996localized,eshuis2005granular,eshuis2010onset}. Together, these phenomena make vibrated granular systems a useful setting for linking dissipative particle dynamics to collective organization.

While the resulting state in granular systems may often look equilibrium-like in some respect, finer observables typically reveal the non-equilibrium nature of the system: velocity distributions are non-Gaussian~\cite{olafsen1999velocity,losert1999velocity,rouyer2000velocity,kudrolli2000nongaussian,reis2007forcing,tatsumi2009experimental}, molecular chaos~\cite{PhysRevE.65.011303, blair2001velocity,blair2003collision,baxter2003gaussian,baxter2007experimental} or equipartition~\cite{GD99b,feitosa2002breakdown,wildman2002coexistence} may not hold, and although individual grain trajectories can resemble Brownian motion~\cite{Sarracino_2010,reis2007caging,danna2003observing}, their collective dynamics depart from equilibrium by showing strong violations of the fluctuation--dissipation theorem~\cite{MARCONI2008111,gradenigo2011nonequilibrium,puglisi2012structure}. 

These observations called for a simple model suitable for numerical and theoretical studies of these nonequilibrium effects. A direct approach is to resolve theoretically the experimental geometry: grains confined between vibrating plates receive energy through vertical motion and transfer part of it to horizontal motion through oblique grain--grain collisions~\cite{mcnamara1997energy,kumaran1998temperature,barrat2002molecular,brey2017kinetic,maynar2019homogeneous,mayo2023confined}. Although this energy transfer controls the homogeneous state and its instabilities~\cite{grossman1997towards,khain2011hydrodynamics,maynar2019understanding}, retaining the full vertical dynamics complicates a hydrodynamic description. The effective two-dimensional $\Delta$-model offers a simpler approach by representing this transfer directly in the collision rule~\cite{BRS13}. Particles move in a two-dimensional plane and receive a separating velocity increment $\Delta$ at each collision, so that energy injection competes with inelastic dissipation while conserving pair momentum~\cite{BGMB13,BMGB14}. The complex 3D vibrations along the vertical axis are replaced by effective collisions that can inject energy in the two-dimensional plane.
\begin{figure*}[t]
\centering
\fullwidthgraphic{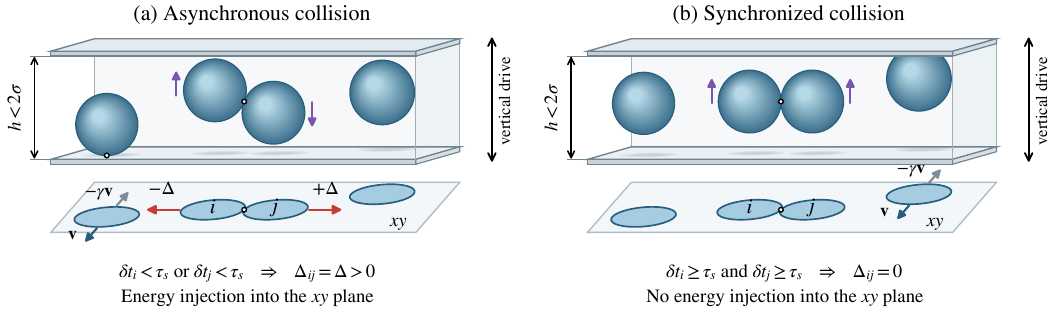}
\caption{Schematic illustration of a confined granular monolayer between vertically vibrating plates. (a) Asynchronous collision: if at least one particle has collided within the synchronization time $\tau_{\rm s}$, energy is injected through velocity increments $\pm\Delta\hat{\boldsymbol{\sigma}}$. (b) Synchronized collision: if both particles have remained collision-free for longer than $\tau_{\rm s}$, energy injection is suppressed ($\Delta_{ij}=0$). Purple arrows indicate the vertical particle motion. The lower sketches represent the effective horizontal dynamics in the $xy$ plane. In each lower sketch, the isolated particle illustrates the viscous drag $-\gamma\mathbf{v}$, opposing the horizontal velocity $\mathbf{v}$.}
\label{fig:diagram}
\end{figure*}
This simple collision rule allows for computations of the transport coefficients of the resulting fluid, including in the dense regime~\cite{soto2014shear,brey2015hydrodynamics,GBS18}, as summarized in the recent review~\cite{brito2026dynamic}. Its homogeneous state and stability have been studied in detail~\cite{brey2016stability,garzo2021stability}, and extensions address equipartition, diffusion, and segregation in mixtures~\cite{BSG20,garzo2021navierstokes,garzo2024applications,gomezgonzalez2024diffusion,gonzalezmendez2026transport}. Variations of this model have also been used to investigate phase separation, nonequilibrium interfaces or effective chiral interactions~\cite{risso2018effective,b7s9-4l4q,10.1063/5.0332479,maire2026chiraldynamicsintruderdilute,maire2026kinetichydrodynamictheorieschiral}. 

The $\Delta$-model nevertheless neglects interactions between grains and the confining plates in three-dimensional systems. Besides supplying vertical energy, the plates dissipate horizontal motion through friction. At low density, grain--grain collisions become too rare to replenish this lost horizontal energy, and the grains can stop moving horizontally despite continued vertical vibration~\cite{neel2014dynamics}. To capture this effect, the model can be supplemented with a drag term $-\gamma(\mathbf{v}-\mathbf U_{\rm w})$, which relaxes the grain velocity $\mathbf{v}$ toward the horizontal plate velocity $\mathbf U_{\rm w}$, usually set to zero~\cite{MPSF25,maire2024interplay}. Plate roughness also introduces fluctuations in the horizontal velocity at contact~\cite{prevost2002forcing,gradenigo2011nonequilibrium,puglisi2012structure}, which can be represented by weak stochastic forcing. Both damping and stochastic forcing modify the hydrodynamic behavior of the system, as shown in Ref.~\onlinecite{GGG26}, in which the transport coefficients of this model were derived. 

A final ingredient absent from the $\Delta$-model is synchronization with the vibrating plate. Under suitable driving conditions, grains can settle into periodic vertical motion locked to the plate oscillation~\cite{Pieraski1983,Pieraski1985,Chastaing2015,mujica2016dynamics,schindler2023phase,rivas2011sudden}. Simple dynamical models of a bouncing ball explain this locking through transitions between chaotic and periodic motion~\cite{everson1986chaotic,Metha1990,Luck1993}, as the driving parameters are varied. Such a synchronization mechanism can have dramatic effects on the dynamics of the system~\cite{rivas2011sudden}. Recent numerical studies have shown that it plays a central role in the absorbing transition exhibited by quasi-two-dimensional vibrated granular systems~\cite{maire2024interplay,MPSF25}. In particular, Maire et al.~\cite{maire2024interplay} demonstrated that synchronization can qualitatively alter the nature of the transition between horizontally active and absorbing states. Indeed, synchronized grains collide at nearly equal heights. This suppresses vertical-to-horizontal energy transfer, which in turn makes the horizontal motion stop even without drag~\cite{maire2024interplay,MPSF25}. Frequent collisions at high density disrupt this synchronization, whereas longer collision intervals let the vertical motion synchronize. To represent this mechanism in the 2D $\Delta$-model described above, we assume that a grain synchronizes with the plate after a collision-free interval of duration $\tau_\text{s}$, and that each collision disrupts this synchronization. The velocity increment $\Delta_{ij}$ then depends on the state of the two grains just before they collide: it is zero if both are synchronized and takes a finite value otherwise. A schematic illustration of the system is shown in Fig.~\ref{fig:diagram}.

Within the kinetic description proposed in Ref.~\onlinecite{MPSF25}, this
history dependence was treated at a mean-field level by replacing the
collision-dependent injection amplitude $\Delta_{ij}$ by an effective quantity
$\overline{\Delta}_{ij}$ controlled by the competition between the collision frequency and the synchronization time. This approximation provides a simple description of the homogeneous active branch and captures, at a mean-field level, the synchronization mechanism underlying the termination of the active branch and the consequent discontinuous absorbing transition observed in this system. In this regime, a decrease of the collision frequency favors synchronization, which reduces the effective horizontal energy injection and therefore lowers the granular temperature. The resulting decrease in the collision frequency further enhances synchronization, producing a positive feedback that eventually terminates the finite-temperature active branch. Thus, the discontinuous transition is associated with the loss of the
finite-temperature active solution rather than with a progressive vanishing of the active population.

In the absence of synchronization, a continuous transition can instead emerge through a distinct mechanism.
In this case, the system approaches the absorbing state through a progressive change in the population dynamics. Close to criticality, the system develops a heterogeneous population of active and inactive particles, and the absorbing transition occurs through the progressive vanishing of the
active fraction rather than through the disappearance of the kinetic energy carried by active particles. Consequently, a homogeneous description in which all particles are treated as active cannot reproduce this critical regime
quantitatively.

A complementary issue concerns the hydrodynamic description of the active state. In Ref.~\onlinecite{MPSF25}, the transport coefficients entering the hydrodynamic theory were approximated by their equilibrium hard-disk expressions. However, in a recent work,\cite{GGG26} the Navier--Stokes transport coefficients of the $\Delta$-model at moderate densities in the presence of viscous drag and stochastic forcing were derived, showing that external dissipation and collisional injection can substantially modify transport. Thus, the question arises as to whether, and if so to what extent, the conclusions drawn from Ref.\ \onlinecite{MPSF25} may be altered when the correct forms of the Navier--Stokes transport coefficients are accounted for.

While the role of synchronization in the homogeneous energy balance and in the absorbing transition has been characterized in previous studies~\cite{maire2024interplay,MPSF25}, a consistent transport theory for the synchronized active state is still lacking. In particular, synchronization introduces an additional feedback that is absent from the conventional $\Delta$-model: local variations of density and temperature modify the collision frequency and therefore the degree of synchronization, which in turn changes the amount of energy injected through collisions. In the present work, we incorporate this feedback into an Enskog kinetic description of the $\Delta$ model based on the Chapman--Enskog expansion\cite{CC70} by treating the synchronization-dependent injection amplitude within the mean-field closure introduced in Ref.~\onlinecite{MPSF25}. However, an important difference from Ref.~\onlinecite{MPSF25} is that here the parameter $\overline{\Delta}_{ij}(\mathbf{r}_1,\mathbf{r}_2)$ for a pair of colliding particles takes into account delocalization in collisions. In other words, it accounts for the fact that the particles are located at different positions $\mathbf{r}_1$ and $\mathbf{r}_2$. This feature introduces additional complexities compared to the conventional $\Delta$-model.~\cite{GBS18}

The extension of the Chapman--Enskog method to dissipative dynamics has been employed in this paper to explicitly derive the expressions of Navier--Stokes transport coefficients. As usual,\cite{G19} these forms have been approximately determined by considering the leading Sonine approximations to their kinetic contributions. In spite of this sort of approximation, the forms of the transport coefficients account for inelasticity, viscous drag, finite-density effects, and synchronization. To assess the accuracy of the theoretical predictions, they are compared with event-driven molecular-dynamics (EDMD) simulations, and the limitations of the synchronization mean-field closure close to the absorbing transition are examined through the fourth velocity cumulant and collisional synchronization statistics. In general, the comparison between the theory and the simulations shows good agreement in the active branch, indicating the reliability of the kinetic theory derived here. Importantly, an overall very good agreement is obtained when the
theoretical transport coefficients are evaluated using the steady granular temperature measured in the simulations. This observation supports the accuracy of the Chapman--Enskog description of transport and indicates that a significant part of the remaining discrepancies originates instead from the mean-field prediction of the homogeneous reference state, particularly from its estimate of the granular temperature. By using the simulated temperature, we largely remove this source of discrepancy, which is mainly associated with correlations and with the heterogeneous coexistence of active and inactive particle populations not explicitly represented in the present mean-field theory.

The paper is organized as follows. Section~\ref{sec:model} introduces the model and its Enskog kinetic description, and Sec.~\ref{sec3} characterizes the homogeneous state. Sections~\ref{sec:chapman-enskog} and~\ref{sec:transport} develop the Chapman--Enskog expansion and derive the forms of the Navier--Stokes transport coefficients in the leading Sonine approximations. Some representative cases are considered to compare theory with simulations. 
Section~\ref{sec:discussion} discusses the main findings and future directions. Technical details of the Chapman--Enskog calculation, the origin of the deviations in the fourth cumulant, and the event-driven molecular-dynamics simulations are given in the Appendices.

\section{Enskog kinetic theory for confined dragged granular gases with synchronization}\label{sec:model}

\subsection{Collision rules in the \texorpdfstring{$\Delta$}{Delta}-model}

We consider a set of identical solid particles, modeled as \emph{two-dimensional} hard disks of diameter $\sigma$ and mass $m$. Collisions between grains are inelastic and are characterized by a constant coefficient of normal restitution $0< \al\leq 1$, where $\al=1$ corresponds to elastic collisions. In the context of the so-called $\Delta$-model~\cite{BRS13}, the $z$-to-$xy$ energy-transfer mechanism is effectively represented through modified collision rules. Let us consider a collision between a particle $i$ and a particle $j$. The relation between their respective precollisional velocities $(\mathbf v_1,\mathbf v_2)$ and their postcollisional velocities $(\mathbf v_1',\mathbf v_2')$ reads

\begin{align}
\mathbf v_1'={}&\mathbf v_1-\frac{1}{2}(1+\alpha)
(\widehat{\boldsymbol\sigma}\cdot\mathbf g)\widehat{\boldsymbol\sigma}
-\Delta_{ij}\widehat{\boldsymbol\sigma}, \label{1.1}\\
\mathbf v_2'={}&\mathbf v_2+\frac{1}{2}(1+\alpha)
(\widehat{\boldsymbol\sigma}\cdot\mathbf g)\widehat{\boldsymbol\sigma}
+\Delta_{ij}\widehat{\boldsymbol\sigma}. \label{1.1b}
\end{align}
Here, $\mathbf g=\mathbf v_1-\mathbf v_2$ is the relative velocity, $\widehat{\boldsymbol\sigma}$ is the unit collision vector joining the centers of the two colliding disks and pointing from particle $1$ to particle $2$, and particles are approaching when $\widehat{\boldsymbol\sigma}\cdot\mathbf g>0$. In Eqs.~\eqref{1.1} and \eqref{1.1b}, $\Delta_{ij}\geq0$ is a collision-dependent non-negative velocity increment. At this stage, $\Delta_{ij}$ is deliberately left unspecified. The conventional $\Delta$-model is recovered for $\Delta_{ij}\equiv\Delta$~\cite{BRS13}, whereas the synchronization-dependent rule considered here is introduced in Sec.~\ref{subsec:sync}. Writing the mechanical collision rule in this general form allows the collision dynamics to be analyzed independently of the subsequent coarse-graining of the synchronization history.

The additional velocity increment is directed along the line of centers, so that the collision rule is also consistent with angular-momentum conservation~\cite{L04}. The relative velocity after collision is
\begin{equation}
\label{1.2}
\mathbf g'=\mathbf g-(1+\al)(\widehat{\boldsymbol\sigma}\cdot\mathbf g)
\widehat{\boldsymbol\sigma}-2\Delta_{ij}\widehat{\boldsymbol\sigma},
\end{equation}
so that
\begin{equation}
\label{1.3}
\widehat{\boldsymbol\sigma}\cdot\mathbf g'
=-\al(\widehat{\boldsymbol\sigma}\cdot\mathbf g)-2\Delta_{ij}.
\end{equation}
Momentum is conserved by Eqs.~\eqref{1.1} and \eqref{1.1b}, whereas kinetic energy is not. The energy change in one collision is
\begin{align}\label{1.4n}
\Delta E\equiv{}&\frac{m}{2}\left(v_1'^2+v_2'^2-v_1^2-v_2^2\right) \nonumber \\
={}&m\left[\Delta_{ij}^2+\alpha\Delta_{ij}
(\widehat{\boldsymbol\sigma}\cdot\mathbf g)
-\frac{1-\alpha^2}{4}(\widehat{\boldsymbol\sigma}\cdot\mathbf g)^2\right].
\end{align}
Thus, for $\alpha<1$, energy is gained or lost depending on whether
$\widehat{\boldsymbol\sigma}\cdot\mathbf g$ is smaller than or larger than
$2\Delta_{ij}/(1-\alpha)$, respectively. In particular, for $\alpha=1$ and $\Delta_{ij}=0$ the collision is elastic. For later use, we also introduce the restituting collision $(\mathbf v_1'',\mathbf v_2'')\to(\mathbf v_1,\mathbf v_2)$ with the same collision vector $\widehat{\boldsymbol\sigma}$:
\begin{align}
\mathbf v_1''={}&\mathbf v_1-\frac{1}{2}(1+\alpha^{-1})
(\widehat{\boldsymbol\sigma}\cdot\mathbf g)\widehat{\boldsymbol\sigma}
-\alpha^{-1}\Delta_{ij}\widehat{\boldsymbol\sigma}, \label{1.5}\\
\mathbf v_2''={}&\mathbf v_2+\frac{1}{2}(1+\alpha^{-1})
(\widehat{\boldsymbol\sigma}\cdot\mathbf g)\widehat{\boldsymbol\sigma}
+\alpha^{-1}\Delta_{ij}\widehat{\boldsymbol\sigma}. \label{1.5b}
\end{align}
For a fixed value of $\Delta_{ij}$, which is assumed not to depend explicitly
on the precollisional velocities, the corresponding velocity-space Jacobians are
\begin{equation}
\dd\mathbf v_1'\dd\mathbf v_2'=\al\,\dd\mathbf v_1\dd\mathbf v_2,
\qquad
\dd\mathbf v_1''\dd\mathbf v_2''=\al^{-1}\dd\mathbf v_1\dd\mathbf v_2.
\end{equation}

\subsection{Effective synchronization-dependent \texorpdfstring{$\Delta$}{Delta} model}\label{subsec:sync}

As discussed in the Introduction, a key ingredient of quasi-two-dimensional vibrated granular layers is \emph{synchronization}: after a sufficiently long free flight, an isolated particle can relax toward a periodic vertical motion in phase with the vibrating plates, whereas a grain--grain collision disrupts this synchronized motion. When both colliding grains are synchronized, their vertical coordinates are nearly identical and the transfer from vertical to horizontal motion is strongly suppressed. Conversely, a collision involving at least one asynchronized particle is typically non-planar and can transfer a finite amount of energy to the horizontal degrees of freedom.

Following Ref.~\onlinecite{MPSF25}, we describe this mechanism in our purely two-dimensional model through a single synchronization time $\tau_{\rm s}$. Each particle $i$ carries a ``time since last collision'' $\delta t_i$: a free flight longer than $\tau_{\rm s}$ ‘‘synchronizes'' that particle, whereas every collision resets its clock. The microscopic collision-dependent injection amplitude is therefore taken as
\begin{equation}
\label{1.23}
\Delta_{ij}=
\begin{cases}
\Delta>0, & \text{if }\ \delta t_i<\tau_{\rm s}\ \text{or}\ \delta t_j<\tau_{\rm s},\\[1mm]
0, & \text{otherwise}.
\end{cases}
\end{equation}
In the limit $\tau_{\rm s}\to\infty$, particles never synchronize between successive collisions and Eq.~\eqref{1.23} reduces to the conventional $\Delta$-model, $\Delta_{ij}\equiv\Delta$. As in the synchronization-dependent kinetic treatment of Ref.~\onlinecite{MPSF25}, we focus on the active branch and do not introduce an independent active-particle fraction. In the notation of that work, this amounts to setting $P_{\rm a}=1$. Hence, the present theory is intended to describe the synchronization-controlled active state and the mechanism associated with the discontinuous transition; the activation--inactivation population dynamics required for the continuous critical transition is not retained.

An exact kinetic description of Eq.~\eqref{1.23} would require the joint statistics of velocities and synchronization histories. To close the theory in terms of the one-particle distribution, we first neglect correlations between these two sets of variables. Thus, for an arbitrary function $h(\mathbf v_1,\mathbf v_2)$, we assume
\begin{equation}
\label{1.24}
\left\langle h(\mathbf v_1,\mathbf v_2)\Delta_{ij}\right\rangle_{\rm coll}
\simeq
\left\langle h(\mathbf v_1,\mathbf v_2)\right\rangle_{\rm coll}
\left\langle\Delta_{ij}\right\rangle_{\rm coll},
\end{equation}
where $\langle\cdots\rangle_\text{coll}$ denotes an average performed over all colliding pairs. We then introduce a second mean-field approximation at the level of the collision rule. Assuming statistically independent Poissonian
collision histories for the two particles, and a local, quasi-stationary approximation for the collision frequency, the 
probability that both particles have remained collision-free for a time longer than $\tau_{\rm s}$ is approximated by
\begin{equation} \label{Poisson_prob}
\exp\left\{-\tau_{\rm s}\left[\omega_{\rm e}(\mathbf r_1,t)+\omega_{\rm e}(\mathbf r_2,t)\right]\right\},
\end{equation}
where
\begin{equation}
    \omega_\text{e}(\mathbf{r},t) = \sqrt{\frac{4\pi T(\mathbf{r},t)}{m}}\chi(\phi) n(\mathbf{r},t)\sigma
\end{equation}
is the Maxwellian approximation for the (local) Enskog equilibrium collision frequency. Here, $n$ is the number density of particles, $T$ is the granular temperature, $\chi$ is the pair correlation function and $\phi=\pi n\sigma^2/4$ is the packing fraction. The hydrodynamic fields $n$ and $T$ will be defined in subsection \ref{subsec2c}. 

Thus, according to Eq.\ \eqref{Poisson_prob}, for a pair whose centers are at $\mathbf r_1$ and $\mathbf r_2$, the history-dependent random variable $\Delta_{ij}$ is replaced by the deterministic pairwise amplitude
\begin{equation}
\label{1.25}
\overline{\Delta}_{ij}(\mathbf r_1,\mathbf r_2,t)
=
\Delta\left\{1-
\exp\left[-\tau_{\rm s}\left(\omega_{\rm e}(\mathbf r_1,t)+\omega_{\rm e}(\mathbf r_2,t)\right)\right]\right\}.
\end{equation}
For spatially homogeneous steady states, the Poisson approximation
underlying Eq.~\eqref{1.25} is found to describe properly the
collision statistics obtained from molecular dynamics simulations of the collision rule with the microscopic rule Eq.~\eqref{1.23} over the range investigated here, as
shown in Appendix~\ref{app:a2deviations}. Nevertheless, in contrast to the choice made by Ref. \onlinecite{MPSF25}, note that the definition of effective injection $\overline{\Delta}_{ij}$ proposed here goes beyond the homogeneous description. Within the closure $\Delta_{ij} \to \overline{\Delta}_{ij}$,
\begin{equation}
\left\langle\Delta_{ij}\right\rangle_{
{\rm coll}\,|\,\mathbf r_1,\mathbf r_2}
\simeq
\overline{\Delta}_{ij}(\mathbf r_1,\mathbf r_2,t).
\end{equation}
Equation~\eqref{1.25} is symmetric under exchange of the two colliding particles,
\begin{equation}
\overline{\Delta}_{ij}(\mathbf r_1,\mathbf r_2,t)
=
\overline{\Delta}_{ij}(\mathbf r_2,\mathbf r_1,t),
\end{equation}
and, within the present mean-field description, it depends on the hydrodynamic state sampled by the two particles but not explicitly on their velocities. Both properties will be important below.

Replacing the fluctuating collision amplitude by Eq.~\eqref{1.25} is an additional mean-field step. More precisely, for each colliding pair at positions $\mathbf r_1$ and
$\mathbf r_2$, the fluctuating history-dependent variable $\Delta_{ij}$ is
replaced by the deterministic effective amplitude
$\overline{\Delta}_{ij}(\mathbf r_1,\mathbf r_2,t)$ before the corresponding integrations over particle velocities and collision geometries are performed. Averaging over all colliding pairs then yields 
\begin{equation}
\left\langle F(\Delta_{ij})\right\rangle_{\rm coll}
\simeq \left\langle F(\overline{\Delta}_{ij}(\mathbf r_1,\mathbf r_2,t))\right\rangle_{\rm coll},
\end{equation}
where it must be emphasized that the right-hand-side average is performed using the effective collision kernel.
This approximation is not an exact consequence of the
Poisson statistics, but rather an additional closure
assumption that neglects fluctuations of the injection
amplitude at fixed particle positions. 
The present closure therefore neglects both velocity--synchronization correlations and fluctuations of the collision-dependent injection amplitude around its effective pairwise value.

\subsection{Enskog kinetic equation for a dragged confined granular fluid} \label{subsec2c}

After the synchronization histories have been eliminated through the closure above, the kinetic state is again described solely by the one-particle velocity distribution $f(\mathbf r,\mathbf v;t)$. At moderate densities and for a two-dimensional system (hard disks), $f$ obeys the Enskog kinetic equation
\begin{equation}
\label{1.6}
\frac{\partial f}{\partial t}+\mathbf v\cdot\bm\nabla f+\mathcal Ff
=J_{\rm E}[\mathbf r,\mathbf v|f,f].
\end{equation}
For a collision between particles whose centers are at $\mathbf r$ and $\mathbf r+\boldsymbol\sigma$, the collision rules \eqref{1.1}--\eqref{1.5b} are now understood with the replacement
\begin{equation}
\Delta_{ij}\to
\overline{\Delta}_{ij}(\mathbf r,\mathbf r+\boldsymbol\sigma,t).
\end{equation}
Since $\overline{\Delta}_{ij}$ depends on the pair positions but not on the velocities, the direct and restituting velocity transformations retain the same Jacobians as in the conventional $\Delta$-model. The resulting Enskog collision operator is
\begin{widetext}
\begin{align}
\label{1.7}
J_{\rm E}[\mathbf r,\mathbf v_1|f,f]
={}&\sigma\int\dd\mathbf v_2\int\dd\widehat{\boldsymbol\sigma}\,
\Theta\left[-\widehat{\boldsymbol\sigma}\cdot\mathbf g
-2\overline{\Delta}_{ij}(\mathbf r,\mathbf r+\boldsymbol\sigma,t)\right]
\left[-\widehat{\boldsymbol\sigma}\cdot\mathbf g
-2\overline{\Delta}_{ij}(\mathbf r,\mathbf r+\boldsymbol\sigma,t)\right]\alpha^{-2}
\nonumber\\
&\times\chi(\mathbf r,\mathbf r+\boldsymbol\sigma)
 f(\mathbf r,\mathbf v_1'',t)
 f(\mathbf r+\boldsymbol\sigma,\mathbf v_2'',t)
\nonumber\\
&-\sigma\int\dd\mathbf v_2\int\dd\widehat{\boldsymbol\sigma}\,
\Theta(\widehat{\boldsymbol\sigma}\cdot\mathbf g)
(\widehat{\boldsymbol\sigma}\cdot\mathbf g)
\chi(\mathbf r,\mathbf r+\boldsymbol\sigma)
 f(\mathbf r,\mathbf v_1,t)
 f(\mathbf r+\boldsymbol\sigma,\mathbf v_2,t).
\end{align}
\end{widetext}
Here $\chi[\mathbf r,\mathbf r\pm\boldsymbol\sigma|n(t)]$ is the equilibrium pair correlation function at contact, considered as a functional of the nonequilibrium density field
\begin{equation}
\label{1.8}
n(\mathbf r,t)=\int\dd\mathbf v\,f(\mathbf r,\mathbf v,t).
\end{equation}

Besides receiving energy from the vertical driving through collisions, the horizontal motion is damped by the interaction with the confining plates. Following Ref.~\onlinecite{MPSF25}, this effect is represented by a viscous drag acting during free flight. The same deterministic drag operator was considered in Ref.~\onlinecite{GGG26}, where it was supplemented by a stochastic Langevin term to model an interstitial bath. Here no stochastic forcing is included, consistently with the dry absorbing-state model of Ref.~\onlinecite{MPSF25}; a finite additive noise would continuously inject horizontal energy even in the absence of grain--grain collisions. The drag operator is
\begin{equation}
\label{1.9}
\mathcal Ff=
-\gamma\frac{\partial}{\partial\mathbf v}\cdot
\left[(\mathbf v-\mathbf U_{\rm w})f\right],
\end{equation}
where $\gamma$ is the scalar drag coefficient, taken as a free parameter independent of the packing fraction $\phi$, and $\mathbf U_{\rm w}$ is the velocity of the plates. Hence Eq.~\eqref{1.6} becomes
\begin{equation}
\label{1.10}
\frac{\partial f}{\partial t}+\mathbf v\cdot\bm\nabla f
-\gamma\hat{\bf U}\cdot\frac{\partial f}{\partial\mathbf v}
-\gamma\frac{\partial}{\partial\mathbf v}\cdot\mathbf V f
=J_{\rm E}[\mathbf r,\mathbf V|f,f],
\end{equation}
where $\mathbf V=\mathbf v-\mathbf U$ is the peculiar velocity and $\hat{\bf U}=\mathbf U-\mathbf U_{\rm w}$. In terms of the distribution function $f$, the mean flow velocity $\mathbf U$ and granular temperature $T$ are defined as
\begin{equation}
\label{1.11}
\mathbf U(\mathbf r,t)
=\frac{1}{n(\mathbf r,t)}\int\dd\mathbf v\,\mathbf v f(\mathbf r,\mathbf v,t),
\end{equation}
and
\begin{equation}
\label{1.12}
T(\mathbf r,t)
=\frac{m}{2n(\mathbf r,t)}\int\dd\mathbf v\,V^2f(\mathbf r,\mathbf v,t).
\end{equation}

A useful identity for the Enskog collision operator is obtained by multiplying it by an arbitrary test function $\psi(\mathbf v)$ and integrating over velocity. The result is
\begin{align}\label{1.13weak}
I_\psi\equiv{}&\int\dd\mathbf v_1\,\psi(\mathbf v_1)
J_{\rm E}[\mathbf r,\mathbf v_1|f,f]\nonumber\\
={}&\sigma\int\dd\mathbf v_1\int\dd\mathbf v_2
\int\dd\widehat{\boldsymbol\sigma}\,
\Theta(\widehat{\boldsymbol\sigma}\cdot\mathbf g)
(\widehat{\boldsymbol\sigma}\cdot\mathbf g)
\chi(\mathbf r,\mathbf r+\boldsymbol\sigma)
\nonumber\\
&\times f(\mathbf r,\mathbf v_1,t)
 f(\mathbf r+\boldsymbol\sigma,\mathbf v_2,t)
\left[\psi(\mathbf v_1')-\psi(\mathbf v_1)\right],
\end{align}
where $\mathbf v_1'$ is the direct postcollisional velocity obtained from Eq.~\eqref{1.1} with
$\Delta_{ij}\to\overline{\Delta}_{ij}(\mathbf r,\mathbf r+\boldsymbol\sigma,t)$. The derivation of Eq.~\eqref{1.13weak} is not reproduced here, but it can be carried out formally by the same change from restituting to direct collision variables as in the conventional $\Delta$-model.\cite{GBS18} The position-dependent quantity $\overline{\Delta}_{ij}(\mathbf r,\mathbf r+\boldsymbol\sigma,t)$ remains fixed during this change of variables in velocity space and therefore does not modify the Jacobian. Thus, the weak form retains the same formal structure as in the conventional Enskog description~\cite{BP04,G19,GBS18}. In addition, the symmetry of Eq.~\eqref{1.25} under exchange of the pair positions permits the usual particle-exchange symmetrization of the weak identity.

Since the collision rule conserves particle number and total momentum, the balance equations retain the standard form
\begin{equation}
\text{D}_{t} n+n\bm\nabla\cdot\mathbf U=0,
\label{1.14}
\end{equation}
\begin{equation}
\text{D}_{t}\mathbf U=-\rho^{-1}\bm\nabla\cdot\mathsf P-\gamma\hat{\bf U},
\label{1.15}
\end{equation}
and
\begin{equation}
\text{D}_{t}T+n^{-1}\left(\bm\nabla\cdot\mathbf q+\mathsf P:\bm\nabla\mathbf U\right)
=-2\gamma T-\zeta T,
\label{1.16}
\end{equation}
where $\text{D}_{t}=\partial_t+\mathbf U\cdot\bm\nabla$ and $\rho=mn$ is the mass density. The cooling rate $\zeta$ accounts for the net collisional energy change: it contains the usual inelastic loss proportional to $1-\alpha^2$ together with the synchronization-controlled injection terms, and can therefore become negative when collisional injection exceeds dissipation. The pressure tensor and heat flux are written as $\mathsf P=\mathsf P_{\rm k}+\mathsf P_{\rm c}$ and $\mathbf q=\mathbf q_{\rm k}+\mathbf q_{\rm c}$. As usual, the kinetic contributions $\mathsf P_{\rm k}$ and $\mathbf q_{\rm k}$ are defined, respectively, as 
\begin{align}
\label{1.17}
\mathsf P_{\rm k}(\mathbf r,t)&=\int\dd\mathbf v\,m\mathbf V\mathbf V f(\mathbf r,\mathbf v,t),\\
\label{1.18}
\mathbf q_{\rm k}(\mathbf r,t)&=\int\dd\mathbf v\,\frac{m}{2}V^2\mathbf V f(\mathbf r,\mathbf v,t).
\end{align}

The forms of the collisional transfer contributions $\mathsf P_{\rm c}$ and $\mathbf q_{\rm c}$ can be obtained by following similar mathematical steps as those made in the Appendix A of Ref.\ \onlinecite{GBS18}. Although the intermediate algebra is not displayed, one point is worth emphasizing because the derivation here differs from the constant-$\Delta$ case. In converting the spatial difference between the two descriptions of a colliding pair into a line integral, the identity
\begin{align}
&F(\mathbf r,\mathbf r+\boldsymbol\sigma)-F(\mathbf r-\boldsymbol\sigma,\mathbf r)\nonumber\\
&\qquad=\bm\nabla\cdot\left[\boldsymbol\sigma\int_0^1\dd x\;
F\left(\mathbf r-x\boldsymbol\sigma,\mathbf r+(1-x)\boldsymbol\sigma\right)\right]
\end{align}
must be applied to the complete product
\[
[\psi(\mathbf v_1')-\psi(\mathbf v_1)]
\chi(\mathbf r_1,\mathbf r_2)
f(\mathbf r_1,\mathbf v_1,t)
f(\mathbf r_2,\mathbf v_2,t).
\]
This is because the postcollisional velocity $\mathbf v_1'$ itself depends on the positions of the pair through $\overline{\Delta}_{ij}$. Consequently, the pairwise amplitude remains inside the line integral. The resulting collisional-transfer expressions therefore retain the formal structure of those of the conventional $\Delta$-model, with the constant injection amplitude replaced by the corresponding pairwise quantity along the collision link.\cite{GBS18}

The exact expression of the collisional pressure tensor $\mathsf P_{\rm c}$ is
\begin{widetext}
\begin{align}
\label{1.19}
\mathsf P_{\rm c}={}&\frac{1+\alpha}{4}m\sigma^2
\int\dd\mathbf v_1\int\dd\mathbf v_2\int\dd\widehat{\boldsymbol\sigma}\,
\Theta(\widehat{\boldsymbol\sigma}\cdot\mathbf g)
(\widehat{\boldsymbol\sigma}\cdot\mathbf g)
\widehat{\boldsymbol\sigma}\widehat{\boldsymbol\sigma}
\nonumber\\
&\times\int_0^1\dd x\,
\left[(\widehat{\boldsymbol\sigma}\cdot\mathbf g)
+\frac{2\overline{\Delta}_{ij}
\left(\mathbf r-x\boldsymbol\sigma,\mathbf r+(1-x)\boldsymbol\sigma,t\right)}{1+\al}\right]
\nonumber\\
&\times
\chi\left(\mathbf r-x\boldsymbol\sigma,
\mathbf r+(1-x)\boldsymbol\sigma\right)
f\left(\mathbf r-x\boldsymbol\sigma,\mathbf v_1,t\right)
f\left(\mathbf r+(1-x)\boldsymbol\sigma,\mathbf v_2,t\right).
\end{align}

The collisional contribution $\mathbf q_{\rm c}$ to the heat flux is
\begin{align}
\label{1.20}
\mathbf q_{\rm c}={}&\frac{1+\alpha}{4}m\sigma^2
\int\dd\mathbf v_1\int\dd\mathbf v_2\int\dd\widehat{\boldsymbol\sigma}\,
\Theta(\widehat{\boldsymbol\sigma}\cdot\mathbf g)
(\widehat{\boldsymbol\sigma}\cdot\mathbf g)^2
(\widehat{\boldsymbol\sigma}\cdot\mathbf G)
\widehat{\boldsymbol\sigma}\int_0^1\dd x\,
\nonumber\\
&\times
\chi\left(\mathbf r-x\boldsymbol\sigma,
\mathbf r+(1-x)\boldsymbol\sigma\right)
f\left(\mathbf r-x\boldsymbol\sigma,\mathbf v_1,t\right)
f\left(\mathbf r+(1-x)\boldsymbol\sigma,\mathbf v_2,t\right)
\nonumber\\
&-\frac{m\sigma^2}{4}
\int\dd\mathbf v_1\int\dd\mathbf v_2\int\dd\widehat{\boldsymbol\sigma}\,
\Theta(\widehat{\boldsymbol\sigma}\cdot\mathbf g)
(\widehat{\boldsymbol\sigma}\cdot\mathbf g)
\widehat{\boldsymbol\sigma}
\int_0^1\dd x\;
\overline{\Delta}_{ij}
\left(\mathbf r-x\boldsymbol\sigma,\mathbf r+(1-x)\boldsymbol\sigma,t\right)
\nonumber\\
&\times\left[
\overline{\Delta}_{ij}
\left(\mathbf r-x\boldsymbol\sigma,\mathbf r+(1-x)\boldsymbol\sigma,t\right)
+\al(\widehat{\boldsymbol\sigma}\cdot\mathbf g)
-2(\widehat{\boldsymbol\sigma}\cdot\mathbf G)
\right]
\nonumber\\
&\times
\chi\left(\mathbf r-x\boldsymbol\sigma,
\mathbf r+(1-x)\boldsymbol\sigma\right)
f\left(\mathbf r-x\boldsymbol\sigma,\mathbf v_1,t\right)
f\left(\mathbf r+(1-x)\boldsymbol\sigma,\mathbf v_2,t\right),
\end{align}
where $\mathbf G=\frac12(\mathbf V_1+\mathbf V_2)$. Finally, the cooling rate is
\begin{align}
\label{1.22}
\zeta={}&-\frac{m\sigma}{2nT}
\int\dd\mathbf v_1\int\dd\mathbf v_2\int\dd\widehat{\boldsymbol\sigma}\,
\Theta(\widehat{\boldsymbol\sigma}\cdot\mathbf g)
(\widehat{\boldsymbol\sigma}\cdot\mathbf g)
\nonumber\\
&\times\left\{
\left[\overline{\Delta}_{ij}(\mathbf r,\mathbf r+\boldsymbol\sigma,t)\right]^2
+\al\overline{\Delta}_{ij}(\mathbf r,\mathbf r+\boldsymbol\sigma,t)
(\widehat{\boldsymbol\sigma}\cdot\mathbf g)
-\frac{1-\al^2}{4}(\widehat{\boldsymbol\sigma}\cdot\mathbf g)^2
\right\}
\chi(\mathbf r,\mathbf r+\boldsymbol\sigma)
f(\mathbf r,\mathbf v_1,t)
f(\mathbf r+\boldsymbol\sigma,\mathbf v_2,t).
\end{align}
\end{widetext}

Equations~\eqref{1.19}--\eqref{1.22} provide the  expressions of the collisional transfer contributions to the momentum and heat fluxes and the cooling rate 
within the pairwise mean-field closure. At this stage, no homogeneous reduction of $\overline{\Delta}_{ij}$ has been made: the collision amplitude remains attached to the two members of the colliding pair. This distinction is essential in an inhomogeneous state and must be retained before any gradient expansion is performed. In the particular case in which $\overline{\Delta}_{ij} = \Delta \equiv \text{constant}$, Eqs.\ \eqref{1.19}--\eqref{1.22} reduce to those obtained for the conventional $\Delta$-model.\cite{GBS18}

As will be shown later, in the Chapman--Enskog analysis the pairwise
dependence of Eq.~\eqref{1.25} generates an additional first-order
contribution to the vector kinetic equations and, consequently, to the
kinetic parts of the thermal conductivity and diffusive heat coefficient.
By contrast, the explicit first-order contribution arising from the spatial
dependence of $\overline{\Delta}_{ij}$ inside the collisional-transfer line
integrals vanishes after integration over the collision link.

\section{Homogeneous state}\label{sec3}

Before addressing spatially inhomogeneous states, we characterize the homogeneous dynamics of the synchronization-controlled active fluid. As in the previous kinetic-theory treatment of the $\Delta$-model with external forcing analyzed in Ref.~\onlinecite{GGG26}, the time-dependent homogeneous solution will later provide the reference state for the Chapman--Enskog expansion~\cite{CC70}. For homogeneous time-dependent states, the density $n$ and temperature $T$ are spatially uniform, and with an appropriate choice of reference frame the mean flow velocity vanishes, $\mathbf U=\mathbf U_{\rm w}=\mathbf 0$.

Homogeneity has an immediate consequence for the pairwise synchronization closure introduced in Eq.~\eqref{1.25}. Since the local Enskog collision frequency is the same at the positions of the two colliding particles,
\begin{equation}
\omega_{\rm e}(\mathbf r_1,t)=\omega_{\rm e}(\mathbf r_2,t)=\omega_{\rm e}(t),
\end{equation}
the pair-dependent amplitude becomes independent of the particle positions. We denote this homogeneous effective injection amplitude by
\begin{equation}
\label{eq:barDelta_local}
\overline{\Delta}(t)
\equiv
\Delta\left(1-\text{e}^{-2\omega_{\rm e}(t)\tau_{\rm s}}\right).
\end{equation}
Thus, in a homogeneous state, the pairwise mean-field closure reduces to a deterministic $\Delta$-model with a state-dependent effective injection amplitude $\overline{\Delta}(t)$. We recall that the factor multiplying $\Delta$ in Eq.~\eqref{eq:barDelta_local} is the mean-field probability that, at the instant of collision, at least one of the two particles has experienced another collision during the preceding interval $\tau_{\rm s}$. Synchronization is therefore controlled by the competition between the collision frequency $\omega_{\rm e}$ and the synchronization rate $1/\tau_{\rm s}$: for $\omega_{\rm e}\tau_{\rm s}\ll1$, one has $\overline{\Delta}\ll\Delta$, whereas for $\omega_{\rm e}\tau_{\rm s}\gg1$ the conventional $\Delta$-model limit $\overline{\Delta}\simeq\Delta$ is recovered.

With Eq.~\eqref{eq:barDelta_local}, the Enskog equation~\eqref{1.10} becomes
\begin{equation}
\label{2.1}
\frac{\partial f}{\partial t}
-\gamma\frac{\partial}{\partial\mathbf v}\cdot\mathbf v f
=
J_{\rm E}^{\rm H}[\mathbf V|f,f],
\end{equation}
where the homogeneous Enskog collision operator is
\begin{align}
\label{2.2}
J_{\rm E}^{\rm H}[f,f]
={}&
\sigma\chi
\int\dd\mathbf v_2
\int\dd\widehat{\boldsymbol\sigma}\,
\Theta(-\widehat{\boldsymbol\sigma}\cdot\mathbf g-2\overline{\Delta})
(-\widehat{\boldsymbol\sigma}\cdot\mathbf g-2\overline{\Delta})
\nonumber\\
&\times
\alpha^{-2}
f(\mathbf v_1'',t)f(\mathbf v_2'',t)
\nonumber\\
&-
\sigma\chi
\int\dd\mathbf v_2
\int\dd\widehat{\boldsymbol\sigma}\,
\Theta(\widehat{\boldsymbol\sigma}\cdot\mathbf g)
(\widehat{\boldsymbol\sigma}\cdot\mathbf g)
f(\mathbf v_1,t)f(\mathbf v_2,t).
\end{align}
Here, $\chi$ is the pair correlation function evaluated at the homogeneous density $n$. The operator~\eqref{2.2} has the same formal structure as the homogeneous collision operator of the conventional $\Delta$-model, but with the constant injection amplitude replaced by the homogeneous synchronization-controlled quantity $\overline{\Delta}(t)$. 

For homogeneous time-dependent states, the only nontrivial balance equation is that of the temperature, Eq.~\eqref{1.16},
\begin{equation}
\label{2.3}
\partial_t T=-(2\gamma+\zeta)T.
\end{equation}
Following the same strategy as in previous works,\cite{GGG26} we do not take the stationary distribution as the reference state. Instead, we first consider the time-dependent homogeneous solution and only impose stationarity after its temperature dependence has been characterized. For times longer than the mean free time, one expects the system to reach a hydrodynamic regime in which the distribution \(f\) becomes a \emph{normal} solution, namely, its explicit time dependence occurs only through the granular temperature \(T(t)\). In this regime, $\partial_t f=(\partial_T f)(\partial_t T)$ and Eq.\ \eqref{2.1} reads
\beq
\label{2.4}
-(2\gamma +\zeta)T\frac{\partial f}{\partial T}-\gamma\frac{\partial}{\partial\mathbf{v}}\cdot\mathbf{v}f=J_{\text{E}}^{\text{H}}[f,f],
\eeq
where use has been made of Eq.\ \eqref{2.3}. In addition, for homogeneous states, Eq.\ \eqref{1.22}  gives the following form for the cooling rate
\begin{equation}\label{2.5}
\begin{aligned}
\zeta(t)={}&-\frac{m\sigma\chi}{nT}
\int \dd\mathbf v_1\int \dd\mathbf v_2
\Bigg[g\overline\Delta^{\,2}
+\frac{\pi}{4}\alpha\overline\Delta g^2\\
&\hspace{26mm}-\frac{1}{6}(1-\alpha^2)g^3\Bigg]
 f(\mathbf v_1,t)f(\mathbf v_2,t).
\end{aligned}
\end{equation}
where use has been made of the result \cite{NE98}
\label{2.5.1}
\begin{align}
B_k&\equiv\int \dd\widehat{\boldsymbol\sigma}\,
\Theta(\widehat{\boldsymbol\sigma}\cdot\mathbf g)
(\widehat{\boldsymbol\sigma}\cdot\widehat{\mathbf g})^k \nonumber \\
&=\sqrt{\pi}\,
\frac{\Gamma\!\left[(k+1)/2\right]}
{\Gamma\!\left(k/2+1\right)},
\qquad \widehat{\mathbf g}=\mathbf g/g,\quad k\geq0.
\end{align}
Now, dimensional analysis suggests that the solution to Eq. \eqref{2.4} can be written in the scaled form \cite{BGMB13,BMGB14,GBS18}
\beq
\label{2.8}
f(\mathbf{v},\gamma,\overline{\Delta}, \phi)=n v_\text{th}^{-2}\varphi(\mathbf{c},\gamma^*, \overline{\Delta}^*, \phi)\equiv n v_\text{th}^{-2}\varphi(\mathbf{c},\gamma^*,\Delta^*, \tau_\text{s}^*, \phi),
\eeq
where $\mathbf{c}\equiv \mathbf{v}/v_\text{th}$ is a dimensionless velocity,
$v_\text{th}=\sqrt{2T/m}$ being the thermal speed. In Eq.\ \eqref{2.8}, the reduced drag $\gamma^*$ and the reduced parameters $\Delta^*$ and $\overline{\Delta}^*$ are defined as
\begin{equation} \label{eq:36}
    \gamma^*=\frac{\ell\gamma}{v_\text{th}}, \qquad \Delta^* = \frac{\Delta}{v_\text{th}}, \qquad \overline{\Delta}^* = \Delta^*\left(1-\text{e}^{-2\omega_\text{e}^*\tau_\text{s}^*}\right),
\end{equation}
where $\ell=1/(n \sigma)$ is proportional to the mean free path of hard disks. Moreover, we define the (reduced) quantities
\begin{equation} \label{eq:37}
    \omega_\text{e}^*(\phi) = \frac{\ell \omega_\text{e}}{v_\text{th}} =  \sqrt{2\pi}\chi(\phi), \qquad \tau_\text{s}^* = \frac{v_\text{th}}{\ell} \tau_\text{s}.
\end{equation}
In the absence of synchronization and for non-dragged systems, the consistency of the (scaled) solution \eqref{2.8} has been tested against computer simulations for monocomponent \cite{BGMB13,BMGB14} and multicomponent \cite{BSG20} granular fluids. We assume that this consistency remains valid when both synchronization and drag are present. To express Eq.\ \eqref{2.4} in dimensionless form, we first note that
\beq
\label{2.10}
T\frac{\partial f}{\partial T} = -\frac{n}{2}v_\text{th}^{-2} \left[ \frac{\partial}{\partial \mathbf{c}} \cdot (\mathbf{c}\varphi) -2 \frac{\partial \varphi}{\partial \theta} \right],
\eeq
where we have introduced the operator
\begin{equation}
\frac{\partial}{\partial \theta} := -\frac{1}{2}\left(\gamma^* \frac{\partial }{\partial \gamma^*} + \Delta^* \frac{\partial }{\partial \Delta^*} -\tau_\text{s}^* \frac{\partial}{\partial \tau_\text{s}^*}\right),
\end{equation}
which accounts for the implicit temperature dependence of the scaled distribution $\varphi$ through $\gamma^*$, $\Delta^*$, and $\tau_\text{s}^*$. Accordingly, Eq.\ \eqref{2.4} can be rewritten as
\begin{equation}
\label{2.11}
\frac{\Lambda}{2}\left[ \frac{\partial}{\partial \mathbf{c}} \cdot (\mathbf{c}\varphi) -2\frac{\partial\varphi}{\partial \theta} \right]
-\gamma^*\frac{\partial}{\partial\mathbf{c}}\cdot(\mathbf{c}\varphi) = J_{\text{E}}^*[\varphi,\varphi],
\end{equation}
where $\Lambda\equiv 2\gamma^* + \zeta^*$, $\zeta^*=\ell \zeta/v_\text{th}$, and
$J_{\text{E}}^*=\ell v_\text{th}J_{\text{E}}^\text{H}/n$. In the steady state, $\Lambda=0$.

Since the distribution function $f$ is isotropic in velocity space in the homogeneous state, the heat flux vanishes ($\mathbf q=\mathbf 0$) and the pressure tensor is isotropic, i.e., $P_{ij}=p\delta_{ij}$. According to Eqs.\ \eqref{1.17} and \eqref{1.19}, the (reduced) hydrostatic pressure $p^* = p/(nT)$ is given by
\begin{align}
\label{2.11.2}
p^* = 1+\chi\phi\left[
1+\al+\frac{4}{\pi}\overline{\Delta}^*
\int \dd\mathbf{c}_1\int \dd\mathbf{c}_2\;
g^* \,\varphi(\mathbf{c}_1)\varphi(\mathbf{c}_2)
\right].
\end{align}

Although the exact form of $\varphi$ is not known, information about its deviation from the Maxwellian distribution can be obtained from the kurtosis, or fourth cumulant,
\begin{equation}
\label{2.12}
a_{2}=\frac{1}{2}\int\; \dd{\bf c}\; c^4 \varphi(c)-1.
\end{equation}
The cumulant $a_{2}$ measures the deviation of $\varphi$ from its Maxwellian form $\pi^{-1}e^{-c^2}$. The coefficient $a_2$ can be obtained by multiplying Eq.\ \eqref{2.11} by $c^4$ and integrating over velocity. The result is
\beq
\label{2.13}
-\frac{\Lambda}{2}\left(1+a_2+\frac{1}{2}\frac{\partial a_2}{\partial \theta}\right)+\gamma^*\left(1+a_2\right)=\frac{\mu_4}{8},
\eeq
where
\beq
\label{2.14}
\mu_4=\int \dd\mathbf{c}\; c^4\; J_{\text{E}}^*[\varphi,\varphi].
\eeq

To solve the differential equation \eqref{2.13} one needs to know $\zeta^*$ and $\mu_4$. Since these quantities are defined in terms of the unknown scaled distribution $\varphi(\mathbf{c})$, one may estimate them by using the first Sonine approximation to $\varphi(\mathbf{c})$:
\beq
\label{2.15}
\varphi(\mathbf{c}) \approx \pi^{-1}e^{-c^2}\Bigg[1+\frac{a_2(\gamma^*,\overline{\Delta}^*)}{2}\left(c^4-4c^2+2\right)\Bigg].
\eeq
The reduced cooling rate $\zeta^*$ and the collisional fourth moment $\mu_4$ can be evaluated within the Sonine approximation \eqref{2.15}. For the conventional $\Delta$-model, their explicit expressions were derived in Ref.\ \onlinecite{BMGB14}. For the homogeneous collision operator considered here, synchronization enters the collisional moments through the replacement $\Delta^*\to\overline{\Delta}^*$. Hence, the Sonine expressions previously obtained for the conventional $\Delta$-model can be directly adapted to the present homogeneous problem by carrying out this replacement. Therefore, neglecting quadratic terms in $a_2$, $\zeta^*$ and $\mu_4$ can be written as
\beq
\label{2.16}
\zeta^*\to \zeta_0^{(0)}+\zeta_0^{(1)}a_2, \quad \mu_4\to \mu_4^{(0)}+\mu_4^{(1)}a_2,
\eeq
where
\begin{align}\label{2.17}
\zeta_0^{(0)}={}&\sqrt{2\pi}\,\chi
\left(\frac{1-\alpha^2}{2}-\sqrt{\frac{\pi}{2}}\alpha\overline\Delta^*
-\overline\Delta^{*2}\right),\\ \label{2.17b}
\zeta_0^{(1)}={}&\frac{\sqrt{2\pi}}{16}\,\chi
\left[\frac{3}{2}(1-\alpha^2)+\overline\Delta^{*2}\right].
\end{align}
\beq
\label{2.18}
\mu_4^{(0)}=\frac{C_0}{A_0}\,\zeta_0^{(0)}, 
\qquad 
\mu_4^{(1)}=
\left(\frac{4 A_1+C_1}{A_0}\right)\zeta_0^{(0)}-4\zeta_0^{(1)}.
\eeq
The explicit forms of the quantities $A_0$, $A_1$, $C_0$ and $C_1$ are displayed in the Appendix \ref{ap:a1}.

\begin{figure*}[t]
\centering
\fullwidthgraphic{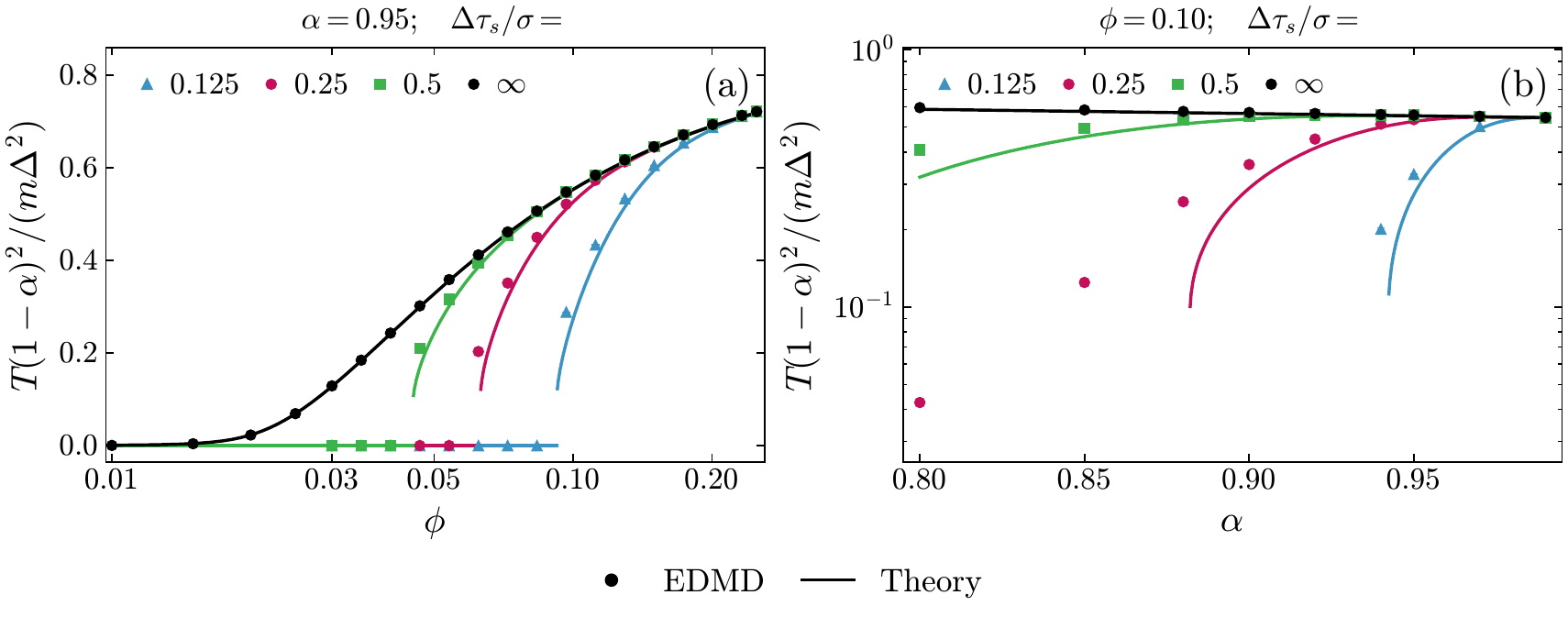}
\caption{Scaled steady granular temperature $T(1-\alpha)^2/(m\Delta^2)$ as a function of (a) the packing fraction $\phi$ for $\alpha=0.95$ and (b) the coefficient of restitution $\alpha$ for $\phi=0.10$. Symbols are EDMD simulation results for $\Delta\tau_{\rm s}/\sigma=0.125$, $0.25$, $0.5$, and $\infty$. Solid lines correspond to the theoretical predictions obtained within the synchronization mean-field approximation. In all cases, $\gamma\sigma/\Delta = 0.04$.}
\label{fig:temperature}
\end{figure*}

\subsection{Homogeneous steady state}

We now focus on the finite-temperature homogeneous steady state corresponding to the active branch of the model. Within the mean-field synchronization closure, the steady state results from the balance between collisional injection, inelastic dissipation, and viscous drag. In the Maxwellian approximation to $f$ (i.e., $a_2=0$), this balance reduces to the homogeneous mean-field description introduced in Ref.~\onlinecite{MPSF25}. The Sonine correction to $f$ considered here provides the non-Gaussian refinement used in the subsequent transport calculation. In the steady state ($\Lambda=0$), Eq.~\eqref{2.13}, consistently linearized in $a_2$, yields
\begin{equation} \label{eq:47}
a_\text{2,ss} = - \dfrac{\zeta_0^{(0)} + \frac{1}{4}\mu_4^{(0)}}{\zeta_0^{(0)} + \zeta_0^{(1)} +  \frac{1}{4}\mu_4^{(1)}}.
\end{equation}
To obtain an analytical estimate of the steady temperature, we determine the stationary value of \(\overline{\Delta}^*\) by considering the Maxwellian approximation to $f$,  and hence neglect the $a_2$ correction
in the steady-state energy balance.  Thus, \(\zeta^*\simeq\zeta_0^{(0)}\), and the condition \(\Lambda=0\) becomes \(\gamma_{\rm ss}^*=-\zeta_0^{(0)}/2\). Here, the subscript ‘‘$\rm ss$'' means that the quantity is evaluated in the steady state. Under the above approximation, $\overline{\Delta}^*$ obeys the quadratic equation
\begin{equation}
    \overline{\Delta}^{*2} + \sqrt{\frac{\pi}{2}}\alpha \overline{\Delta}^* - \left(\frac{2\gamma_{\rm ss}^*}{\sqrt{2\pi}\chi} + \frac{1-\al^2}{2}\right) = 0,
\end{equation}
whose physical solution \footnote{The other solution would lead to $\overline{\Delta}_\text{ss}^* < 0$.} is
\begin{equation}
\label{vic1}
    \overline{\Delta}^*_\text{ss}
    =\frac{-\sqrt{\frac{\pi}{2}}\,\alpha\ +\ 
    \sqrt{\frac{\pi}{2}\alpha^2+2(1-\alpha^2)+\frac{8\gamma_{\rm ss}^*}{\sqrt{2\pi}\,\chi}}}{2}.
\end{equation}
When $\gamma_{\rm ss}^*=0$ and $\tau_\text{s}^* \to \infty$, Eq.\ \eqref{vic1} reduces to the expression of $\Delta^*_\text{ss}$ obtained in the conventional $\Delta$-model.\cite{GBS18}
Equation \eqref{vic1} allows us to determine the reduced steady-state temperature (defined with respect to the energy scale \(m\Delta^2/2\)) as the solution of the following implicit equation:
\begin{align}
\label{tem_ss}
    T_\text{ss}^*& = \left(1-\text{e}^{-2\omega_\text{e}^*\tau_\text{s,ss}^*}\right)^2{\overline{\Delta}_\text{ss}^{*-2}} \nonumber \\
    &= \left[\frac{2\left(1-\text{e}^{-2\omega_\text{e}^*\tau_\text{s,ss}^*}\right)}
    {-\sqrt{\frac{\pi}{2}}\,\alpha
    +\sqrt{\frac{\pi}{2}\alpha^2+2(1-\alpha^2)+\frac{8\gamma_\text{ss}^*}{\sqrt{2\pi}\,\chi}}}\right]^2,
    \end{align}
where \(\gamma_{\rm ss}^*=\gamma^*(T_{\rm ss}^*)\) and \(\tau_{\rm s,ss}^*=\tau_{\rm s}^*(T_{\rm ss}^*)\). Hence, the above relation is implicit in \(T_{\rm ss}^*\), since \(\gamma^*\propto T^{-1/2}\) whereas \(\tau_{\rm s}^*\propto T^{1/2}\). It is worth noting that this implicit equation does not necessarily admit a positive solution for arbitrary model parameters. In such a case, the system falls into an absorbing state $T_{\rm ss}=0$ because collisions do not inject enough energy into the system.
Note that all the quantities appearing in Eq.\ \eqref{tem_ss} are evaluated in the steady state. In addition, $p^*$ can be estimated by introducing Sonine approximation \eqref{2.15} into Eq.\ \eqref{2.11.2}. Up to linear order in $a_2$, the result is
\begin{equation}
\label{vic2}
    p^*=1+\chi\phi\left[1+\al + 2\sqrt{\frac{2}{\pi}} \left(1-\frac{1}{16}a_2\right)\overline{\Delta}^* \right].
\end{equation}

\begin{figure*}[t]
\centering
\fullwidthgraphic{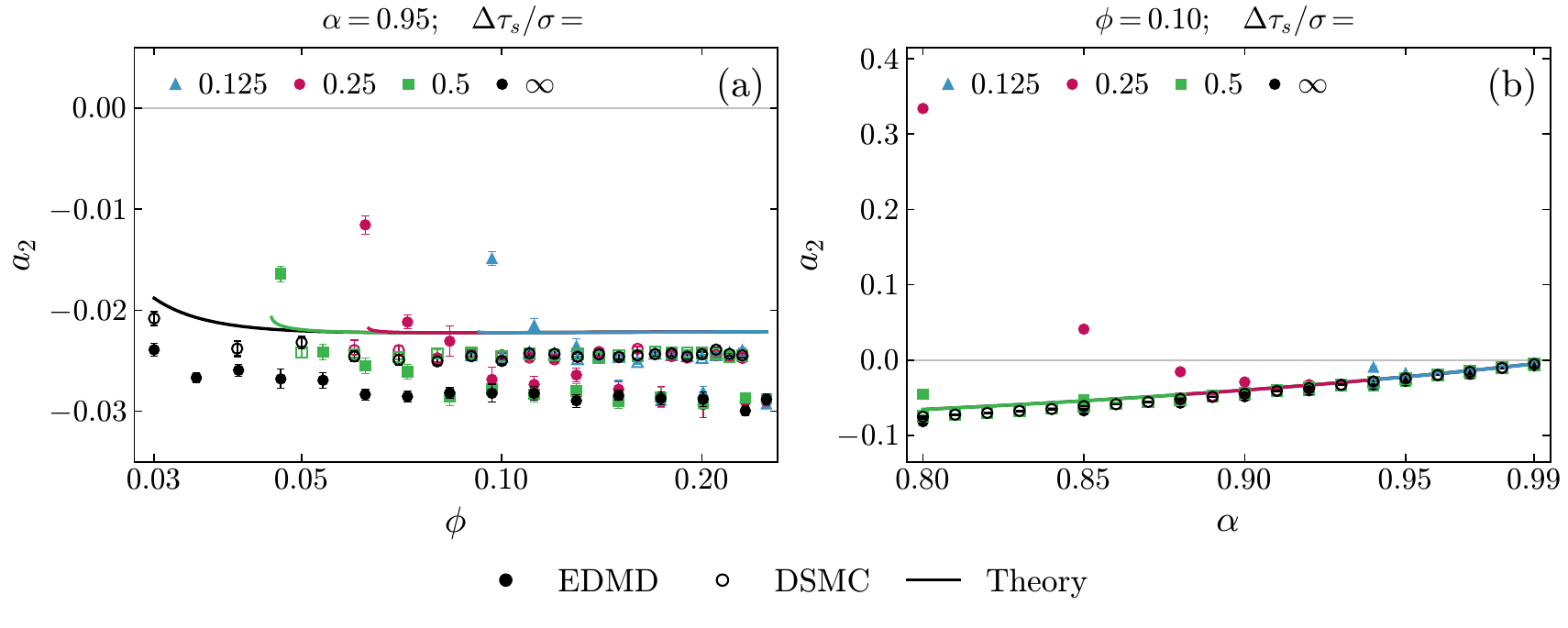}
\caption{Steady-state fourth cumulant $a_2$ as a function of (a) the packing fraction $\phi$ for fixed $\alpha=0.95$ and (b) the coefficient of restitution $\alpha$ for fixed $\phi=0.10$. Symbols denote EDMD and DSMC simulation results, while solid lines correspond to the analytical kinetic-theory predictions within the synchronization mean-field approximation. The theoretical and DSMC results remain close over most of the active branch and predict small values of $|a_2|$, consistently with the linear-in-$a_2$ treatment adopted throughout the theory. Close to the synchronization-controlled transition, however, the microscopic EDMD simulations exhibit a pronounced increase of $a_2$. The origin of this discrepancy, associated with the mean-field suppression of fluctuations of the collision-dependent injection amplitude and with neglected velocity--synchronization correlations, is analyzed in Appendix~\ref{app:a2deviations}. In all cases, the parameters are the same as in Fig~\ref{fig:temperature}. The DSMC code employed here is an adaptation of the implementation reported in Ref.~\onlinecite{GGG26}.}
\label{fig:a2}
\end{figure*}

To close our equations, we approximate the pair correlation function at contact in the granular gas by Henderson's equilibrium expression~\cite{henderson1975harddiscs},
\begin{equation}
\label{1.12.1}
\chi(\phi)=\frac{1-\frac{7}{16}\phi}{(1-\phi)^2}.
\end{equation}

We then test our theory by performing event-driven molecular dynamics (EDMD)
simulations using the microscopic collision rule Eq.~\eqref{1.23}, as
described in Appendix~\ref{app:EDMD}. The results are presented in
Fig.~\ref{fig:temperature}, which compares the (scaled) steady granular
temperature $T(1-\alpha)^2/(m\Delta^2)$ obtained from the synchronization
mean-field theory with these microscopic EDMD simulations.
Overall, the comparison shows that the theory captures the influence of
synchronization on the homogeneous active state and, in particular, the
progressive displacement of the low-temperature edge of the active branch as
the synchronization time is varied. The synchronization effect becomes more
pronounced as $\Delta\tau_{\rm s}/\sigma$ decreases. A shorter synchronization
time allows the particles to reach the synchronized state more rapidly, i.e.,
to lock faster with the vibrating plates in the underlying confined system,
thereby suppressing collisional energy injection more efficiently. As a
consequence, at fixed $\alpha$ the active branch is progressively restricted
to larger packing fractions, whereas at fixed $\phi$ it is restricted to
larger values of the coefficient of restitution, closer to the elastic limit.
Equivalently, as the density is decreased and the collision frequency becomes
comparable to the inverse synchronization time, synchronization is increasingly
favored, the effective injection amplitude is reduced, and the steady
temperature departs from the nonsynchronized limit and progressively decreases
until the finite-temperature active branch terminates at the low-temperature
edge associated with the absorbing transition. Conversely, when the collision
frequency is sufficiently large, collisions continuously disrupt
synchronization, $\overline{\Delta}$ approaches $\Delta$, and the system tends
toward the behavior of the conventional $\Delta$-model. The same limit is
recovered for $\tau_{\rm s}\to\infty$, for which synchronization is completely
suppressed. 

Although the mean-field theory reproduces these trends rather well, small but systematic discrepancies with EDMD emerge close to the termination of the active branch. Their origin can be further assessed by numerically solving the homogeneous Boltzmann--Enskog equation while retaining the synchronization dynamics beyond the present mean-field closure. This calculation improves the agreement with EDMD with respect to the current mean-field description, indicating that the replacement of the collision-dependent injection amplitude by $\overline{\Delta}_{ij}$ constitutes an important source of the discrepancy. In particular, this closure neglects fluctuations of the binary injection amplitude and its correlations with the particle velocities by approximating
$\langle \mathbf v^n\Delta_{ij}^m\rangle_{\rm coll}
\simeq\langle \mathbf v^n\rangle\overline{\Delta}_{ij}^m$, or, more simply,
$\langle\Delta_{ij}^m\rangle_{\rm coll}
\simeq\langle\Delta_{ij}\rangle_{\rm coll}^m$. However, the homogeneous Boltzmann--Enskog solution still does not fully reproduce the EDMD results. This remaining discrepancy points to effects beyond the homogeneous Boltzmann--Enskog description, such as departures from molecular chaos and spatial correlations associated with the formation of synchronized and asynchronized regions. These effects, together with the role of velocity--synchronization correlations, are examined in Appendix~\ref{app:a2deviations}.

To close the study of the homogeneous state, Figure~\ref{fig:a2} compares the steady-state fourth cumulant $a_2$ predicted by the kinetic theory with those obtained by numerically solving the Enskog equation by means of the Direct Simulation Monte Carlo (DSMC) method\cite{B94} and from microscopic molecular dynamics (EDMD) simulations. Over most of the active branch, the theoretical values of $|a_2|$ remain small, and the DSMC results obtained with the same effective mean-field collision rule closely follow the analytical prediction. This supports the consistent approximation adopted throughout the kinetic calculation, in which terms beyond linear order in $a_2$ are neglected. Since the main purpose of the present work is to develop an analytical transport theory, we shall retain this linear-in-$a_2$ description in the remainder of the paper.

Close to the synchronization-controlled transition, however, the microscopic EDMD simulations exhibit a pronounced increase of $a_2$, which is not reproduced either by the kinetic theory or by the corresponding DSMC simulations. As discussed in detail in Appendix~\ref{app:a2deviations}, this discrepancy is again due to the mean-field approximation rather than the linear Sonine treatment. Indeed, $a_2$ and other quantities related to higher moments of the velocity distributions are highly sensitive to approximations such as $\langle \Delta_{ij}^m\rangle_{\rm coll}\simeq \langle \Delta_{ij}\rangle_{\rm coll}^m$, which neglect some sources of correlations and fluctuations, both becoming increasingly important as the transition is approached. 


\section{Chapman--Enskog method. First order approximation}\label{sec:chapman-enskog}

Having characterized the homogeneous state, we now extend the analysis to weakly inhomogeneous states. We solve the Enskog equation~\eqref{1.10} by means of the Chapman--Enskog method~\cite{CC70} (conveniently adapted to dissipative dynamics), following the same general strategy as in previous kinetic-theory analyses of confined and driven granular fluids~\cite{GGG19a,GBS18,GGG26}. We assume that, after a short kinetic transient, the velocity distribution becomes a normal solution whose space and time dependence occurs only through the hydrodynamic fields $n$, $\mathbf U$, and $T$. Synchronization remains slaved to these fields through the pairwise closure~\eqref{1.25} and therefore does not introduce an additional hydrodynamic variable.

As usual, for states with sufficiently small spatial gradients, the velocity distribution is expanded as
\beq
\label{3.0}
f=f^{(0)}+f^{(1)}+\cdots,
\eeq
where $f^{(k)}$ is of $k$th order in the gradients. Analogous expansions are introduced for the fluxes, the cooling rate, and the Enskog collision operator:
\beq
\label{3.0.1}
\mathsf{P}=\mathsf{P}^{(0)}+\mathsf{P}^{(1)}+\cdots,
\quad
\mathbf{q}=\mathbf{q}^{(0)}+\mathbf{q}^{(1)}+\cdots,
\eeq
\beq
\label{3.0.2}
\zeta=\zeta^{(0)}+\zeta^{(1)}+\cdots,
\quad
J_\text{E}=J_\text{E}^{(0)}+J_\text{E}^{(1)}+\cdots.
\eeq
Moreover, the time derivative operator $\partial_t$ is also expanded as
\beq
\label{3.0.3}
\partial_t=\partial_t^{(0)}+\partial_t^{(1)}+\cdots.
\eeq
As in Refs.~\onlinecite{GGG19a,GGG26}, the friction coefficient $\gamma$ is taken to be of zeroth order in gradients because the drag acts locally and remains finite in a homogeneous state. Since the homogeneous reference state is comoving with the walls, the slip velocity is assigned first order, $\hat{\bf U}=\mathbf U-\mathbf U_{\rm w}=\mathcal O(\bm\nabla)$. Accordingly, the zeroth-order slip velocity vanishes and departures from the wall-comoving state enter only at first order. In the following we retain terms through Navier--Stokes order (first-order in the spatial gradients).

A point that requires special care in the present problem is that the effective collision rule is intrinsically pair dependent. At zeroth order in spatial gradients, the spatial separation of the two colliding
particles cannot be resolved at the hydrodynamic level, so that both sample
the same local state. The pairwise amplitude therefore reduces to the local version of the homogeneous quantity introduced in Sec.~\ref{sec3}, i.e., 
\begin{equation}
\label{eq:barDelta-local-CE}
\overline{\Delta}_{ij}^{(0)} \to \overline{\Delta}(\mathbf r,t)
=\Delta\left(1-\mathrm e^{-2\omega_{\rm e}(\mathbf r,t)\tau_{\rm s}}\right).
\end{equation}
Moreover, at this order, the hydrodynamic fields obey
\begin{equation}
\label{3.0.5}
\partial_t^{(0)}n=0,
\qquad
\partial_t^{(0)}\mathbf U=\mathbf 0,
\end{equation}
\beq
\label{3.0.6}
\partial_t^{(0)}T=-(2\gamma+\zeta^{(0)})T,
\eeq
where the zeroth-order cooling rate $\zeta^{(0)}$ is obtained from Eq.~\eqref{1.22} by retaining zeroth-order quantities. An approximate form of $\zeta_0^*=\ell\zeta^{(0)}/v_\text{th}$ is given by Eqs.~\eqref{2.16} and \eqref{2.17}. In addition,
the corresponding zeroth-order distribution function $f^{(0)}$ is the local version of the homogeneous normal solution characterized in Sec.~\ref{sec3}:
\beq
\label{3.0.7}
f^{(0)}(\mathbf r,\mathbf v;t)
=
n(\mathbf r;t)v_\text{th}^{-2}(\mathbf r;t)
\varphi\left(\mathbf c,\gamma^*,\Delta^*,\tau_\text{s}^*,\phi\right).
\eeq
At this order the synchronized collision operator therefore has exactly the homogeneous structure discussed in Sec.~\ref{sec3}, evaluated at the local hydrodynamic fields. The non-Gaussianity of $f^{(0)}$ is characterized by the fourth cumulant $a_2$. The derivatives $\Upsilon_\theta\equiv\partial_\theta a_2$, $\Upsilon_{\gamma^*}\equiv\partial_{\gamma^*}a_2$, $\Upsilon_\chi\equiv\partial_\chi a_2$, and $\Upsilon_{\overline{\Delta}^*}\equiv\partial_{\overline{\Delta}^*}a_2$, evaluated at the steady state, enter the first-order solution and are given in Appendix~\ref{app:CEdetails}. Moreover, $\mathbf{q}^{(0)}=\mathbf{0}$ and $P_{ij}^{(0)}=p\delta_{ij}$, where a good estimate of the hydrostatic pressure $p$ is given by Eq.\ \eqref{vic2}.

To first order, as in the conventional $\Delta$-model,\cite{GBS18} the distribution function $f^{(1)}$ can be written as
\begin{align}\label{3.1}
f^{(1)}(\mathbf V)={}&\bcal{A}(\mathbf V)\cdot\bm\nabla\ln T
+\bcal{B}(\mathbf V)\cdot\bm\nabla\ln n \nonumber \\
&+\frac{1}{2}\mathcal C_{ij}(\mathbf V)
\left(\frac{\partial U_i}{\partial r_j}+\frac{\partial U_j}{\partial r_i}
-\delta_{ij}\bm\nabla\cdot\mathbf U\right) \nonumber \\
&+\mathcal D(\mathbf V)\bm\nabla\cdot\mathbf U.
\end{align}
The unknowns $\bcal{A}$, $\bcal{B}$, $\mathcal C_{ij}$, and $\mathcal D$ satisfy a set of coupled linear integral equations. In the steady state ($\Lambda=0$), these integral equations are given by Eqs.\ \eqref{3.2}--\eqref{3.5}.

There are two distinct ways in which synchronization enters the first-order problem. The first is the local state dependence of the zeroth-order reference state through $\overline{\Delta}^*$. At fixed density,
\begin{equation}
\label{eq:new1}
T\left(\frac{\partial\ln\overline{\Delta}^*}{\partial T}\right)_n
=-\frac{1}{2}
+\frac{\omega_\text{e}^*\tau_\text{s}^*}
{\mathrm e^{2\omega_\text{e}^*\tau_\text{s}^*}-1},
\end{equation}
whereas, at fixed temperature,
\begin{equation}
\label{eq:new2}
n\left(\frac{\partial\ln\overline{\Delta}^*}{\partial n}\right)_T
=
\frac{2\omega_\text{e}^*\tau_\text{s}^*}
{\mathrm e^{2\omega_\text{e}^*\tau_\text{s}^*}-1}
\left(1+\phi\frac{\partial\ln\chi}{\partial\phi}\right).
\end{equation}
These derivatives account for the variation of the local homogeneous reference state when $T$ or $n$ changes.

The second effect is genuinely associated with the finite separation of the colliding pair. For the Enskog pair $(\mathbf r,\mathbf r+\boldsymbol\sigma)$, Taylor expansion of the local collision frequency $\omega_\text{e}(\mathbf r+\boldsymbol\sigma,t)$ gives the result
\begin{equation}
\omega_\text{e}(\mathbf r+\boldsymbol\sigma,t)
=
\omega_\text{e}(\mathbf r,t)
+\sigma\widehat{\boldsymbol\sigma}\cdot\bm\nabla\omega_\text{e}(\mathbf r,t)
+\mathcal O(\bm\nabla^2).
\end{equation}
Insertion into Eq.~\eqref{1.25} yields
\begin{align}
\label{eq:pairwise-delta-expansion}
\overline{\Delta}_{ij}(\mathbf r,\mathbf r+\boldsymbol\sigma,t)
={}&\overline{\Delta}(\mathbf r,t)
+\overline{\Delta}_{ij}^{(1)}(\mathbf r,\widehat{\boldsymbol\sigma},t)
+\mathcal O(\bm\nabla^2),\\
\label{eq:pairwise-delta-firstorder}
\overline{\Delta}_{ij}^{(1)}
={}&
\sigma\overline{\Delta}
\frac{\omega_\text{e}\tau_\text{s}}
{\mathrm e^{2\omega_\text{e}\tau_\text{s}}-1}
\widehat{\boldsymbol\sigma}\cdot\bm\nabla\ln\omega_\text{e}.
\end{align}
For hard disks,
\begin{equation}
\label{eq:grad-omegae}
\bm\nabla\ln\omega_\text{e}
=
\frac{1}{2}\bm\nabla\ln T
+\left(1+\phi\frac{\partial\ln\chi}{\partial\phi}\right)
\bm\nabla\ln n,
\end{equation}
so that Eq.~\eqref{eq:pairwise-delta-firstorder} can also be written as
\begin{align}
\label{eq:pairwise-delta-firstorder-fields}
\overline{\Delta}_{ij}^{(1)}
=
\sigma\overline{\Delta}
\frac{2\omega_\text{e}\tau_\text{s}}
{\mathrm e^{2\omega_\text{e}\tau_\text{s}}-1}
&\Bigg[
\frac{1}{4}\widehat{\boldsymbol\sigma}\cdot\bm\nabla\ln T
\nonumber \\
&+\frac{1}{2}
\left(1+\phi\frac{\partial\ln\chi}{\partial\phi}\right)
\widehat{\boldsymbol\sigma}\cdot\bm\nabla\ln n
\Bigg].
\end{align}
Equations~\eqref{eq:new1}--\eqref{eq:new2} and \eqref{eq:pairwise-delta-firstorder-fields} describe different pieces of the same synchronization feedback and must both be retained. The former enter through the hydrodynamic derivatives of the local zeroth-order state, whereas the latter arises from the explicit dependence of the collision rule on the two positions of the colliding pair. As expected, in the limit $\tau_{\rm s}\to\infty$ the synchronization-induced corrections vanish: Eq.~\eqref{eq:new1} reduces to
the conventional $\Delta$-model result
$T(\partial_T\ln\Delta^*)_n=-1/2$, Eq.~\eqref{eq:new2} vanishes, and the
explicit pairwise contribution
\eqref{eq:pairwise-delta-firstorder-fields} disappears. In the additional
absence of drag ($\gamma=0$), the results of the conventional
$\Delta$-model are recovered.\cite{GBS18} 

As mentioned above, substitution of \(f^{(1)}\) into the Enskog kinetic equation leads to a set of coupled linear integral equations for the coefficients \(\bcal{A}\), \(\bcal{B}\), \(\mathcal C_{ij}\), and \(\mathcal D\). Their derivation follows the usual Chapman--Enskog procedure employed in the conventional $\Delta$-model.\cite{GBS18} However, in contrast to the results given in Ref.\ \onlinecite{GBS18}, when Eq.~\eqref{eq:pairwise-delta-firstorder-fields} is inserted into the first-order Enskog operator, it generates an additional vector source, which we will denote by $\bcal{R}$. Since $\overline{\Delta}_{ij}^{(1)}$ is proportional only to $\bm\nabla T$ and $\bm\nabla n$, this contribution modifies the integral equations for $\bcal{A}$ and $\bcal{B}$, but does not enter directly into the integral equations of the tensorial quantity $\mathcal C_{ij}$ and the scalar quantity $\mathcal D$. Technical details of the Chapman--Enskog expansion to first order in spatial gradients are provided in Appendix~\ref{app:CEdetails}.


It is also useful to distinguish the kinetic contributions to the fluxes from their corresponding collisional transfer contributions. In Eqs.~\eqref{1.19} and \eqref{1.20}, the pair amplitude is sampled along the collision link, with centers at $\mathbf r-x\boldsymbol\sigma$ and $\mathbf r+(1-x)\boldsymbol\sigma$. Its first-order expansion is proportional to $1-2x$, and therefore its new direct contribution to the line integral vanishes because $\int_0^1\dd x\,(1-2x)=0$. Consequently, at Navier--Stokes order the collisional-transfer expressions retain the same formal structure as in the conventional $\Delta$-model after the replacement $\Delta\to\overline{\Delta}$. The heat flux vector nevertheless inherits the correction to $\bcal{A}$ and $\bcal{B}$ through the kinetic contributions to the transport coefficients. The detailed argument is also given in Appendix~\ref{app:CEdetails}.

The resulting first-order solution determines the Navier--Stokes constitutive relations presented in the following section.

\section{Navier--Stokes transport coefficients}\label{sec:transport}

The first-order Chapman--Enskog solution \eqref{3.1} yields the Navier--Stokes constitutive equations for the pressure tensor 
\beq 
\label{4.1}
P_{ij}^{(1)}=-\eta \left(\frac{\partial U_i}{\partial r_j}+\frac{\partial U_j}{\partial r_i}-\delta_{ij}\bm{\nabla}\cdot\mathbf{U}\right)
-\eta_\text{b}\delta_{ij}\bm{\nabla} \cdot \mathbf{U},
\eeq
the heat flux 
\begin{equation} \label{eq:heat-flux}
\mathbf{q}^{(1)} = -\kappa \bm{\nabla} T - \mu\bm{\nabla} n,
\end{equation}
and the cooling rate
\begin{equation}
\label{3.13}
\zeta^{(1)}=\zeta_U\bm{\nabla} \cdot {\bf U}.
\end{equation}
Here, \(\eta\) and \(\eta_{\mathrm b}\) are the shear and bulk viscosities, respectively, \(\kappa\) is the thermal conductivity, \(\mu\) is the diffusive heat coefficient, and \(\zeta_U\) is the first-order contribution to the cooling rate.

These constitutive relations have the same hydrodynamic structure as those employed in Ref. \onlinecite{MPSF25}. However, in that work the transport coefficients were approximated by their equilibrium hard-disk expressions. In the present paper the transport coefficients are obtained consistently from the synchronized Enskog kinetic equation and therefore incorporate the combined effects of inelasticity, viscous drag, finite density, and synchronization. The detailed derivation and the Sonine evaluation of the corresponding integral equations are also presented in Appendix \ref{app:CEdetails}. In the following, we only display the resulting forms for the transport coefficients. Unless otherwise stated, all quantities are evaluated at the homogeneous steady state.

\begin{figure*}[t]
\centering
\fullwidthgraphic{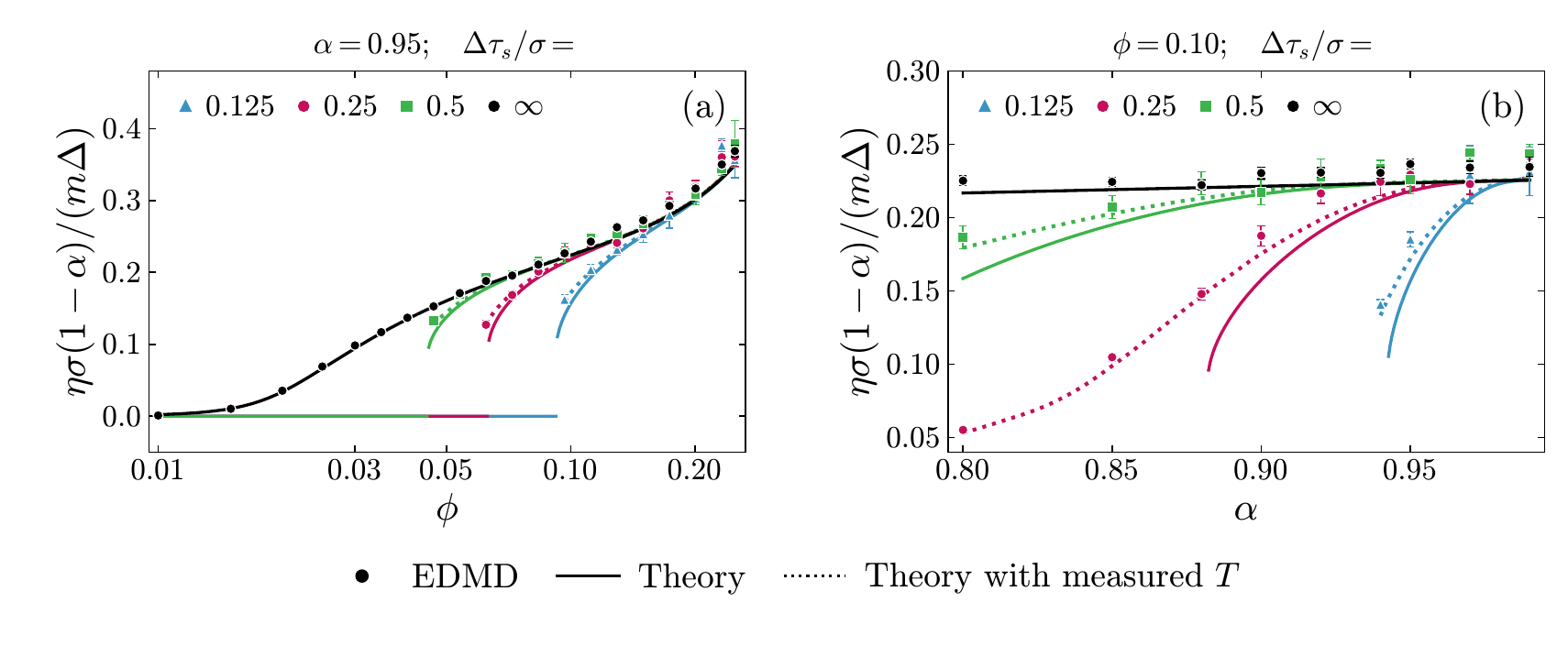}
\caption{Scaled shear viscosity $\eta\sigma(1-\alpha)/(m\Delta)$ as a function of (a) the packing fraction $\phi$ for $\alpha=0.95$ and (b) the coefficient of restitution $\alpha$ for $\phi=0.10$. Symbols are EDMD simulation results, solid lines correspond to the theoretical predictions obtained within the synchronization mean-field approximation, and dotted lines show the same theoretical expressions evaluated using the steady granular temperature measured in the simulations. In all cases, the parameters are the same as in Fig.~\ref{fig:temperature}.}
\label{fig:viscosity}
\end{figure*}

\subsection{Shear and bulk viscosities}

The shear viscosity has kinetic and collisional contributions,
\begin{equation}
    \eta = \eta_\text{k}+\eta_\text{c},
\end{equation}
whereas the bulk viscosity $\eta_\text{b}$ is purely collisional and therefore vanishes in the dilute limit. As shown in Appendix \ref{app:CEdetails}, the explicit pairwise variation of $\overline{\Delta}_{ij}$ does not generate an additional direct contribution to the tensorial or scalar parts at Navier--Stokes order. Consequently, no new synchronization-gradient term arises explicitly
in the shear or bulk viscosities. Hence, the shear and bulk viscosities retain the same formal structure as in the conventional $\Delta$-model after the replacement $\Delta\to\overline{\Delta}$, together with $a_2$ evaluated for the synchronization-controlled homogeneous
reference state.

The bulk viscosity $\eta_\text{b}$ can be written as
\beq
\label{4.6}
\eta_\text{b}=\eta_\text{b}'+\eta_\text{b}'',
\eeq
where
\begin{align}\label{4.7}
\eta_\text{b}^{\prime}=
\frac{32\sqrt{2\pi}}{\pi^2}\chi\phi^2
\left[\frac{1+\alpha}{4}\sqrt{\frac{\pi}{2}}
\left(1-\frac{a_2}{16}\right)
+\frac{\pi}{8}\overline\Delta^*\right]\eta_0,
\end{align}
\begin{align} \label{4.7b}
\eta_\text{b}^{\prime\prime}=
\frac{1}{2}\chi\phi\overline\Delta^*e_D^*
\left(1+\frac{15}{32}a_2\right)\eta_0.
\end{align}
The collisional contribution to the shear viscosity $\eta_\text{c}$ is
\beq
\label{4.5}
\eta_\text{c}= \chi\phi \left[\frac{1+\al}{2}+\frac{1}{\sqrt{2\pi}}
\overline{\Delta}^*\left(1+\frac{3}{32}\,a_2\right)
\right]\eta_\text{k}+\frac{1}{2}\eta_\text{b}',
\eeq
In Eqs.\ \eqref{4.7} and \eqref{4.7b},
\begin{equation}
    \eta_0=\frac{1}{2\sqrt{\pi}}\;\frac{\sqrt{mT}}{\sigma}
\end{equation}
is the low-density value of the shear viscosity for an ordinary gas of hard disks ($\alpha = 1$ and $\Delta = 0$). Here, \(e_D^*\equiv n\sigma v_{\rm th}e_D\), where \(e_D\) is the leading Sonine coefficient associated with the scalar function \(\mathcal D(\mathbf V)\) entering the first-order distribution function \eqref{3.1}. This coefficient is determined in Appendix \ref{ap:a4} from the integral equation for \(\mathcal D\).

Within the so-called first Sonine approximation for the transport coefficients, the kinetic contribution to the shear viscosity $\eta_\text{k}$ is
\begin{align}
\label{vic3}
\eta_\text{k}={}&\frac{\sqrt{2\pi}}{\nu_\eta^*+2\gamma^*}
\Bigg\{1-\frac{1}{4}\chi\phi(1+\alpha)(1-3\alpha) \nonumber\\
&+\chi\phi\overline\Delta^*
\left[\sqrt{\frac{2}{\pi}}(1+2\alpha)
\left(1-\frac{a_2}{16}\right)+\overline\Delta^*\right]\Bigg\}\eta_0.
\end{align}
where
\begin{align}
\nu_\eta^*={}&\frac{\sqrt{2\pi}}{8}\chi\Bigg[
(1+\alpha)(7-3\alpha)\left(1-\frac{a_2}{32}\right) \nonumber \\
&+2\sqrt{2\pi}(1-\alpha)\overline\Delta^*
-2\overline\Delta^{*2}\left(1+\frac{3a_2}{32}\right)\Bigg].
\end{align}

\begin{figure*}[t]
\centering
\fullwidthgraphic{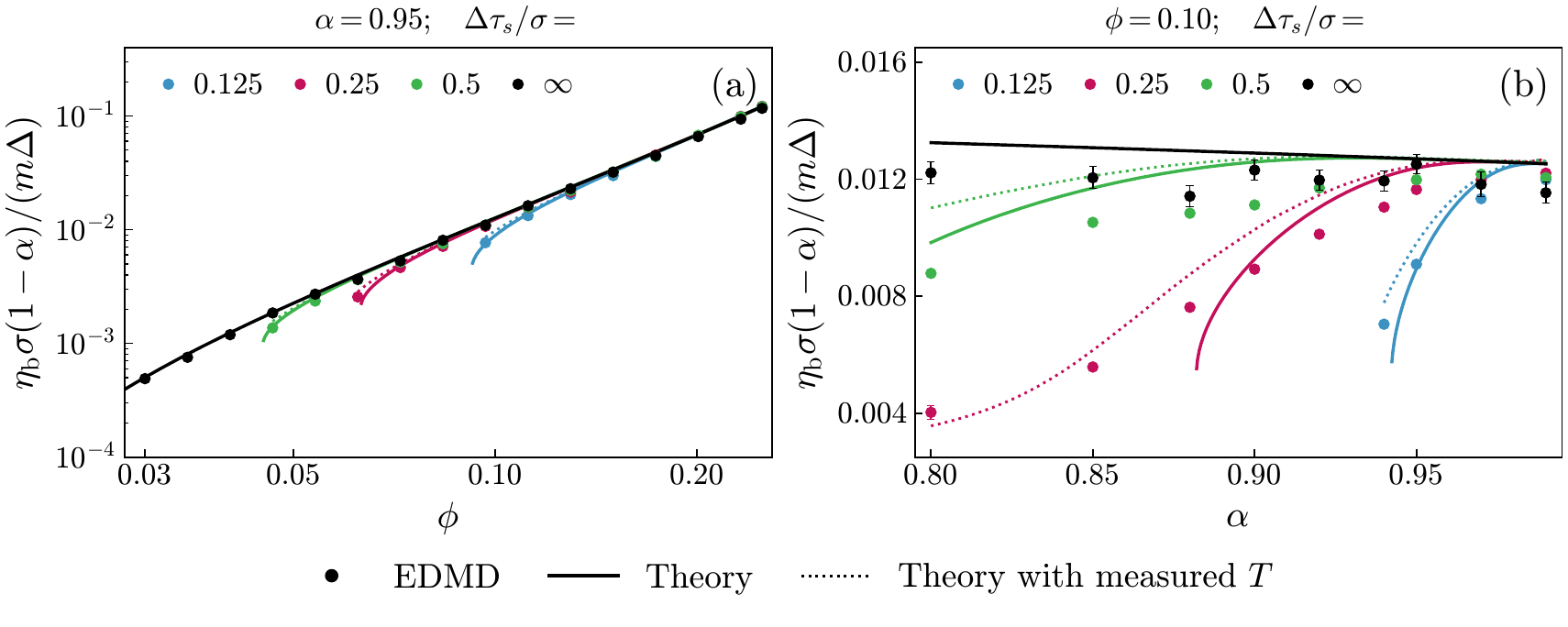}
\caption{Scaled bulk viscosity $\eta_{\rm b}\sigma(1-\alpha)/(m\Delta)$ as a function of (a) the packing fraction $\phi$ for $\alpha=0.95$ and (b) the coefficient of restitution $\alpha$ for $\phi=0.10$. Symbols are EDMD simulation results, solid lines correspond to the theoretical predictions obtained within the synchronization mean-field approximation, and dotted lines show the same theoretical expressions evaluated using the steady granular temperature measured in the simulations. In all cases, the parameters are the same as in Fig.~\ref{fig:temperature}.}
\label{fig:bulk_viscosity}
\end{figure*}

\subsection{Thermal conductivity and diffusive heat coefficient}

Similarly, the thermal conductivity $\kappa$ and the diffusive heat coefficient $\mu$ have both kinetic and collisional contributions:
\begin{equation}
    \kappa = \kappa_\text{k} + \kappa_\text{c}, \qquad \mu = \mu_\text{k} + \mu_\text{c}.
\end{equation}
As shown in Appendix \ref{app:CEdetails}, the collisional contributions $\kappa_\text{c}$ and $\mu_\text{c}$ can be obtained, at Navier--Stokes order, by using the approximations \eqref{4.9} as in the usual $\Delta$-model and replacing $\Delta \to \overline{\Delta}$. The results are
\begin{align}
\kappa_\text{c}={}&\frac{3}{4}\chi\phi
\left[1+\alpha+\sqrt{\frac{2}{\pi}}\overline\Delta^*
\left(1-\frac{a_2}{32}\right)\right]\kappa_\text{k} \nonumber \\
&+\frac{2}{\pi}\chi\phi^2
\left[(1+\alpha)\left(1+\frac{7a_2}{16}\right)
+\sqrt{\frac{\pi}{2}}\overline\Delta^*\right]\kappa_0,
\end{align}

\begin{equation}
    \mu_\text{c} = \frac{3}{4}\chi\phi \left[ 1+ \alpha + \sqrt{\frac{2}{\pi}}\overline{\Delta}^*\left(1-\frac{1}{32}a_2\right) \right]\mu_\text{k},
\end{equation}
where
\begin{equation}
    \kappa_0 = 4\frac{\eta_0}{m}
\end{equation}
is the low-density value of the thermal conductivity for an ordinary gas of hard disks ($\alpha = 1$ and $\Delta = 0$).

\begin{figure*}[t]
\centering
\fullwidthgraphic{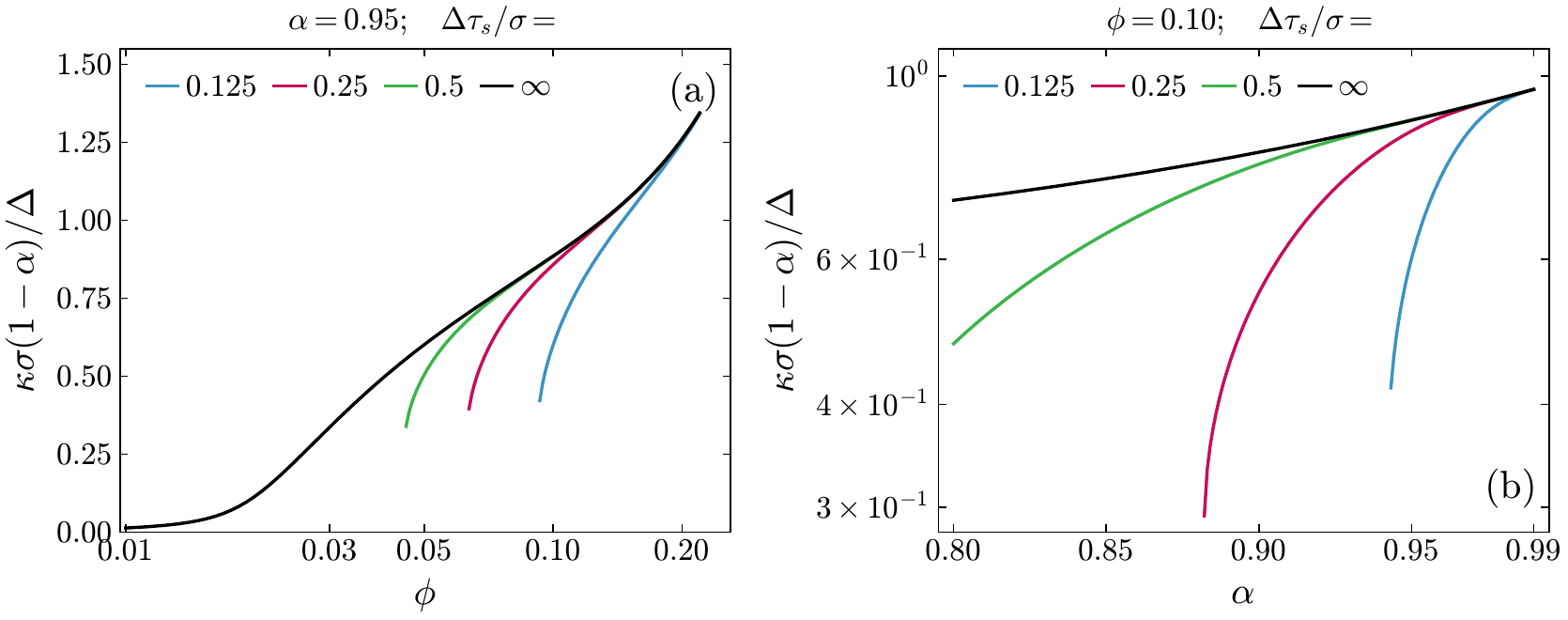}
\caption{Scaled thermal conductivity $\kappa\sigma(1-\alpha)/\Delta$ as a function of (a) the packing fraction $\phi$ at $\alpha=0.95$ and (b) the coefficient of restitution $\alpha$ at $\phi=0.10$. Solid lines show the theoretical predictions within the synchronization mean-field approximation. In all cases, the parameters are the same as in Fig.~\ref{fig:temperature}.}
\label{fig:thermal_conductivity}
\end{figure*}
The kinetic contribution $\kappa_\text{k}$ to the thermal conductivity is
\begin{widetext}
\begin{align}
        \kappa_{\mathrm{k}}={}&\left(\nu_\kappa^*+3\gamma^*-\frac{1}{2}\zeta^{*}_0
        -\zeta^{*}_0\frac{\partial\ln\zeta_0^*}{\partial\theta}\right)^{-1}\sqrt{\frac{\pi}{2}}
        \Bigg\{1+2a_2+\Upsilon_\theta \nonumber \\
        &-\chi\phi\Bigg[\frac{3}{8}(1+\alpha)^2\big(1-2\alpha-a_2(1+\alpha)\big)
        +\frac{1}{\sqrt{2\pi}}\overline{\Delta}^*\left[ \left(\frac{3}{4}-\frac{97}{64}a_2\right)
        +3(1+\alpha)\left(1-\frac{\sqrt{2\pi}}{2}\,\overline{\Delta}^*\right)\right.
        \nonumber \\ & \left. -\frac{a_2}{8}\left(21+\frac{27}{4}\alpha\right)\alpha
        -\frac{9}{2}(1+\alpha)^2
        -\overline{\Delta}^{*2}\left(1-\frac{1}{16}a_2\right) \right] \Bigg] \nonumber \\
        &+\frac{3}{16}\chi\phi\,\Upsilon_\theta
        \Bigg[(1+\alpha)^3+\frac{1}{2\sqrt{2\pi}}\overline{\Delta}^*
        \left(13+2\alpha(8+3\alpha)-\frac{4}{3}\overline{\Delta}^{*2}\right)\Bigg] \nonumber \\
        &+\chi\phi\,\overline{\Delta}^*
        \frac{2\omega_\text{e}^*\tau_\text{s}^*}
        {\mathrm e^{2\omega_\text{e}^*\tau_\text{s}^*}-1}
        \Bigg[\frac{3}{4}\left(
        \frac{2\alpha^2+4\overline{\Delta}^{*2}-1}{\sqrt{2\pi}}
        +2\alpha\overline{\Delta}^*\right)
        +\frac{a_2}{64\sqrt{2\pi}}
        \left(47+18\alpha^2-12\overline{\Delta}^{*2}\right)\Bigg]
        \Bigg\}\kappa_0,
\end{align}
where
\begin{equation}
\label{vic5}
    \nu_\kappa^* =\frac{\sqrt{2\pi}}{16}\,\chi\left[
        (1+\alpha)(19-15\alpha)
        +\frac{(1+\alpha)(7-3\alpha)}{32}\,a_2
        -2\sqrt{\frac{\pi}{2}}\,(5\alpha-1)\,\overline{\Delta}^*
        -\left(10-\frac{3}{16}a_2\right)\overline{\Delta}^{*2}
        \right].
\end{equation}
\noindent Finally, the kinetic part $\mu_\text{k}$ of the heat flux diffusion coefficient is given by \footnote{During the present calculation, we identified a sign typo in the denominator of the expression for $\mu_{\rm k}$ reported in
Ref.~\onlinecite{GGG26}. The expression given here includes the corresponding
correction.}
\begin{align}
        \mu_\text{k} = {} &\left(\nu^*_\kappa+3\gamma^*\right)^{-1}\sqrt{\frac{\pi}{8}}\Bigg\{2 \left(a_2 + \phi \frac{\partial a_2}{\partial \phi}\right) \nonumber \\ 
        & + 4\left(1+\frac{1}{2}\phi\frac{\partial \ln \chi}{\partial \phi}\right)\chi \phi \Bigg[ \frac{3}{8}(1+\alpha)\left[\alpha(\alpha-1)+\frac{a_2}{6}(14-3\alpha + 3\alpha^2)\right] \nonumber \\ 
        & +\frac{1}{\sqrt{2\pi}}\overline{\Delta}^* \left[2\overline{\Delta}^{*2}-3\left(\frac{1}{2}-\alpha^2\right)+\frac{3}{2}\sqrt{2\pi}\,\alpha\,\overline{\Delta}^* + \frac{a_2}{32}(47 + 18\alpha^2 - 4\overline{\Delta}^{*2})\right] \Bigg]  \nonumber \\
        & + \frac{3}{8}\chi\phi^2 \frac{\partial a_2}{\partial \phi}\Bigg[ \left(1+\alpha\right)^3
        +\frac{1}{2\sqrt{2\pi}}\,\overline{\Delta}^*
        \left[13+2\alpha(8+3\alpha)-\frac{4}{3}\overline{\Delta}^{*2}\right] \Bigg] \nonumber \\
        & +4\left(1+\phi\frac{\partial\ln\chi}{\partial\phi}\right)
        \chi\phi\,\overline{\Delta}^*
        \frac{2\omega_\text{e}^*\tau_\text{s}^*}
        {\mathrm e^{2\omega_\text{e}^*\tau_\text{s}^*}-1}
        \Bigg\{\frac{3}{4}\left[
        \frac{2\alpha^2+4\overline{\Delta}^{*2}-1}{\sqrt{2\pi}}
        +2\alpha\overline{\Delta}^*\right]
        +\frac{a_2}{64\sqrt{2\pi}}
        \left(47+18\alpha^2-12\overline{\Delta}^{*2}\right)\Bigg\} \nonumber \\
        & + \sqrt{\frac{8}{\pi}}\Bigg[ \zeta_0^*\left(1+\phi\frac{\partial\ln\chi}{\partial\phi}\right)
    \left(1+\frac{2\omega_\text{e}^*\tau_\text{s}^*}{e^{2\omega_\text{e}^*\tau_\text{s}^*}-1}\,
    \overline{\Delta}^*\frac{\partial \ln \zeta_0^*}{\partial \overline{\Delta}^*}\right)
    +\chi\phi\frac{\partial\chi}{\partial\phi}\frac{\partial}{\partial\chi}\left(\frac{\zeta_0^*}{\chi}\right) -\zeta_0^*\,\gamma^*\frac{\partial \ln \zeta_0^*}{\partial \gamma^*}
    \Bigg] \frac{\kappa_\text{k}}{\kappa_0}\Bigg\}\kappa_0 \frac{T}{n}.
\end{align}
\end{widetext}

\begin{figure*}[t]
\centering
\fullwidthgraphic{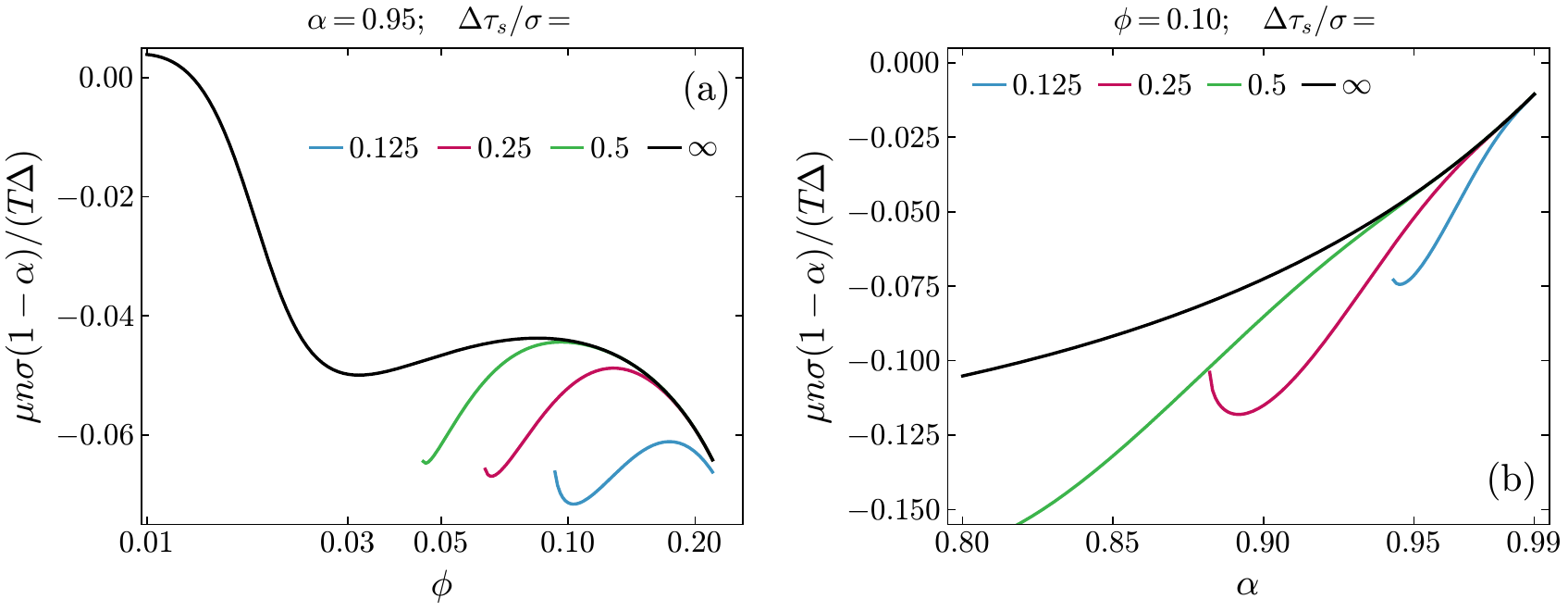}
\caption{Scaled diffusive heat coefficient $\mu n\sigma(1-\alpha)/(T\Delta)$ as a function of (a) the packing fraction $\phi$ for $\alpha=0.95$ and (b) the coefficient of restitution $\alpha$ for $\phi=0.10$. Solid lines are the theoretical predictions within the synchronization mean-field approximation. In all cases, the parameters are the same as in Fig.~\ref{fig:temperature}.}
\label{fig:diffusive_heat}
\end{figure*}

It is worth emphasizing that the terms proportional to
\begin{equation}
\frac{2\omega_{\rm e}^*\tau_{\rm s}^*}
{\exp(2\omega_{\rm e}^*\tau_{\rm s}^*)-1}
\end{equation}
in $\kappa_{\rm k}$ and $\mu_{\rm k}$ originate from the explicit pairwise
spatial dependence of the synchronization-controlled collision rule. These
terms are absent in the conventional $\Delta$-model and constitute the
direct kinetic contribution associated with the nonlocal synchronization
closure. They vanish in the limit $\tau_{\rm s}\to\infty$, where the
conventional $\Delta$-model is recovered.

\subsection{First-order contribution to the cooling rate}

The first-order contribution to the cooling rate $\zeta^{(1)}$ is defined by Eq.\ \eqref{3.13}. Following the approximations presented in Appendix \ref{app:CEdetails}, the scalar coefficient $\zeta_U$ defining this contribution can be written as
\begin{widetext}
\begin{equation}
\begin{aligned}
\zeta_U={}&
\chi\phi
\left[
2\overline{\Delta}^{*2}
+
4\sqrt{\frac{2}{\pi}}
\left(1-\frac{a_2}{16}\right)
\alpha\overline{\Delta}^{*}
-\frac{3}{2}(1-\alpha^2)
\right]
\\[1mm]
&+
\frac{\sqrt{2\pi}}{512}
\chi
\left[
96(1-\alpha^2)
+
64\overline{\Delta}^{*2}
+
a_2
\left(
9(1-\alpha^2)
+
30\overline{\Delta}^{*2}
\right)
\right]e_D^*,
\end{aligned}
\end{equation}
\begin{figure*}[t]
\centering
\fullwidthgraphic{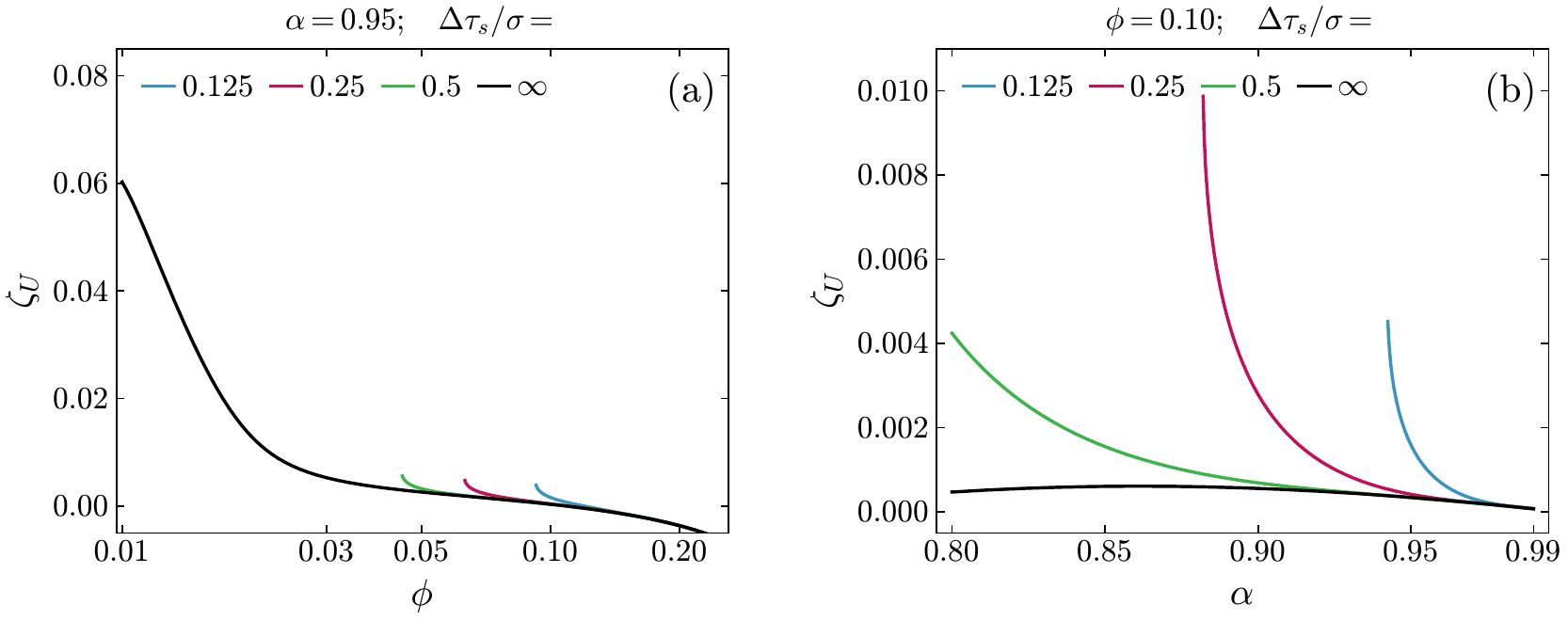}
\caption{First-order contribution to the (active-branch) cooling-rate coefficient $\zeta_U$ as a function of (a) the packing fraction $\phi$ for $\alpha=0.95$ and (b) the coefficient of restitution $\alpha$ for $\phi=0.10$. Solid lines correspond to the theoretical predictions obtained within the synchronization mean-field approximation. In all cases, the parameters are the same as in Fig.~\ref{fig:temperature}.}
\label{fig:zetaU}
\end{figure*}

\noindent where
\begin{equation}\label{eq:eD}
    e_D^* = \left[\nu_\zeta^* + 4\gamma^*-\overline{\zeta}^{(1,1)*}\left(a_2+\frac{1}{2}\Upsilon_\theta\right)\right]^{-1}\left[-a_2+\left(\zeta^{(1,0)}+p^*\right)\left(a_2 + \frac{1}{2}\Upsilon_\theta\right)+\frac{1}{2}\phi\frac{\partial a_2}{\partial \phi}+\frac{1}{8n}I_\zeta\right].
\end{equation}
In Eq.\ \eqref{eq:eD}, we have introduced the dimensionless quantity $\overline{\zeta}^{(1,1)*}=\ell \overline{\zeta}^{(1,1)}/v_\text{th}$. In the leading Sonine approximation, $ \nu_\zeta^*$ is given by 
\begin{align} 
        \nu_\zeta^* = {} &
        -\sqrt{\frac{\pi}{2}}\,
        \frac{(1+\alpha)}{96}\,\chi\,
        \left(30\alpha^3-30\alpha^2+153\alpha-185\right) - \frac{\sqrt{2\pi}}{16384}\chi\Bigg\{ -4 \overline{\Delta}^{*2}\left(32+15 a_2\right)\nonumber\\& +36\al^2\overline{\Delta}^{*2}\left(32+3a_2\right)+6 \overline{\Delta}^{*2}\left(416-33 a_2\right) \nonumber \\
        &+ 512 \sqrt{\frac{\pi}{2}}\overline{\Delta}^{*}\left(-2+11\al+3\al^3\right) + 2048\, \overline{\Delta}^{*2}\left(1+\frac{15}{32}a_2\right)\Bigg\}.
\end{align}
\end{widetext}
The explicit expressions of $I_\zeta$ and the derivatives of $a_2$ are displayed in Appendix \ref{app:CEdetails}.

\subsection{Some illustrative examples}\label{subsec:illustrative_examples}

To illustrate the influence of synchronization on the Navier--Stokes transport coefficients, Figs.~\ref{fig:viscosity}--\ref{fig:zetaU} collect representative results for the shear and bulk viscosities, the thermal conductivity, the diffusive heat coefficient, and the first-order contribution to the cooling rate. The same two paths through parameter space are considered throughout: variation of the packing fraction at fixed $\alpha=0.95$ and variation of the coefficient of restitution at fixed $\phi=0.10$. When available, we also provide numerical measurements of these quantities via EDMD simulations as described in Appendix~\ref{app:EDMD}.

A common feature of all these quantities is the qualitative distinction between the synchronized and nonsynchronized cases. In the absence of synchronization, transport coefficients vary smoothly as the density or the coefficient of restitution is varied. By contrast, in the presence of synchronization, the transport coefficients exhibit an abrupt jump at finite values of $\phi$ or $\alpha$, reflecting the discontinuous drop of the granular temperature to zero at the absorbing transition.  

The shear viscosity and the thermal conductivity show very similar trends: both increase along the active branch and approach the nonsynchronized result when collisions are frequent enough to suppress synchronization. For the shear viscosity, the EDMD data also show that the theory performs well over most of the active branch. The deviations of the theory from simulations become more apparent close to the transition. However, the agreement between theory and simulations can be clearly improved when one evaluates the shear viscosity $\eta$ from Eqs.\ \eqref{4.5} and \eqref{vic3} but replaces the theoretical value of the granular temperature $T$ with its value $T_\text{sim}$ measured in the simulations. This behavior indicates that the largest discrepancy close to the transition is associated with the mean-field description of the homogeneous state rather than with the Sonine approximation considered to evaluate the transport coefficient itself.

The bulk viscosity $\eta_\text{b}$ is purely collisional and therefore displays a particularly strong density dependence, consistent with its vanishing in the dilute limit. As in the case of the shear viscosity $\eta$, we observe that the replacement $T\to T_\text{sim}$ in the theoretical expressions \eqref{4.6}--\eqref{4.7b} defining $\eta_\text{b}$ improves its agreement with simulations. In fact, the resulting agreement is very good over most of the range
investigated. This observation provides a useful separation between two
different approximations entering the theory. On the one hand, it supports the Chapman--Enskog description of the transport response around the homogeneous reference state. On the other hand, it indicates that an important part of the remaining discrepancy close to the end of the active branch originates from the mean-field prediction of that reference state, and in particular from its prediction of the steady granular temperature. Replacing
$T$ by $T_{\rm sim}$ largely removes this source of error and thereby reduces the effect of the active--inactive population heterogeneity that is not explicitly represented in the present single-population mean-field closure.

The diffusive heat coefficient $\mu$ remains negative over the range displayed and is especially sensitive to synchronization close to the termination of the active branch. Since $\mu$ couples the heat flux directly to a density gradient, this enhanced sensitivity reflects the additional density dependence introduced by $\overline{\Delta}(\phi,T,\tau_{\rm s})$. The same feedback is visible in $\zeta_U$, which measures the first-order response of the collisional energy balance to a local expansion or compression. Close to the synchronized transition, a small change of density or temperature produces a comparatively large change in the synchronization probability and hence in the effective injection, making both $\mu$ and $\zeta_U$ particularly sensitive to the proximity of the termination of the active branch.

\section{Discussion}
\label{sec:discussion}
In this work, we have developed an Enskog kinetic description of
Navier--Stokes transport of a confined granular
monolayer whose collisional energy injection is controlled by
synchronization with the vibrating boundaries. Here, synchronization
refers to the tendency of particles that remain collision-free for a
sufficiently long time to lock their vertical motion to that of the
confining plates. Collisions disrupt this synchronized state, so that energy
is injected in a subsequent collision unless both colliding particles have
already synchronized with the boundaries.
The results have been obtained within the context of the
$\Delta$-model\cite{BRS13},
a coarse-grained collisional model that attempts to account in an effective
way for the effect of confinement on grain dynamics. The $\Delta$-model
introduces the parameter $\Delta>0$ in the collisional rules to mimic
the transfer of kinetic energy from the vertical degrees of freedom of
granular particles to the horizontal ones. The main appeal of the present
description is its simplicity. The microscopic collision rule depends on
the collision histories of the two particles, but this history dependence
is reduced, at the mean-field level, to an effective pairwise injection
amplitude controlled by the competition between the local Enskog collision
frequency and the synchronization time. The resulting kinetic theory
admits a finite-temperature active branch, in which collisions occur
frequently enough to continuously disrupt synchronization and sustain a
nonzero collisional energy injection. Despite this strong coarse-graining,
the resulting theory captures remarkably well the behavior observed in
microscopic EDMD simulations over a broad region of this active branch.

The homogeneous steady temperature provides the most direct test of this
mean-field description and, at the same time, reflects the feedback mechanism
introduced by synchronization: a reduction of the collision frequency gives
particles more time to synchronize with the vibrating plates, thereby
decreasing the effective injection amplitude. The associated reduction of
the granular temperature further decreases the collision frequency and
reinforces synchronization. Eventually, this positive feedback causes the finite-temperature active
solution to disappear, providing within the present mean-field description
the signature of the synchronization-controlled discontinuous transition.
In the present theory we set the fraction of active particles $P_{\rm a}$ to one, so that all particles are
assumed to belong to the active population. The absorbing state, by contrast,
corresponds to a state in which the collisional dynamics is no longer able to
sustain the granular motion. Thus, the theory describes how the
finite-temperature active state evolves as synchronization becomes increasingly
important and where this active solution ceases to exist, but it does not
describe the subsequent dynamics toward the absorbing state.
This synchronization-induced feedback is absent in the conventional
$\Delta$-model and constitutes one of the main physical consequences of
introducing synchronization into the kinetic description.

The comparison of the homogeneous state with EDMD also helps to delineate
the accuracy and limitations of the present closure. Over most of the active
branch, the mean-field theory reproduces the steady granular temperature
rather well, whereas small but systematic deviations become progressively
more visible on approaching its low-temperature end. At fixed coefficient
of restitution, decreasing the packing fraction reduces the collision
frequency and drives the system toward the synchronization-controlled transition. In this regime, the assumptions underlying the mean-field closure become less accurate. The analysis of Appendix~\ref{app:a2deviations} shows that fluctuations of the binary collision-dependent amplitude $\Delta_{ij}$ (which are suppressed by the deterministic replacement $\Delta_{ij}\rightarrow\overline{\Delta}$) become increasingly
important. At the same time, the synchronization state becomes
correlated with the pre-collisional velocities. The physical origin of these
correlations also suggests that their importance should increase with viscous damping. Indeed, the synchronization state is determined by the time elapsed since the previous collision, while the same free-flight time controls the decay of the particle velocity through the drag
dynamics. A larger $\gamma$ therefore makes the pre-collisional velocity more sensitive to the collision history and may enhance the correlation between synchronization and velocity. A systematic exploration of this dependence on the drag strength would provide a useful additional test of the mean-field factorization. The effects associated with fluctuations in $\Delta_{ij}$ and velocity--synchronization correlations remain small
well inside the active phase but grow markedly close to the transition, where the collision history becomes increasingly relevant.

Further insight into the origin of these deviations is provided by
numerical solutions of the homogeneous Boltzmann--Enskog equation in which the synchronization dynamics is retained beyond the deterministic mean-field replacement $\Delta_{ij}\rightarrow\overline{\Delta}_{ij}$. These calculations
improve the agreement with EDMD, indicating that the mean-field closure is an important source of the observed discrepancy. However, the homogeneous Boltzmann--Enskog description still does not completely reproduce the microscopic results. This residual difference suggests that additional effects, such as departures from molecular chaos and spatial correlations associated with the formation of synchronized and asynchronized domains, may also play a role.

The discrepancies are even more pronounced upon increasing the
inelasticity (decreasing $\alpha$) at fixed density. Stronger inelasticity lowers the kinetic energy and reduces the collision frequency, thereby favoring longer collision-free intervals and increasing the tendency toward synchronization. The dynamics consequently becomes more sensitive to the detailed collision history, enhancing the importance of injection-amplitude fluctuations and velocity--synchronization correlations. Spatial correlations associated with synchronized and asynchronized domains may also become more relevant in this regime.

A distinct limitation arises in the regime of the continuous absorbing transition discussed in Ref.~\onlinecite{MPSF25}. In that case, the transition occurs through the progressive vanishing of the active fraction, which becomes an additional relevant dynamical variable. Since the present theory explicitly sets $P_{\rm a}=1$, such activation--inactivation population dynamics lies outside its scope. This mechanism should therefore be distinguished from the synchronized--asynchronized dynamics considered here along the finite-temperature active branch.

Having established the homogeneous reference state and the main
limitations of its mean-field description, we now turn to the consequences of synchronization for Navier--Stokes transport. An important consequence of synchronization is that the effective collision rule is intrinsically pair-dependent in an inhomogeneous state. This fact introduces new conceptual and technical difficulties that are not present in the conventional
$\Delta$-model.\cite{GBS18} In particular, at Navier--Stokes order, this produces a new vector source $\bcal{R}$ in the
Chapman--Enskog solution. This source modifies the kinetic contributions to the thermal conductivity and the diffusive heat coefficient, whereas it does not enter directly in the evaluation of the shear and bulk viscosities. Interestingly, the explicit pairwise correction to the collisional transfer vanishes at first order after integration along the collision link. Hence, the synchronization-induced spatial nonlocality affects heat transport through the kinetic contributions to the thermal conductivity and the diffusive heat coefficient rather than through an additional direct collisional-transfer term. This illustrates how the spatial dependence generated by synchronization enters the Enskog transport problem differently from the standard nonlocality associated with the finite separation of colliding particles.

The comparison with EDMD of the transport coefficients also helps to delineate the accuracy and the limitations of the present closure. For the quantities that can be measured directly in the simulations, the theory performs well over most of the active branch. In particular, the shear $\eta$ and bulk $\eta_\text{b}$ viscosities follow the microscopic results with good accuracy away from the transition. Moreover, when in the analytical expressions of $\eta$ and $\eta_\text{b}$ one replaces the mean-field theoretical temperature by the steady granular temperature measured in EDMD simulations, the agreement between theory and simulations improves systematically. In fact, a very good overall agreement is then obtained over most of the active branch. A similar improvement is observed for the Poisson estimate of the injecting-collision probability. Together with the homogeneous-state analysis discussed above, these observations strongly suggest that an important part of the remaining discrepancy in the transport coefficients originates from the mean-field prediction of the homogeneous reference state, rather than from the Chapman--Enskog expressions for the transport coefficients themselves. In other words, once the correct local temperature is supplied, the transport theory derived within the present closure provides a very accurate description of the response of the active fluid.

These considerations also clarify the scope of the present theory.
Within the synchronized active branch, the pairwise mean-field closure provides a tractable kinetic framework that describes the homogeneous state and its Navier--Stokes transport with comparatively little microscopic information. Its main limitations arise precisely in the regimes where additional slow statistical variables---collision history and synchronization state---can no longer be slaved to the density and temperature fields. Developing a kinetic description in which these variables are retained explicitly would therefore constitute the natural next step.

Two extensions of the present framework are currently being pursued. First, the linear stability of the homogeneous steady state within the
same synchronization closure is being analyzed and compared with EDMD simulations and DSMC results. This will establish whether the transport coefficients obtained here provide a quantitatively consistent hydrodynamic description of the onset of spatial instabilities. Second, the corresponding tracer problem is being investigated in order to determine the diffusion coefficients of an impurity in the synchronized host fluid and to characterize the consequences of synchronization for mass transport and segregation.

On a longer time scale, a more complete kinetic theory should treat synchronization and activation--inactivation dynamics on the same footing. A possible route is to introduce additional kinetic or hydrodynamic fields describing the synchronized and active fractions, rather than replacing their dynamics by instantaneous mean-field probabilities. Such a theory could be confronted with microscopic simulations and experiments not only at the level of homogeneous temperatures and transport coefficients, but also through spatial and temporal correlations close to the absorbing transition. In particular, if the continuous transition is correctly retained, the extended theory could provide a microscopic starting point for studying its critical behavior. Combined with finite-size scaling of simulations and experimental measurements, this would make it possible to test the critical exponents and ultimately identify the universality class of
the activation--inactivation transition.

\begin{acknowledgments}

DGM and VG acknowledge the financial support from Grant No. PID2024-156352NB-I00 funded by MCIU/AEI/10.13039/501100011033/ and by ERDF/EU. DGM, RGG, and VG also acknowledge financial support from Grant No. GR24022 funded by Junta de Extremadura (Spain) and by the European Regional Development Fund (ERDF) ``A way of making Europe.'' The research of David Gonz\'alez M\'endez has been supported by the predoctoral Fellowship No. FPU24/01056 from the Spanish Government. IP acknowledges support from Ministerio de Ciencia, Innovación y Universidades MCIU/AEI/FEDER for ﬁnancial support under grant agreement PID2024-156516NB-10, and Generalitat de Catalunya for ﬁnancial support under Program Icrea Acad\`emia and project
2021SGR-673
\end{acknowledgments}

\section*{AUTHOR DECLARATIONS}

\subsection*{Conflict of Interest}
The authors have no conflicts to disclose.

\subsection*{Author Contributions}

\noindent
\textbf{DGG:} Conceptualization, Methodology, Formal analysis,
Investigation, Software, Validation, Data curation, Visualization,
Writing -- original draft, Writing -- review \& editing.

\noindent
\textbf{RGG:} Conceptualization, Methodology, Formal analysis,
Investigation, Validation,
Writing -- review \& editing.

\noindent
\textbf{RM:} Conceptualization, Methodology, 
Investigation, Software, Validation, Data curation, Visualization, Writing -- review \& editing.

\noindent
\textbf{VG:} Conceptualization, Methodology, Formal analysis,
Investigation, Validation, 
Writing -- review \& editing.

\noindent
\textbf{IP:} Supervision, Project administration, Writing -- review \& editing.

\section*{DATA AVAILABILITY}
The data that support the findings of this study are available from the corresponding author upon reasonable request..

\appendix

\section{Details of the Chapman--Enskog calculation}
\label{app:CEdetails}

In this Appendix, we provide the technical details underlying the Chapman--Enskog solution summarized in Sec.~\ref{sec:chapman-enskog}. The calculation follows the standard Enskog construction used for the conventional $\Delta$-model with external friction~\cite{GGG26}, but the pairwise synchronization closure~\eqref{1.25} introduces one additional first-order ingredient: the collision rule itself varies across the finite separation of the colliding pair. We keep this contribution separate from the conventional Enskog nonlocality in order to identify unambiguously which parts of the transport calculation are modified.

Throughout this Appendix, ``exact'' means exact within the Enskog kinetic equation and the deterministic pairwise mean-field closure of Sec.~\ref{subsec:sync}, before any Sonine truncation is made. The linear integral equations are therefore exact to Navier--Stokes order within that framework. Closed analytical expressions are obtained subsequently by retaining the leading Sonine approximation for the first-order distribution functions, while the zeroth-order distribution is kept to linear order in the fourth cumulant $a_2$.

\subsection{Derivatives of the homogeneous cumulant} \label{ap:a1}

Before obtaining the derivatives of $a_2$ with respect to $\theta$, $\gamma^*$, $\chi$ and $\overline{\Delta}^*$, we provide here the forms of the quantities $A_0$, $A_1$, $C_0$ and $C_1$. They are given by
\beq
\label{2.19}
A_0(\al,\overline{\Delta}^{*})=4\left(\frac{1-\al^2}{2}-\sqrt{\frac{\pi}{2}}
\al \overline{\Delta}^{*}-\overline{\Delta}^{*2}\right),
\eeq
\beq
\label{2.20}
A_1(\al,\overline{\Delta}^{*})=\frac{1}{4}\left[\frac{3}{2}(1-\al^2)+
\overline{\Delta}^{*2}\right],
\eeq

\begin{equation}
\label{2.21}
\begin{aligned}
C_0(\alpha,\overline\Delta^*)={}&
\sqrt{2\pi}\,\alpha\overline\Delta^*
\left(5+3\alpha^2+4\overline\Delta^{*2}\right)\\
&-7+5\alpha^2+2\alpha^4
+2(5+6\alpha^2)\overline\Delta^{*2}
+4\overline\Delta^{*4}.
\end{aligned}
\end{equation}
\begin{align}\label{2.22}
C_1(\alpha,\overline\Delta^*)={}&
\sqrt{\frac{\pi}{2}}(-2+11\alpha+3\alpha^3)\overline\Delta^* \nonumber \\
&-\frac{1}{16}\Big[209+32\alpha-147\alpha^2-30\alpha^4 \nonumber \\
&\hspace{15mm}-(78+36\alpha^2)\overline\Delta^{*2}
+4\overline\Delta^{*4}\Big].
\end{align}

In the absence of spatial gradients, $f^{(0)}$ satisfies the Enskog equation
\begin{equation}
\label{3.0.4}
    \partial_t^{(0)} f^{(0)}-\gamma\frac{\partial}{\partial \mathbf{v}}\cdot \mathbf{v} f^{(0)}=J_\text{E}^{(0)}[\mathbf{V}|f^{(0)},f^{(0)}],
\end{equation}
where $J_\text{E}^{(0)}[\mathbf{V}|f^{(0)},f^{(0)}]$ is given by the homogeneous operator defined in Eq.\ \eqref{2.2} with the replacement $f(\mathbf{v}; t)\to f^{(0)}(\mathbf{r}, \mathbf{v}; t)$ and all hydrodynamic quantities evaluated locally. The zeroth-order distribution is therefore the local version of the homogeneous normal solution. Using the scaled representation of $f^{(0)}$ introduced in Eq.\ \eqref{3.0.7}, the corresponding distribution $\varphi$ obeys the \textit{unsteady} equation
\beq
\label{3.0.8}
-
\Lambda
\frac{\partial\varphi}{\partial\theta} + \frac{\zeta_0^*}{2}
\frac{\partial}{\partial\mathbf{c}}\cdot
\left(\mathbf{c}\varphi\right)
=
J_\text{E}^{*}[\mathbf{c}|\varphi,\varphi],
\eeq
where the dependence of $\overline{\Delta}^*$ on the local fields is the local homogeneous dependence defined by Eq.\ \eqref{eq:barDelta-local-CE}. The explicit pair-separation correction to $\overline{\Delta}_{ij}$ is first order and is treated separately in Sec.\ \ref{ap:a2}.

Since an exact solution to Eq.\ \eqref{3.0.8} is unknown, the leading deviations of the unsteady distribution $\varphi$ can be characterized through the kurtosis $a_2(T)$, whose evolution equation reads
\begin{equation} \label{eq:50}
    \frac{\partial a_2}{\partial \theta} = -\dfrac{\mu_4 + 4\zeta^*(1+a_2)}{2\Lambda},
\end{equation}
where $\Lambda$ keeps its previous definition, with the sole replacement $\zeta^* \to \zeta_0^*$. In the steady state, $a_\text{2,ss}$ is given by Eq.\ \eqref{eq:47} and hence, the numerator and denominator of Eq.\ \eqref{eq:50} vanish. Thus, the derivative $\partial_\theta a_2$ becomes indeterminate. Equation \eqref{eq:50} can be more explicitly written when one expands $\mu_4$ and $\zeta^{*}$ to first order in $a_2$. The result is
\begin{widetext}
\beq
\label{2.29}
\frac{\partial a_2}{\partial \theta}=-\dfrac{1}{2}\dfrac{\mu_4^{(0)}+4\zeta_0^{(0)}+\left(\mu_4^{(1)} + 4\zeta_0^{(0)}+4\zeta_0^{(1)}\right)a_2}{2\gamma^* + \zeta^{(0)}_0 + \zeta_0^{(1)}a_2}\equiv -\frac{1}{2}\frac{X_1+X_2 a_2}{Y_1+Y_2 a_2},
\eeq
where the expressions of $X_1$, $X_2$, $Y_1$, and $Y_2$ can be easily identified from Eq.\ \eqref{2.29}. The derivative $\Upsilon_\theta\equiv \partial_\theta a_2$ can be determined by applying l'H\^opital's rule to Eq.\ \eqref{2.29}. The final result leads to a quadratic equation for $\Upsilon_\theta$:
\beq
\label{2.30}
2 Y_2 \Upsilon_\theta^2+\Big[2\left(Y_{1,\theta}+Y_{2,\theta}a_{2,\text{ss}}\right)+X_2\Big]\Upsilon_\theta+\Big(X_{1,\theta}+X_{2,\theta} a_{2,\text{ss}}\Big)=0,
\eeq
where $X_{i,\theta}\equiv \partial_\theta X_i$ and $Y_{i,\theta}\equiv \partial_\theta Y_i$. However, since the magnitude of $\Upsilon_\theta$ has been shown to be small, one can neglect the quadratic term in $\Upsilon_\theta$ in \eqref{2.30} and obtain the simple expression
\beq
\label{2.31}
\Upsilon_\theta=-\frac{X_{1,\theta}+X_{2,\theta} a_{2,\text{ss}}}{2\left(Y_{1,\theta}+Y_{2,\theta}a_{2,\text{ss}}\right)+X_2}.
\eeq
The derivatives $\Upsilon_{\gamma^*}    \equiv \partial_{\gamma^*} a_2$, $\Upsilon_\chi\equiv \partial_\chi a_2$ and $\Upsilon_{\overline{\Delta}^*} \equiv \partial_{\overline{\Delta}^*}a_2$ can be easily derived from Eq.\ \eqref{2.29}, with the results
\beq
\label{2.32}
\Upsilon_{\gamma^*} = -\dfrac{4\Upsilon_\theta}{X_2 + 2Y_2 \Upsilon_\theta}, \qquad 
\Upsilon_\chi=-\frac{2\left(Y_{1,\chi}+Y_{2,\chi}a_{2,\text{ss}}\right)\Upsilon_\theta+X_{1,\chi}+X_{2,\chi}a_{2,\text{ss}}}
{X_2+2Y_2 \Upsilon_\theta},
\eeq
\beq
\label{2.32b}
\Upsilon_{\overline{\Delta}^*} = -\frac{2\left(Y_{1,\overline{\Delta}^*}+Y_{2,\overline{\Delta}^*}a_{2,\text{ss}}\right)\Upsilon_\theta+X_{1,\overline{\Delta}^*}+X_{2,\overline{\Delta}^*}a_{2,\text{ss}}}
{X_2+2Y_2 \Upsilon_\theta}.
\eeq
The density derivative entering the transport coefficients is therefore
\beq
\label{3.31}
\phi\frac{\partial a_2}{\partial \phi}=\phi \frac{\partial a_2}{\partial \gamma^*}\frac{\partial \gamma^*}{\partial \phi}+\phi \frac{\partial a_2}{\partial \chi}\frac{\partial \chi}{\partial \phi} + \phi\frac{\partial a_2}{\partial \overline{\Delta}^*}\frac{\partial\overline{\Delta}^*}{\partial \phi},
\eeq
with
\begin{align}
    \phi\frac{\partial \gamma^*}{\partial \phi}=-\gamma^*, \qquad \phi\frac{\partial\overline{\Delta}^*}{\partial \phi} = \dfrac{2\omega^*_\text{e}\tau_\text{s}^*}{\text{e}^{2\omega^*_\text{e}\tau_\text{s}^*}-1}\overline{\Delta}^*\left(1+\phi\frac{\partial \ln \chi}{\partial\phi}\right).
\end{align}
\end{widetext}

\subsection{First-order Chapman--Enskog approximation} \label{ap:a2}

We now turn to the first-order Chapman--Enskog expansion of the Enskog kinetic equation around the local homogeneous reference state. The aim is to determine the contributions linear in the hydrodynamic gradients, which ultimately yield the Navier--Stokes transport coefficients. The analysis follows the conventional Enskog procedure, while consistently accounting for the additional spatial dependence introduced by the synchronization-dependent collision rule.

The first-order Enskog operator contains three conceptually distinct spatial contributions. The first two are the conventional Enskog nonlocality associated with the displaced distribution function and with the pair correlation function. The third one is specific to the present synchronization closure and comes from the spatial dependence of the collision rule itself through $\overline{\Delta}_{ij}$. To separate these contributions, we first write
\begin{equation}
\label{eq:f0-spatial-expansion}
f^{(0)}(\mathbf r+\boldsymbol\sigma,\mathbf v_2)
=
f^{(0)}(\mathbf r,\mathbf v_2)
+\sigma\widehat{\sigma}_i\partial_i f^{(0)}(\mathbf r,\mathbf v_2)
+\mathcal O(\bm\nabla^2),
\end{equation}
and, within the same local-contact approximation used in the conventional Enskog calculation,
\begin{equation}
\label{eq:chi-spatial-expansion}
\chi(\mathbf r,\mathbf r+\boldsymbol\sigma)
=
\chi
+\frac{\sigma}{2}\widehat{\sigma}_i\partial_i\chi
+\mathcal O(\bm\nabla^2).
\end{equation}
The pairwise injection amplitude is expanded according to Eqs.~\eqref{eq:pairwise-delta-expansion}--\eqref{eq:pairwise-delta-firstorder-fields}. In particular, its zeroth-order value is $\overline{\Delta}$, whereas its first-order part is proportional to $\widehat{\boldsymbol\sigma}\cdot\bm\nabla\ln T$ and $\widehat{\boldsymbol\sigma}\cdot\bm\nabla\ln n$.

With these expansions, the first-order collision operator can be organized as
\begin{widetext}
\begin{align}
\label{eq:JE-firstorder-decomposition}
J_\text{E}^{(1)}={}&-\mathcal L f^{(1)}
-\mathcal K_i\!\left[\partial_i f^{(0)}\right]
-\frac{1}{2}(\partial_i\ln\chi)\,\mathcal K_i\!\left[f^{(0)}\right]
+\mathcal R_i\,\partial_i\ln T
+2\left(1+\phi\frac{\partial\ln\chi}{\partial\phi}\right)
\mathcal R_i\,\partial_i\ln n.
\end{align}
Here, the linearized collision operator $\mathcal L f^{(1)}$ is
\beq
\label{3.6}
\mathcal L f^{(1)}
=-\left(J_\text{E}^{(0)}[f^{(0)},f^{(1)}]
+J_\text{E}^{(0)}[f^{(1)},f^{(0)}]\right),
\eeq
and $\mathcal K_i$ is the usual Enskog operator generated by the spatial displacement of the second member of the colliding pair,
\begin{align}
\label{3.12}
\mathcal K_i[X]={}&-\sigma^2\chi
\int\dd\mathbf v_2\int\dd\widehat{\boldsymbol\sigma}\,
\Theta(-\widehat{\boldsymbol\sigma}\cdot\mathbf g-2\overline{\Delta})
(-\widehat{\boldsymbol\sigma}\cdot\mathbf g-2\overline{\Delta})
\widehat\sigma_i\alpha^{-2}
 f^{(0)}(\mathbf v_1'')X(\mathbf v_2'')
\nonumber\\
&+\sigma^2\chi
\int\dd\mathbf v_2\int\dd\widehat{\boldsymbol\sigma}\,
\Theta(\widehat{\boldsymbol\sigma}\cdot\mathbf g)
(\widehat{\boldsymbol\sigma}\cdot\mathbf g)
\widehat\sigma_i f^{(0)}(\mathbf v_1)X(\mathbf v_2).
\end{align}
\end{widetext}
Thus, $\mathcal K_i$ retains exactly the conventional role: it accounts for the Enskog spatial nonlocality of the distribution functions and, through Eq.~\eqref{eq:chi-spatial-expansion}, of the pair correlation function. The explicit spatial variation of the synchronized collision rule is contained separately in $\bcal{R}$. The Cartesian components of this new source are
\begin{widetext}
\begin{align}
\label{eq:Rsync-strong}
\mathcal R_i(\mathbf V_1)
={}&
\frac{\sigma^2\chi\overline{\Delta}}{2}
\frac{\omega_{\rm e}\tau_{\rm s}}
{\mathrm e^{2\omega_{\rm e}\tau_{\rm s}}-1}
\left.
\frac{\partial}{\partial\overline{\Delta}_{ij}}
\Bigg\{
\int \dd\mathbf v_2
\int \dd\widehat{\boldsymbol\sigma}\,
\widehat{\sigma}_i\,
\Theta\!\left(
-\widehat{\boldsymbol\sigma}\cdot\mathbf g
-2\overline{\Delta}_{ij}
\right) \left(
-\widehat{\boldsymbol\sigma}\cdot\mathbf g
-2\overline{\Delta}_{ij}
\right)
\alpha^{-2}
f^{(0)}(\mathbf v_1'')
f^{(0)}(\mathbf v_2'')
\Bigg\}
\right|_{\overline{\Delta}_{ij}=\overline{\Delta}} .
\end{align}
\end{widetext}
Within the derivative in Eq.~\eqref{eq:Rsync-strong}, the restituting velocities
$\mathbf v_1''$ and $\mathbf v_2''$ are those obtained from the collision rule with
the pair amplitude $\overline{\Delta}_{ij}$. Thus, the derivative acts on the
explicit dependence of the gain kernel on $\overline{\Delta}_{ij}$, both through
the collision-rate factor and through the restituting velocity arguments.
The parametric dependence of the local zeroth-order distribution $f^{(0)}$ on
the hydrodynamic state is held fixed. This state dependence is already accounted
for by the hydrodynamic derivatives of $f^{(0)}$ appearing elsewhere in the
Chapman--Enskog expansion. After taking the derivative, the pair amplitude is
evaluated at its local zeroth-order value
$\overline{\Delta}_{ij}=\overline{\Delta}$.

A useful equivalent representation is obtained by multiplying
Eq.~\eqref{eq:Rsync-strong} by an arbitrary velocity function
$\psi(\mathbf v_1)$ and integrating over $\mathbf v_1$. For fixed
$\overline{\Delta}_{ij}$, the usual change from restituting to direct collision
variables can first be performed. In the resulting integral the explicit
dependence on $\overline{\Delta}_{ij}$ is entirely contained in
$\psi(\mathbf v_1')$. Hence,
\begin{equation}
\frac{\partial}{\partial\overline{\Delta}_{ij}}
\psi(\mathbf v_1')
=
-\widehat{\boldsymbol\sigma}\cdot
\frac{\partial\psi(\mathbf v_1')}{\partial\mathbf v_1'}.
\end{equation}
The resulting weak form for the new source $\bcal{R}$ then reads
\begin{widetext}
\begin{align}
\label{eq:Rsync-weak}
\int\dd\mathbf v_1\,\psi(\mathbf v_1)\mathcal R_i(\mathbf V_1)
={}&-
\frac{\sigma^2\chi\overline{\Delta}}{2}
\frac{\omega_\text{e}\tau_\text{s}}
{\mathrm e^{2\omega_\text{e}\tau_\text{s}}-1}
\int\dd\mathbf v_1\int\dd\mathbf v_2
\int\dd\widehat{\boldsymbol\sigma}\,
\Theta(\widehat{\boldsymbol\sigma}\cdot\mathbf g)
(\widehat{\boldsymbol\sigma}\cdot\mathbf g)
f^{(0)}(\mathbf v_1)f^{(0)}(\mathbf v_2)
\widehat\sigma_i\,
\left.\widehat{\boldsymbol\sigma}\cdot
\frac{\partial\psi(\mathbf v_1')}
{\partial\mathbf v_1'}\right|_{\overline{\Delta}_{ij}=\overline{\Delta}}.
\end{align}
\end{widetext}
Equation~\eqref{eq:Rsync-weak} will be useful below for checking the Chapman--Enskog solubility conditions and for evaluating the heat flux transport coefficients.

Using
\begin{equation}
\label{eq:spatial-derivative-f0}
\partial_i f^{(0)}
=
T\frac{\partial f^{(0)}}{\partial T}\,\partial_i\ln T
+n\frac{\partial f^{(0)}}{\partial n}\,\partial_i\ln n
-\frac{\partial f^{(0)}}{\partial V_j}\,\partial_iU_j,
\end{equation}
and the first-order balance equations, Eq.~\eqref{eq:JE-firstorder-decomposition} leads, in the steady state ($\Lambda=0$), to the following coupled linear integral equations:
\begin{widetext}
\begin{equation}
\label{3.2}
-\frac{1}{2}\zeta^{(0)}\left(1+2\frac{\partial\ln\zeta_0^*}{\partial\theta}\right)
\bcal{A}
-\gamma\frac{\partial}{\partial\mathbf v}\cdot\mathbf V\bcal{A}
+\mathcal L\bcal{A}
=\mathbf A,
\end{equation}
\begin{align}
\label{3.3}
-\gamma\frac{\partial}{\partial\mathbf v}\cdot\mathbf V\bcal{B}
+\mathcal L\bcal{B}
={}&\mathbf B
+\Bigg[
\zeta^{(0)}\left(1+\phi\frac{\partial\ln\chi}{\partial\phi}\right)
\left(1+
\frac{2\omega_\text{e}^*\tau_\text{s}^*}
{\mathrm e^{2\omega_\text{e}^*\tau_\text{s}^*}-1}
\overline{\Delta}^*\frac{\partial\ln\zeta_0^*}{\partial\overline{\Delta}^*}
\right)
\nonumber\\
&+\chi\phi\frac{\partial\chi}{\partial\phi}
\frac{\partial}{\partial\chi}\left(\frac{\zeta^{(0)}}{\chi}\right)
-\zeta^{(0)}\gamma^*\frac{\partial\ln\zeta_0^*}{\partial\gamma^*}
\Bigg]\bcal{A},
\end{align}
\beq
\label{3.4}
-\gamma\frac{\partial}{\partial\mathbf v}\cdot\mathbf V\mathcal C_{ij}
+\mathcal L\mathcal C_{ij}=C_{ij},
\eeq
\beq
\label{3.5}
-\gamma\frac{\partial}{\partial\mathbf v}\cdot\mathbf V\mathcal D
+\mathcal L\mathcal D=D.
\eeq
\end{widetext}
In Eqs.\ \eqref{3.2}--\eqref{3.5}, the functions $\mathbf A(\mathbf V)$, $\mathbf B(\mathbf V)$, $C_{ij}(\mathbf V)$, and $D(\mathbf V)$ retain the same formal expressions as in the conventional local $\Delta$ problem (with $\Delta$ replaced by the local zeroth-order value $\overline{\Delta}$), except that $\mathbf A(\mathbf V)$ and $\mathbf B(\mathbf V)$ now contain additional terms associated with the new vector source $\bcal{R}$. Their explicit forms are:
\begin{align}\label{3.8}
\mathbf A(\mathbf V)=-\mathbf VT\frac{\partial f^{(0)}}{\partial T}
-\frac{p}{\rho}\frac{\partial f^{(0)}}{\partial\mathbf V}
\left(1+\frac{\partial\ln p^*}{\partial\theta}\right)
\nonumber \\-\bcal{K}\left[T\frac{\partial f^{(0)}}{\partial T}\right] +\bcal{R},
\end{align}
\begin{align}\label{3.9}
\mathbf B(\mathbf V)=-\mathbf Vn\frac{\partial f^{(0)}}{\partial n}
-\frac{p}{\rho}\left(1+\phi\frac{\partial\ln p^*}{\partial\phi}\right)
\frac{\partial f^{(0)}}{\partial\mathbf V}
-\nonumber \\ \bcal{K}\left[n\frac{\partial f^{(0)}}{\partial n}\right] -\frac{1}{2}\phi\left(\frac{\partial\ln\chi}{\partial\phi}\right)
\bcal{K}\left[f^{(0)}\right] \nonumber \\
 + 2\left(1+\phi\frac{\partial\ln\chi}{\partial\phi}\right)
\bcal{R},
\end{align}
\begin{equation}
\label{3.10}
C_{ij}(\mathbf V)
=V_i\frac{\partial f^{(0)}}{\partial V_j}
+\mathcal K_i\left[\frac{\partial f^{(0)}}{\partial V_j}\right],
\end{equation}
\begin{align}\label{3.11}
D(\mathbf V)=\frac{1}{2}\frac{\partial}{\partial\mathbf V}\cdot
\left(\mathbf Vf^{(0)}\right)
+\left(\zeta_U+p^*\right)T\frac{\partial f^{(0)}}{\partial T}
\nonumber \\ - f^{(0)}+n\frac{\partial f^{(0)}}{\partial n} +\frac{1}{2}\mathcal K_i\left[\frac{\partial f^{(0)}}{\partial V_i}\right].
\end{align}
It is worth noting that, in the limit where synchronization is absent ($\tau_\text{s}\to\infty$), we have $\bcal{R}\to \bm{0}$, and one recovers (in the absence of drag) the corresponding integral equations for the conventional $\Delta$ model.\cite{GBS18} Note that in Eqs.\ \eqref{3.8}--\eqref{3.11} the temperature derivative of $f^{(0)}$ has been written as
\beq
\label{3.12.1}
T\frac{\partial f^{(0)}}{\partial T}
=-\frac{1}{2}\,\frac{\partial}{\partial\mathbf V}\cdot\left(\mathbf V f^{(0)}\right)
+nv_\text{th}^{-2}\frac{\partial \varphi}{\partial\theta}.
\eeq

For later use, it is also important to distinguish derivatives of the dimensional and reduced injection amplitudes. From Eq.\ \eqref{eq:barDelta-local-CE},
\begin{align}\label{eq:A29}
    T\left(\frac{\partial \ln{\overline{\Delta}}}{\partial T }\right)_n &= \frac{\omega_\text{e}\tau_\text{s}}{\text{e}^{2\omega_\text{e}\tau_\text{s}}-1}, \\
    \label{eq:A30}
    n\left(\frac{\partial \ln{\overline{\Delta}}}{\partial n }\right)_T &= \frac{2\omega_\text{e}\tau_\text{s}}{\text{e}^{2\omega_\text{e}\tau_\text{s}}-1}\left(1+\phi\frac{\partial \ln{\chi}}{\partial \phi}\right).
\end{align}
Note that the $-1/2$ term in Eq.\ \eqref{eq:new1} therefore solely reflects the explicit thermal-velocity scaling of $\overline{\Delta}^*=\overline{\Delta}/v_\text{th}$; it is not part of the derivative of the dimensional collision rule.

The consistency of these new-sourced results can be checked independently from the Chapman--Enskog solubility conditions. Since the hydrodynamic fields are defined entirely through $f^{(0)}$, the higher-order corrections obey
\begin{equation}
\label{eq:CE-subsidiary}
\int\dd\mathbf V\,
\left\{1,\mathbf V,V^2\right\}f^{(k)}=0,
\qquad k\geq1.
\end{equation}
In particular,
\begin{equation}
\int\dd\mathbf V\,mV_i\mathcal A_j=0.
\end{equation}
Taking the first velocity moment of Eq.~\eqref{3.2}, the left-hand side vanishes under this condition. The source must therefore satisfy
\begin{equation}
\label{eq:A-source-consistency}
\int\dd\mathbf V\,mV_i\,A_j(\mathbf V)=0.
\end{equation}
This cancellation can be shown before any Sonine approximation is made. The first moment of the source $A_j - \mathcal{R}_j$ contains the explicit temperature derivative of the collision impulse and can be written as
\begin{equation}
\label{eq:A-firstmoment-exact}
\int\dd\mathbf V\,mV_i(A_j-\mathcal{R}_j)
=
T\left(\frac{\partial\overline{\Delta}}{\partial T}\right)_n
\left(\frac{\partial P_{\mathrm c,ij}^{(0)}}{\partial\overline{\Delta}}\right)_{f^{(0)}},
\end{equation}
where the derivative on the right-hand side acts only on the explicit collision-rule dependence of the zeroth-order collisional pressure. On the other hand, Eq.~\eqref{eq:Rsync-weak} with $\psi=mV_i$ gives exactly
\begin{equation}
\label{eq:R-firstmoment-exact}
\int\dd\mathbf V\,mV_i\mathcal R_j
=-
T\left(\frac{\partial\overline{\Delta}}{\partial T}\right)_n
\left(\frac{\partial P_{\mathrm c,ij}^{(0)}}{\partial\overline{\Delta}}\right)_{f^{(0)}}.
\end{equation}
Hence Eq.~\eqref{eq:A-source-consistency} is identically fulfilled. The same consistency holds in the density-gradient equation because Eqs.\ \eqref{eq:A29} and \eqref{eq:A30} imply
\begin{equation}
\label{eq:barDelta-derivative-relation}
n\left(\frac{\partial\overline{\Delta}}{\partial n}\right)_T
=
2\left(1+\phi\frac{\partial\ln\chi}{\partial\phi}\right)
T\left(\frac{\partial\overline{\Delta}}{\partial T}\right)_n,
\end{equation}
which is precisely the factor multiplying $\bcal{R}$ in Eq.~\eqref{3.9}.

The first-order contribution $\zeta_U$ to the cooling rate can still be written as
\beq
\label{3.14}
\zeta_U=\zeta^{(1,0)}+\zeta^{(1,1)},
\eeq
where

\begin{equation}\label{zeta10}
\begin{aligned}
\zeta^{(1,0)}={}&\chi\phi
\left[2\overline\Delta^{*2}-\frac{3}{2}(1-\alpha^2)\right]\\
&+\frac{8}{\pi}\chi\phi\alpha\overline\Delta^*
\int \dd\mathbf c_1\int \dd\mathbf c_2\,
 g^*\varphi(\mathbf c_1)\varphi(\mathbf c_2).
\end{aligned}
\end{equation}
\begin{equation}
\label{3.16_d2}
\begin{aligned}
\zeta^{(1,1)}={}&-\frac{m\sqrt{\pi}}{nT}\,\chi\sigma
\int \dd\mathbf v_1\int \dd\mathbf v_2\\
&\times\left[\frac{2}{\sqrt{\pi}}\overline\Delta^2 g
+\frac{\sqrt{\pi}}{2}\alpha\overline\Delta g^2
-\frac{1-\alpha^2}{3\sqrt{\pi}}g^3\right]\\
&\times f^{(0)}(\mathbf V_1)\mathcal D(\mathbf V_2).
\end{aligned}
\end{equation}
With this result Eq.\ \eqref{3.5} can be rewritten as
\begin{equation}
\label{3.17}
\begin{aligned}
-\gamma\frac{\partial}{\partial\mathbf v}\cdot\mathbf V\mathcal D
-\zeta^{(1,1)}T\frac{\partial f^{(0)}}{\partial T}
+\mathcal L\mathcal D=D'.
\end{aligned}
\end{equation}
where
\begin{equation}
\label{3.18}
\begin{aligned}
D'(\mathbf V)={}&\frac{1}{2}\frac{\partial}{\partial\mathbf V}\cdot
\left(\mathbf V f^{(0)}\right)
+\left(\zeta^{(1,0)}+p^*\right)T\frac{\partial f^{(0)}}{\partial T}\\
&-f^{(0)}+n\frac{\partial f^{(0)}}{\partial n}
+\frac{1}{2}\mathcal K_i\left[\frac{\partial f^{(0)}}{\partial V_i}\right].
\end{aligned}
\end{equation}

\subsection{Pairwise synchronization and collisional transfer at Navier--Stokes order}
\label{ap:collisional-transfer-pairwise}

The exact expressions~\eqref{1.19} and \eqref{1.20} for $\mathsf{P}_\text{c}$ and $\mathbf{q}_\text{c}$ contain $\overline{\Delta}_{ij}$ evaluated continuously along the collision link. It is therefore useful to check explicitly whether the spatial dependence of the collision rule generates an additional direct transfer term at first order. Along the link,
\begin{equation}
\mathbf r_1=\mathbf r-x\boldsymbol\sigma,
\qquad
\mathbf r_2=\mathbf r+(1-x)\boldsymbol\sigma,
\end{equation}
and the two local collision frequencies are
\begin{align}
\omega_\text{e}(\mathbf r-x\boldsymbol\sigma)
={}&\omega_\text{e}-x\sigma\widehat{\boldsymbol\sigma}\cdot\bm\nabla\omega_\text{e}
+\mathcal O(\bm\nabla^2),\\
\omega_\text{e}(\mathbf r+(1-x)\boldsymbol\sigma)
={}&\omega_\text{e}+(1-x)\sigma\widehat{\boldsymbol\sigma}\cdot\bm\nabla\omega_\text{e}
+\mathcal O(\bm\nabla^2).
\end{align}
Hence,
\begin{align}
\label{eq:barDelta-link-expansion}
&\overline{\Delta}_{ij}
\left(\mathbf r-x\boldsymbol\sigma,\mathbf r+(1-x)\boldsymbol\sigma,t\right)
\nonumber\\
&\quad=\overline{\Delta}
+(1-2x)\sigma\overline{\Delta}
\frac{\omega_\text{e}^*\tau_\text{s}^*}
{\mathrm e^{2\omega_\text{e}^*\tau_\text{s}^*}-1}
\widehat{\boldsymbol\sigma}\cdot\bm\nabla\ln\omega_\text{e}
+\mathcal O(\bm\nabla^2).
\end{align}
The new pairwise correction is therefore antisymmetric with respect to the midpoint $x=1/2$. Since
\begin{equation}
\int_0^1\dd x\,(1-2x)=0,
\end{equation}
its direct contribution to the collisional pressure tensor vanishes. The same conclusion holds for the terms in the heat flux that are linear or quadratic in $\overline{\Delta}_{ij}$, because
\begin{equation}
\overline{\Delta}_{ij}^2
=\overline{\Delta}^2
+2\overline{\Delta}\left[\overline{\Delta}_{ij}-\overline{\Delta}\right]
+\mathcal O(\bm\nabla^2),
\end{equation}
and the first-order correction again integrates to zero. Products of this correction with the first-order spatial expansion of the Enskog pair product are of second order
and are consistently discarded.

Consequently, the explicit collisional-transfer functionals at Navier--Stokes order retain the same formal structure as in the conventional $\Delta$-model after the replacement $\Delta\to\overline{\Delta}$. No additional direct synchronization-gradient term is generated in $\mathsf P_{\rm c}^{(1)}$ or $\mathbf q_{\rm c}^{(1)}$. This statement concerns the direct transfer contribution only: the collisional heat flux coefficients still depend on the corrected vector solution through $\bcal{A}$ and $\bcal{B}$, whereas the tensorial and scalar contributions determining the shear and bulk viscosities are unchanged directly by $\bcal{R}$.

\subsection{Sonine evaluation of the kinetic transport coefficients}
\label{ap:a3}

We now evaluate the kinetic contributions to the transport coefficients by using the leading Sonine approximations to the solutions of the integral equations given in Sec.~\ref{ap:a2}. The tensorial and scalar contributions retain the same formal structure as in the conventional $\Delta$-model after $\Delta\to\overline{\Delta}$. The vector contribution contains, in addition, the moments of $\bcal{R}$. We keep these contributions explicit below so that the approximation level of each result is clear.

\subsubsection{Shear viscosity}

The kinetic contribution to the shear viscosity is defined as
\beq
\label{3.16.1}
\eta_\text{k}=- \frac{1}{4}\int \dd \mathbf{v}\; D_{ij}(\mathbf{V})\; \mathcal{C}_{ij}(\mathbf{V}),
\eeq
where, for a two-dimensional system,
\begin{equation}
D_{ij}(\mathbf{V})
=
m\left(V_iV_j-\frac{1}{2}V^2\delta_{ij}\right).
\end{equation}
To evaluate $\eta_\text{k}$, we approximate the unknown function $\mathcal{C}_{ij}(\mathbf{V})$ by its leading Sonine form,
\begin{equation}
\label{vic4}
\mathcal{C}_{ij}(\mathbf{V})
\to
-\frac{\eta_\text{k}}{nT^2}\,
D_{ij}(\mathbf{V})\,f_\text{M}(\mathbf{V}),
\end{equation}
where
\begin{equation}
f_\text{M}(\mathbf{V})
=
n\frac{m}{2\pi T}
\exp\left(-\frac{mV^2}{2T}\right)
\end{equation}
is the Maxwellian distribution for hard disks.

Multiplying both sides of Eq.~\eqref{3.4} by $D_{ij}(\mathbf{V})$ and integrating over velocity yields
\begin{equation}
\begin{aligned}
(\nu_\eta+2\gamma)\eta_\text{k}={}&nT\Bigg\{
1-\frac{1}{4}\chi\phi(1+\alpha)(1-3\alpha)\\
&+\chi\phi\overline\Delta^*
\left[\sqrt{\frac{2}{\pi}}(1+2\alpha)
\left(1-\frac{a_2}{16}\right)+\overline\Delta^*\right]\Bigg\}.
\end{aligned}
\end{equation}
where $\nu_\eta$ is the collision frequency
\begin{equation}
\nu_\eta =
\frac{\displaystyle \int \dd\mathbf{v}\,
D_{ij}(\mathbf{V})\, \mathcal{L}\mathcal{C}_{ij}(\mathbf{V})}
{\displaystyle \int \dd\mathbf{v}\,
D_{ij}(\mathbf{V})\, \mathcal{C}_{ij}(\mathbf{V})}.
\end{equation}
Introducing its reduced form,
\begin{equation}
\nu_\eta^*=\frac{\nu_\eta}{n\sigma v_\text{th}},
\end{equation}
and evaluating the collision frequency within the Sonine approximation \eqref{vic4} gives the expression reported in Sec.~\ref{sec:transport}. The above equation then directly yields the kinetic contribution $\eta_\text{k}$ quoted there.

\subsubsection{Thermal conductivity and diffusive heat coefficient}

For the kinetic contributions to the heat flux coefficients, the leading Sonine approximations to the unknown functions $\bcal{A}$ and $\bcal{B}$ are
\begin{align}\label{4.9}
\bcal{A}(\mathbf{V})&\to
-\frac{1}{2}\frac{m}{nT^2}\kappa_\text{k}
f_\text{M}(\mathbf{V})\mathbf{S}(\mathbf{V}),\\
\bcal{B}(\mathbf{V})&\to
-\frac{1}{2}\frac{m}{T^3}\mu_\text{k}
f_\text{M}(\mathbf{V})\mathbf{S}(\mathbf{V})\label{4.9b}.
\end{align}
where $\kappa_\text{k}$ and $\mu_\text{k}$ are the kinetic contributions to the thermal conductivity and the diffusive heat coefficient, respectively, and
\beq
\label{4.10}
\mathbf{S}(\mathbf{V})
=
\left(\frac{m}{2}V^2-2T\right)\mathbf{V}.
\eeq
For a two-dimensional system, the kinetic coefficients are defined as
\beq
\label{3.23}
\kappa_\text{k}
=
-\frac{1}{2T}
\int \dd{\bf v}\;
{\bf S}({\bf V})\cdot
{\bcal{A}}({\bf V}),
\eeq
and
\beq
\label{3.24}
\mu_\text{k}
=
-\frac{1}{2n}
\int \dd{\bf v}\;
{\bf S}({\bf V})\cdot
{\bcal{B}}({\bf V}).
\eeq

The kinetic thermal conductivity $\kappa_\text{k}$ is obtained by multiplying both sides of Eq.~\eqref{3.2} by $\mathbf{S}(\mathbf{V})$ and integrating over velocity. The result is
\begin{widetext}
\begin{align}
\left(\nu_\kappa+\gamma-\frac{3}{2}\zeta^{(0)}-\zeta^{(0)}\frac{\partial\ln\zeta_0^*}{\partial\theta}\right)\kappa_\text{k}
={}&\frac{2nT}{m}(1+2a_2+\Upsilon_\theta)
-\frac{1}{4T}\int \dd\mathbf V\,\mathbf S(\mathbf V)\cdot\bcal{K}
\left[\frac{\partial}{\partial\mathbf V}\cdot\mathbf V f^{(0)}\right]\nonumber\\
&+\frac{nv_\text{th}^{-2}}{2T}\int \dd\mathbf V\,\mathbf S(\mathbf V)\cdot\bcal{K}
\left[\frac{\partial\varphi}{\partial\theta}\right]
-\frac{1}{2T}\int\dd\mathbf V\,\mathbf S(\mathbf V)\cdot\bcal{R}(\mathbf V).
\end{align}
Up to linear order in $a_2$, the first collision integral on the right-hand side is
\begin{align}
\int \dd\mathbf V\,\mathbf S(\mathbf V)\cdot\bcal{K}
\left[\frac{\partial}{\partial\mathbf V}\cdot(\mathbf V f^{(0)})\right]
={}&8\frac{nT^2}{m}\chi\phi\Bigg\{\frac{3}{8}(1+\alpha)^2[1-2\alpha-a_2(1+\alpha)]\nonumber\\
&+\frac{\overline\Delta^*}{\sqrt{2\pi}}\Bigg[\left(\frac{3}{4}-\frac{97a_2}{64}\right)
+3(1+\alpha)\left(1-\frac{\sqrt{2\pi}}{2}\overline\Delta^*\right)\nonumber\\
&-\frac{a_2}{8}\left(21+\frac{27}{4}\alpha\right)\alpha
-\frac{9}{2}(1+\alpha)^2
-\overline\Delta^{*2}\left(1-\frac{a_2}{16}\right)\Bigg]\Bigg\},
\end{align}
whereas the second one is
\begin{align}
\frac{nv_\text{th}^{-2}}{2T}\int \dd\mathbf V\,\mathbf S(\mathbf V)\cdot\bcal{K}
\left[\frac{\partial\varphi}{\partial\theta}\right]
={}&\frac{3}{8}\chi\phi\frac{nT}{m}\Upsilon_\theta
\Bigg\{(1+\alpha)^3+\frac{\overline\Delta^*}{2\sqrt{2\pi}}
\left[13+2\alpha(8+3\alpha)-\frac{4}{3}\overline\Delta^{*2}\right]\Bigg\}.
\end{align}
The last term is the new contribution generated by the pairwise synchronization source. Using the weak representation~\eqref{eq:Rsync-weak} and the Sonine approximation~\eqref{2.15}, one obtains (to linear order in $a_2$) the result
\begin{align}\label{eq:Rsync-S-moment}
-\frac{1}{2T}\int\dd\mathbf V\,\mathbf S(\mathbf V)\cdot\bcal{R}(\mathbf V)
=\frac{2nT}{m}\,\chi\phi\overline{\Delta}^*
\frac{2\omega_\text{e}^*\tau_\text{s}^*}
{\mathrm e^{2\omega_\text{e}^*\tau_\text{s}^*}-1}
\Bigg\{\frac{3}{4}\left[
\frac{2\alpha^2+4\overline{\Delta}^{*2}-1}{\sqrt{2\pi}}
+2\alpha\overline{\Delta}^*\right]
+\frac{a_2}{64\sqrt{2\pi}}
\left(47+18\alpha^2-12\overline{\Delta}^{*2}\right)\Bigg\}.
\end{align}
This term vanishes for $\tau_\text{s}^*\to\infty$, as it must in the conventional $\Delta$-model.\cite{GBS18}
\end{widetext}

The collision frequency $\nu_\kappa$ is defined as
\begin{equation}
\nu_\kappa
=
\frac{
\displaystyle
\int \dd{\bf v}\;
{\bf S}({\bf V})\cdot
{\cal L}\bcal{A}({\bf V})
}{
\displaystyle
\int \dd{\bf v}\;
{\bf S}({\bf V})\cdot
\bcal{A}({\bf V})
},
\end{equation}
and its reduced form is
\begin{equation}
\nu_\kappa^*
=
\frac{\nu_\kappa}{n\sigma v_\text{th}}.
\end{equation}
Using the leading Sonine approximation \eqref{4.9}, one gets  Eq.\ \eqref{vic5}. Substitution of the above collision integrals then gives the kinetic thermal conductivity $\kappa_\text{k}$ reported there.

The kinetic diffusive heat coefficient $\mu_\text{k}$ follows in the same way from Eq.~\eqref{3.3}. Multiplying that equation by $\mathbf{S}(\mathbf{V})$ and integrating over velocity gives
\begin{widetext}
\begin{align}
\label{3.29}
\left(\nu_\kappa+3\gamma\right)\mu_\text{k}
=
-\frac{1}{2n}
\int \dd\mathbf{v}\;
\mathbf{S}\cdot\mathbf{B}
+
\Bigg[
\zeta^{(0)}
\left(
1+\phi\frac{\partial\ln\chi}{\partial\phi}
\right)
\left(
1+
\frac{2\omega_\text{e}^*\tau_\text{s}^*}
{\mathrm{e}^{2\omega_\text{e}^*\tau_\text{s}^*}-1}
\overline{\Delta}^*
\frac{\partial\ln\zeta_0^*}
{\partial\overline{\Delta}^*}
\right)
\nonumber\\
+
\chi\phi
\frac{\partial\chi}{\partial\phi}
\frac{\partial}{\partial\chi}
\left(
\frac{\zeta^{(0)}}{\chi}
\right)
-
\zeta^{(0)}\gamma^*
\frac{\partial\ln\zeta_0^*}
{\partial\gamma^*}
\Bigg]
\frac{T}{n}\kappa_\text{k},
\end{align}
where the factor involving $\omega_\text{e}^*\tau_\text{s}^*$ in the coupling to $\kappa_\text{k}$ comes from the density dependence of the local zeroth-order amplitude $\overline{\Delta}$. The first term on the right-hand side of Eq.~\eqref{3.29} can be written as
\begin{align}
\label{3.30}
-\frac{1}{2n}\int \dd\mathbf{v}\,\mathbf{S}\cdot \mathbf{B}= \frac{2T^2}{m}\left(a_2+\phi\frac{\partial a_2}{\partial \phi}\right)+ \frac{1}{2n}\left(1+\frac{1}{2}\phi\frac{\partial \ln\chi}{\partial\phi}\right)
\int \dd\mathbf{v}\;\mathbf{S}\cdot \bcal{K}\left[f^{(0)}\right] +\frac{v_\text{th}^{-2}}{2}
\int \dd\mathbf{v}\, \mathbf{S}\cdot \bcal{K}\left[\phi
\frac{\partial \varphi}{\partial \phi}\right] \nonumber \\
-\frac{1}{n}\left(1+\phi\frac{\partial\ln\chi}{\partial\phi}\right)
\int\dd\mathbf V\,\mathbf S(\mathbf V)\cdot\bcal{R}(\mathbf V),
\end{align}
Neglecting nonlinear terms in $a_2$, the collision integrals appearing in Eq.~\eqref{3.30} are
\begin{align}
\int \dd\mathbf{V}\;\mathbf{S}(\mathbf{V})\cdot\bcal{K}\!\left[f^{(0)}\right]
&=
8\,\frac{nT^2}{m}\,\chi\phi\Bigg\{
\frac{3}{8}(1+\alpha)\left[\alpha(\alpha-1)+
\frac{a_2}{6}\left(14-3\alpha+3\alpha^2\right)\right] \nonumber\\
&\quad+ \frac{1}{\sqrt{2\pi}}\,\overline{\Delta}^{*}\Bigg[
2\,\overline{\Delta}^{*2}
-3\left(\frac12-\alpha^{2}\right)
+\frac{3}{2}\sqrt{2\pi}\,\alpha\,\overline{\Delta}^{*}
+ \frac{a_2}{32}\left(47+18\alpha^{2}-4\overline{\Delta}^{*2}\right)\Bigg]\Bigg\},
\end{align}
and
\begin{equation}
\frac{v_{\text{th}}^{-2}}{2}
\int \dd\mathbf{V}\;\mathbf{S}(\mathbf{V})\cdot\bcal{K}\!\left[\phi \frac{\partial \varphi}{\partial \phi}\right]
=
\frac{3}{8}\,\chi\phi^2\,\frac{T^2}{m}\,\frac{\partial a_2}{\partial \phi}
\Bigg\{\left(1+\alpha\right)^3
+\frac{1}{2\sqrt{2\pi}}\,\overline{\Delta}^{*}
\left[13+2\alpha(8+3\alpha)-\frac{4}{3}\overline{\Delta}^{*2}\right]\Bigg\}.
\end{equation}
Finally, the additional synchronization moment follows from Eq.~\eqref{eq:Rsync-S-moment}:
\begin{align}\label{eq:Rsync-S-density}
-\frac{1}{n}\left(1+\phi\frac{\partial\ln\chi}{\partial\phi}\right)
\int\dd\mathbf V\,\mathbf S(\mathbf V)\cdot\bcal{R}(\mathbf V)
=\frac{4T^2}{m}\left(1+\phi\frac{\partial\ln\chi}{\partial\phi}\right)
\chi\phi\overline{\Delta}^*
\frac{2\omega_\text{e}^*\tau_\text{s}^*}
{\mathrm e^{2\omega_\text{e}^*\tau_\text{s}^*}-1}
\nonumber\\
\times\Bigg\{\frac{3}{4}\left[
\frac{2\alpha^2+4\overline{\Delta}^{*2}-1}{\sqrt{2\pi}}
+2\alpha\overline{\Delta}^*\right]+\frac{a_2}{64\sqrt{2\pi}}
\left(47+18\alpha^2-12\overline{\Delta}^{*2}\right)\Bigg\}.
\end{align}
\end{widetext}
Insertion of these results, together with Eq.~\eqref{eq:Rsync-S-density} and the corrected $\kappa_\text{k}$ entering the coupling term in Eq.~\eqref{3.29}, gives the kinetic coefficient $\mu_\text{k}$ displayed in Sec.~\ref{sec:transport}.

\subsection{Sonine evaluation of the first-order contribution to the cooling rate}
\label{ap:a4}

As shown in Eqs.~\eqref{3.14}--\eqref{3.16_d2}, the pairwise synchronization correction does not generate an additional direct contribution to the scalar quantity $\bm\nabla\cdot\mathbf U$. The first-order contribution to the cooling rate can therefore be evaluated from
\[
\zeta_U=\zeta^{(1,0)}+\zeta^{(1,1)}.
\]
On the one hand, inserting the Sonine approximation \eqref{2.15} into the formal expression \eqref{zeta10} and retaining terms up to linear order in \(a_2\), one obtains

\begin{equation}
\label{3.15_d2}
\zeta^{(1,0)}=\chi\phi \left[2 \overline{\Delta}^{*2}
+4\sqrt{\frac{2}{\pi}}\left(1-\frac{1}{16}a_2\right)\alpha \overline{\Delta}^{*}
-\frac{3}{2}(1-\alpha^2)\right],
\end{equation}

On the other hand, $\zeta^{(1,1)}$ depends on the unknown scalar function $\mathcal{D}(\mathbf{V})$. We now evaluate this latter contribution within the leading Sonine approximation. For a two-dimensional system, the function $\mathcal{D}(\mathbf{V})$ is approximated as
\begin{equation}
\label{vic6}
\mathcal{D}(\mathbf{V})
\to
e_D\,E(\mathbf{V})\,f_\text{M}(\mathbf{V}),
\end{equation}
where $e_D$ is a Sonine coefficient and
\begin{equation}
E(\mathbf{V})
=
\left(
\frac{mV^2}{2T}
\right)^2
-
4\left(
\frac{mV^2}{2T}
\right)
+2.
\end{equation}
With this approximation, the contribution $\zeta^{(1,1)}$ is proportional to $e_D$. It is therefore convenient to write
\begin{equation}
\zeta^{(1,1)}
=
\overline{\zeta}^{(1,1)}e_D.
\end{equation}

The coefficient $e_D$ is obtained by multiplying Eq.~\eqref{3.17} by $E(\mathbf{V})$ and integrating over velocity. After some algebra, one obtains

\beq
\label{3.35}
\Bigg[
\nu_\zeta
+4\gamma
-\overline{\zeta}^{(1,1)}
\left(
a_2+\frac{1}{2}\Upsilon_\theta
\right)
\Bigg]e_D
=
\frac{1}{4n}
\int \dd\mathbf{v}\;
E(\mathbf{V})D'(\mathbf{V}),
\eeq

where the collision frequency $\nu_\zeta$ is defined as
\beq
\label{3.36}
\nu_\zeta
=
\frac{
\displaystyle
\int \dd\mathbf{v}\;
E(\mathbf{V})
\mathcal{L}\mathcal{D}(\mathbf{V})
}{
\displaystyle
\int \dd\mathbf{v}\;
E(\mathbf{V})
\mathcal{D}(\mathbf{V})
}.
\eeq

Within the Sonine approximation \eqref{vic6} , the coefficient $\overline{\zeta}^{(1,1)}$ becomes
\begin{widetext}
\begin{equation}
\overline{\zeta}^{(1,1)}
=
\frac{\sqrt{2\pi}}{512}
\frac{v_\text{th}}{\ell}\,
\chi
\left[
96(1-\alpha^2)
+
64\overline{\Delta}^{*2}
+
a_2
\left(
9(1-\alpha^2)
+
30\overline{\Delta}^{*2}
\right)
\right].
\end{equation}
Moreover, introducing the reduced collision frequency
\begin{equation}
    \nu_\zeta^*=\frac{\ell\nu_\zeta}{v_\text{th}},
\end{equation}
one gets the expression displayed in Sec. \ref{sec:transport}. Now, the right-hand side of Eq.~\eqref{3.35} can be expressed as
\beq
\label{3.37-new-d2}
\frac{1}{4n}
\int \dd\mathbf{v}\;
E(\mathbf{V})D'(\mathbf{V})
=
-a_2
+
\left(
\zeta^{(1,0)}+p^*
\right)
\left(
a_2+\frac{1}{2}\Upsilon_\theta
\right)
+
\frac{1}{2}\phi
\frac{\partial a_2}{\partial\phi}
+
\frac{1}{8n}I_\zeta,
\eeq
where, up to linear order in $a_2$,
\begin{align}
    I_\zeta &= \int \dd\mathbf{v}\; E(\mathbf{V})\,\mathcal{K}_i\left[\frac{\partial f^{(0)}}{\partial V_i}\right] \nonumber\\
    & = n\chi\phi\Bigg\{\frac{3}{4} \left(1+\alpha\right)\left[(1-\alpha^2)(5\alpha-1)-
        \frac{a_2}{6}\left(15\alpha^3-3\alpha^2+69\alpha-41\right)\right]- \frac{2}{\sqrt{2\pi}} \overline{\Delta}^* (I_0 + a_2 I_1) \Bigg\} 
\end{align}
with
\begin{align}
I_0
={}&
-6
+
4\alpha
\left(
-3+3\alpha+4\alpha^2
\right)
+
3\sqrt{2\pi}
(1+\alpha)(-1+3\alpha)
\overline{\Delta}^*
+
8(1+2\alpha)
\overline{\Delta}^{*2}
+
2\sqrt{2\pi}\,
\overline{\Delta}^{*3},
\end{align}
\end{widetext}
and
\begin{equation}
I_1
=
\frac{1}{8}
\left[
31
+
2\alpha
\left(
47+9\alpha+12\alpha^2
\right)
-
4(1+2\alpha)
\overline{\Delta}^{*2}
\right].
\end{equation}
These results allow us to give a closed expression \eqref{eq:eD} for the dimensionless Sonine coefficient $e_D^*$.

\section{Assessment of the synchronization mean-field closure}
\label{app:a2deviations}

Most of the kinetic theory developed in this work consistently retains the fourth cumulant $a_2$ only to linear order. This approximation is sufficient over most of the active branch, where $|a_2|\ll 1$, but EDMD simulations show a
pronounced increase of $a_2$ close to the synchronization-controlled absorbing transition.
Interestingly, DSMC simulations based on the same effective
mean-field collision rule remain close to the analytical
prediction in this regime. This observation suggests that
the discrepancy does not primarily originate from the
Sonine evaluation of the effective kinetic equation, but
from the coarse-graining of the microscopic synchronization
dynamics itself.

To further assess the limitations of this coarse-graining, we first compare different levels of approximation for the homogeneous steady temperature. We then examine the two
main approximations underlying the synchronization closure,
namely, the suppression of fluctuations of the
collision-dependent injection amplitude and the neglect
of its correlations with the precollisional velocities.
Throughout this Appendix, all collisional averages and
probabilities refer to the spatially homogeneous steady state.

\begin{figure}[t]
\centering
\includegraphics[width=\columnwidth]{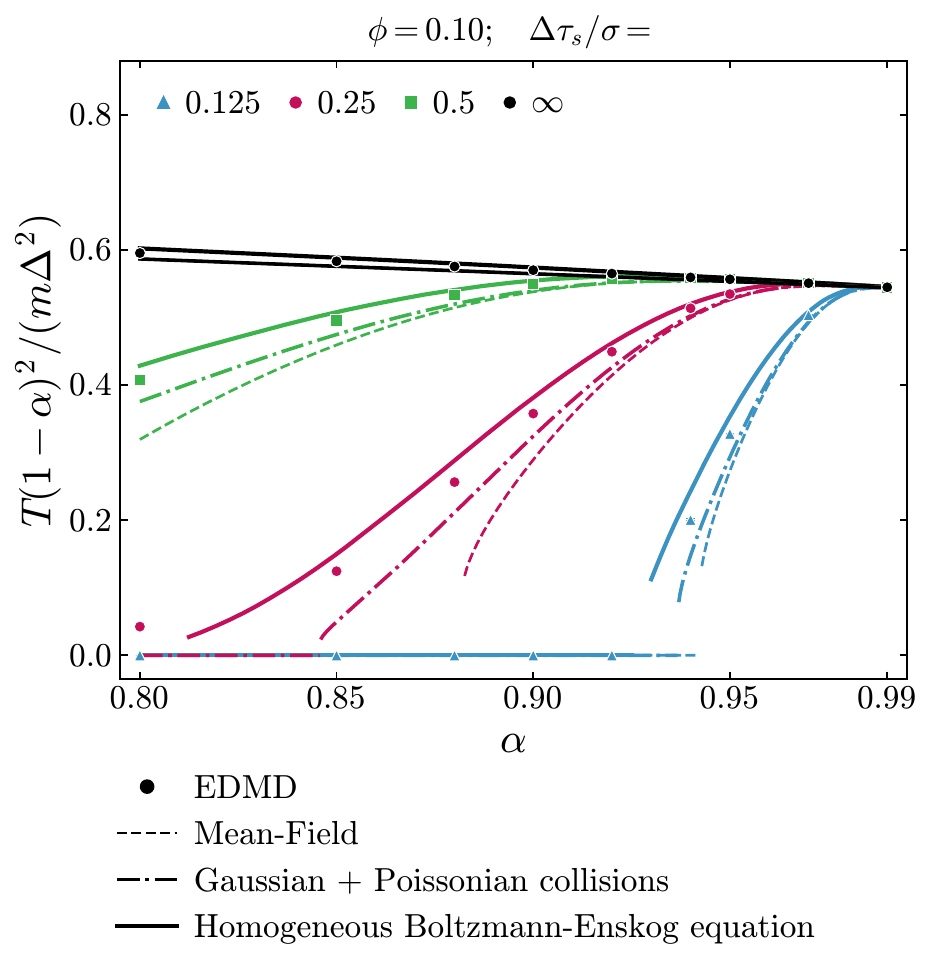}
\caption{Scaled steady granular temperature
$T(1-\alpha)^2/(m\Delta^2)$ as a function of the coefficient
of restitution $\alpha$ for $\phi=0.10$ and different values
of $\Delta\tau_{\rm s}/\sigma$.
Symbols are EDMD simulation results, while dashed,
dash-dotted, and solid lines correspond to the deterministic
mean-field approximation, the Gaussian--Poisson approximation,
and the numerical solution of the homogeneous Boltzmann--Enskog equation,
respectively.
In all cases, $\gamma\sigma/\Delta=0.04$.}
\label{fig:homogeneous-comparison}
\end{figure}

Figure~\ref{fig:homogeneous-comparison} compares the steady
granular temperature obtained from EDMD simulations with
three theoretical descriptions of increasing complexity.
The first is the deterministic synchronization mean-field
theory developed in the main text. The second combines a
Gaussian velocity distribution with Poissonian collision
histories, retaining the binary character of the
collision-dependent injection amplitude. The third is
obtained by numerically solving the homogeneous Boltzmann--Enskog
equation while explicitly accounting for the synchronization
dynamics through the collision histories of the particles. The comparison shows that the deterministic mean-field
approximation tends to underestimate the steady temperature
and predicts a more restricted active branch, particularly
when synchronization becomes important. Retaining the
binary nature of the injection amplitude within the
Gaussian--Poisson approximation already improves the
agreement with EDMD. A further improvement is obtained,
over much of the parameter range, by solving the homogeneous
Boltzmann--Enskog equation with explicit collision-history dynamics.
These results confirm that the mean-field treatment of
synchronization constitutes an important source of the
discrepancies with microscopic simulations. Nevertheless, deviations from EDMD persist even at the homogeneous kinetic level. Such residual differences may originate from precollisional velocity correlations and spatial correlations associated with synchronized and asynchronized regions, neither of which is fully captured by a spatially homogeneous Boltzmann--Enskog description based on molecular chaos. The present comparison does not allow us to disentangle their respective contributions.

To better understand the origin of these discrepancies,
we now examine the statistical approximations entering
the synchronization mean-field closure, focusing first
on the fluctuations of the collision-dependent injection
amplitude.

\subsection{Poisson statistics and fluctuations of the collision-dependent injection}

For the microscopic synchronization rule, it is useful to introduce the binary variable
\begin{equation}
I_{ij}=\begin{cases}
1, & \delta t_i<\tau_{\rm s}\ \text{or}\ \delta t_j<\tau_{\rm s},\\
0, & \text{otherwise},
\end{cases}
\qquad
\Delta_{ij}=I_{ij}\Delta.
\label{eq:C1}
\end{equation}
If $\mathcal{P}\equiv\langle I_{ij}\rangle_{\rm coll}$ denotes the fraction of injecting collisions, the exact binary character of Eq.~\eqref{eq:C1} implies
\begin{equation}
\frac{\langle \Delta_{ij}^{n}\rangle_{\rm coll}}{\Delta^n}=\mathcal{P},
\qquad n\in\mathbb N.
\label{eq:C2}
\end{equation}
For the homogeneous state considered here, Eq.~\eqref{1.25} gives the Poisson estimate adopted in the kinetic theory,
\begin{equation}
\mathcal{P}_{\rm P}=1-\exp[-2\omega_{\rm e}(\phi,T)\tau_{\rm s}].
\label{eq:C3}
\end{equation}

\begin{figure*}[t]
\centering
\fullwidthgraphic{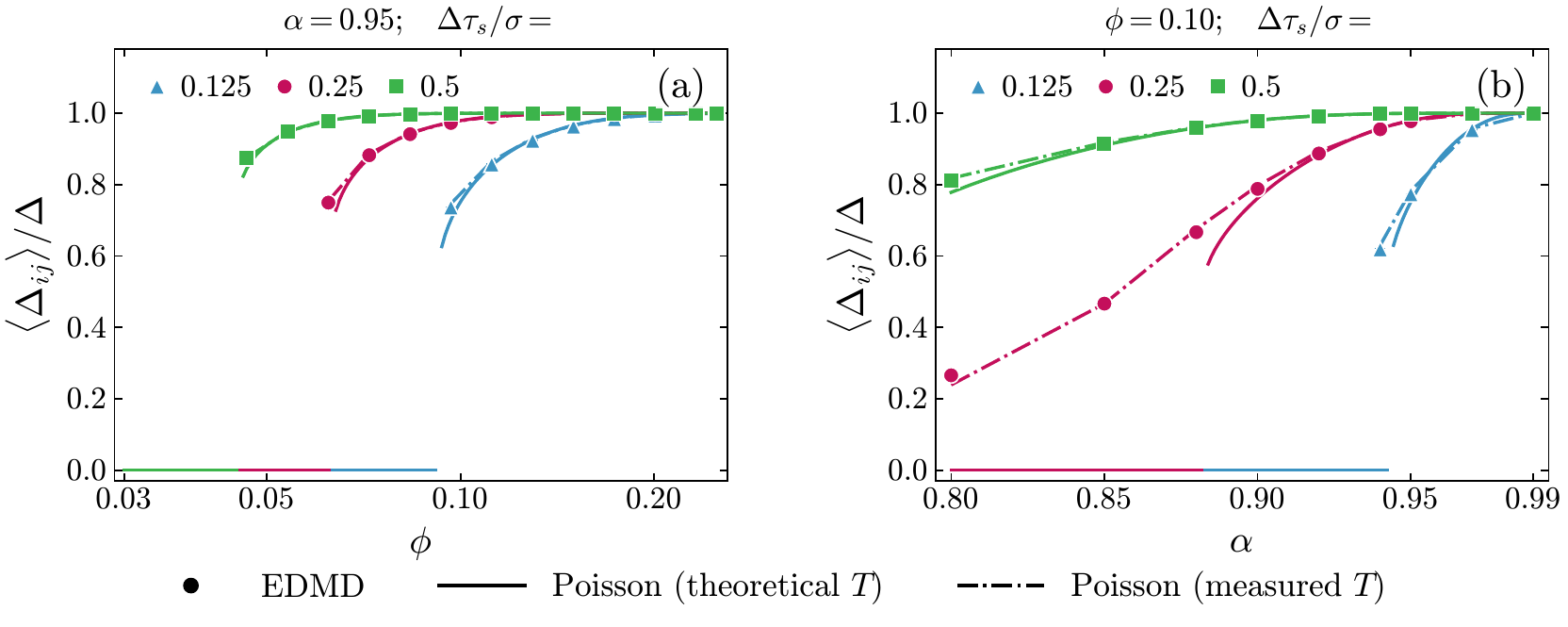}
\caption{Collisional average $\langle\Delta_{ij}\rangle_{\rm coll}/\Delta$ as a function of (a) the packing fraction $\phi$ for $\alpha=0.95$, and (b) the coefficient of restitution $\alpha$ for $\phi=0.10$. Symbols are EDMD results for $\Delta\tau_{\rm s}/\sigma=0.125$, $0.25$, and $0.5$. Lines are the Poisson estimates obtained by using either the theoretical steady temperature or the temperature measured in EDMD. In all cases, $\gamma\sigma/\Delta=0.04$.}
\label{fig:poisson}
\end{figure*}

In order to assess the accuracy of the Poisson approximation in Eq.~\eqref{eq:C3}, Fig.~\ref{fig:poisson} shows the collisional average $\langle\Delta_{ij}\rangle_{\rm coll}/\Delta$, i.e., the fraction $\mathcal{P}$ of collisions for which the microscopic energy injection is active, as a function of $\phi$ at fixed $\alpha=0.95$, and as a function of $\alpha$ at fixed $\phi=0.10$. The EDMD results are compared with the corresponding Poisson predictions. The excellent agreement obtained when the Poisson estimate is evaluated using the measured steady temperature shows that the collision-time statistics are very well described by the Poisson approximation over the range considered. This therefore rules out deviations from Poisson collision statistics as the main origin of the discrepancies observed in $a_2$. As a consequence, in the remainder of this section we will not make any further distinction between the Poisson estimate $\mathcal{P}_{\rm P}$ and its microscopic counterpart $\mathcal{P}$.

Within this homogeneous setting, the subsequent mean-field replacement~\eqref{1.25} is, however, more restrictive. The substitution $\Delta_{ij}\to\overline{\Delta}=\mathcal{P}\Delta$ yields
\begin{equation}
\frac{\overline{\Delta}^{\,n}}{\Delta^n}=\mathcal{P}^n,
\label{eq:C4}
\end{equation}
in contrast with the result~\eqref{eq:C2} obtained from the microscopic binary dynamics. This distinction directly reflects the suppression of the fluctuations of the binary random variable $\Delta_{ij}$, and it is particularly relevant for the fourth cumulant. Since $a_2$ is determined by the fourth velocity moment, and the collision rule depends linearly on $\Delta_{ij}$, the corresponding collisional fourth-moment production contains powers of $\Delta_{ij}$ up to $\Delta_{ij}^4$. In particular, the quartic moment entering the collisional production of the fourth velocity moment is changed from $\mathcal{P}\Delta^4$ to $\mathcal{P}^4\Delta^4$. Thus, whenever $P<1$, the deterministic closure can strongly underestimate the higher-order fluctuations of the collision-dependent injection amplitude, an effect that becomes especially relevant on approaching the synchronized transition.

\begin{figure*}[t]
\centering
\fullwidthgraphic{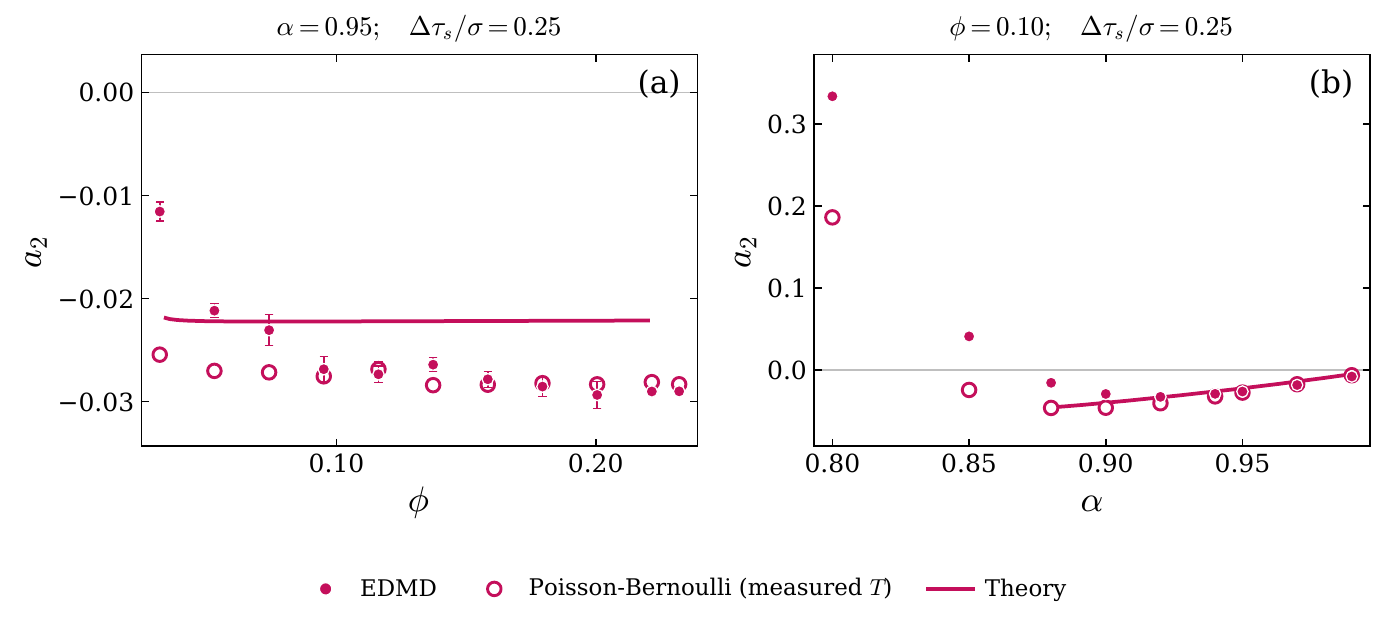}
\caption{Fourth cumulant $a_2$ as a function of (a) the packing fraction $\phi$ for fixed $\alpha=0.95$ and (b) the coefficient of restitution $\alpha$ for fixed $\phi=0.10$, for $\Delta\tau_{\rm s}/\sigma=0.25$. Filled circles correspond to the full microscopic EDMD simulations with collision-history-dependent synchronization. Open circles show the auxiliary Poisson--Bernoulli dynamics, in which the binary character of the injection is retained through $\Delta_{ij}=I_{ij}^{\rm B}\Delta$ with $I_{ij}^{\rm B}$ drawn independently from a Bernoulli distribution of probability $P$, while the correlations between the synchronization state and the pre-collisional velocities are neglected. The solid line is the prediction of the deterministic mean-field kinetic theory, where $\Delta_{ij}$ is replaced by $\overline{\Delta}=\mathcal{P}\Delta$. In the Poisson--Bernoulli simulations, $\mathcal{P}$ is evaluated from the Poisson estimate using the steady temperature measured in EDMD. In all cases, $\gamma\sigma/\Delta = 0.04$.}
\label{fig:poisson-bernoulli}
\end{figure*}

To quantify the effect of these neglected fluctuations, we introduce an auxiliary Poisson--Bernoulli dynamics in which the microscopic collision clocks are removed, while the binary character of the injection is retained. For each collision, an independent Bernoulli variable $I_{ij}^{\rm B}$ is drawn with probability $\mathcal{P}$, and the injection amplitude is set to $\Delta_{ij}=I_{ij}^{\rm B}\Delta$. By construction, this dynamics satisfies $\langle\Delta_{ij}^n\rangle_{\rm coll}=\mathcal{P}\Delta^n$ for every $n\geq1$, but does not retain any correlation between the injection state and the individual collision histories of the colliding particles. Figure~\ref{fig:poisson-bernoulli} compares the resulting $a_2$ with the full microscopic EDMD results and with the deterministic mean-field kinetic theory for $\Delta\tau_{\rm s}/\sigma=0.25$, as a function of $\phi$ at fixed $\alpha=0.95$ and as a function of $\alpha$ at fixed $\phi=0.10$. As revealed by Fig.~\ref{fig:poisson-bernoulli}(b), the Poisson--Bernoulli dynamics significantly improves $a_2$ with respect to the deterministic closure close to the transition, but it does not reproduce the full-clock EDMD result. For example, for $\alpha=0.80$, restoring the binary fluctuations of $\Delta_{ij}$ accounts for about $44\%$ of the discrepancy between both dynamics. Therefore, restoring the fluctuations of $\Delta_{ij}$ is important but is not sufficient to account for the observed fourth cumulant.

\subsection{Velocity--synchronization correlations}

\begin{figure*}[t]
\centering
\fullwidthgraphic{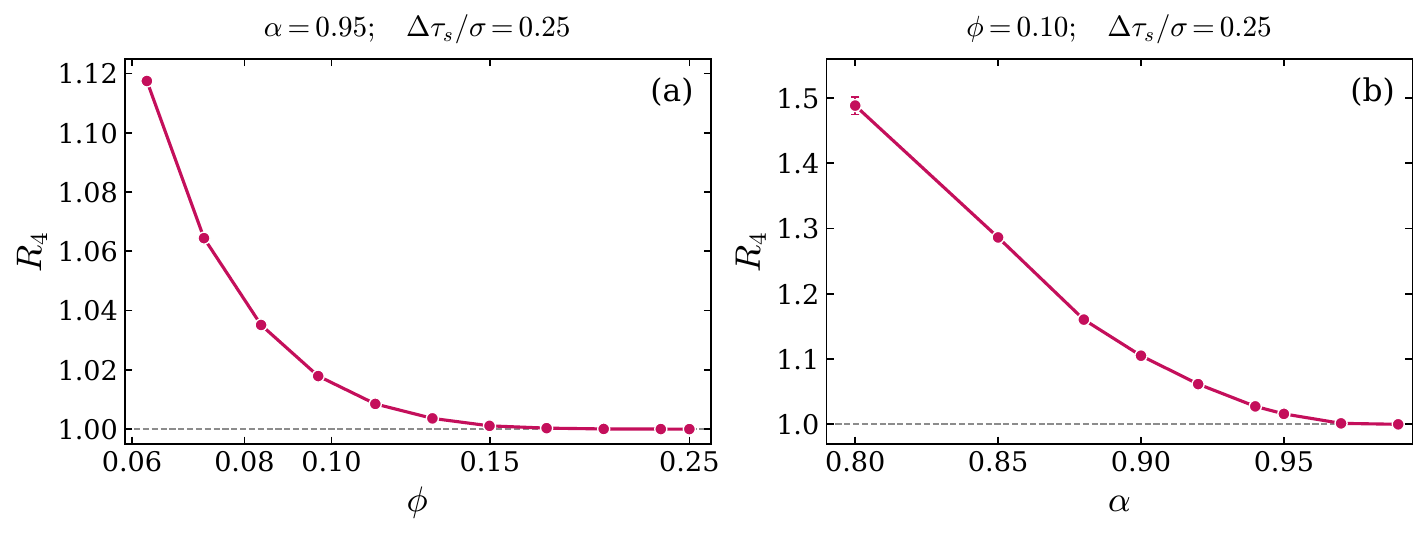}
\caption{Collisional velocity--synchronization correlation ratio $R_4=\langle I_{ij}g_\text{n}^4\rangle_{\rm coll}/[\langle I_{ij}\rangle_{\rm coll}\langle g_\text{n}^4\rangle_{\rm coll}]$ as a function of (a) the packing fraction $\phi$ at fixed $\alpha=0.95$, and (b) the coefficient of restitution $\alpha$ at fixed $\phi=0.10$, for $\Delta\tau_{\rm s}/\sigma=0.25$. Symbols are EDMD simulation results, while the dashed horizontal line indicates the factorized limit $R_4=1$. The results show that the factorization between the synchronization state and the pre-collisional normal relative velocity is very accurate far from the transition, but it breaks down progressively on approaching the low-temperature edge of the synchronized active branch, where $R_4$ departs significantly from unity. In all cases, $\gamma\sigma/\Delta = 0.04$}.
\label{fig:R4}
\end{figure*}

The remaining approximation is the factorization of the synchronization state from the pre-collisional velocities. To test it directly, we define
\begin{equation}
R_q\equiv
\frac{\langle I_{ij}g_\text{n}^q\rangle_{\rm coll}}
{\langle I_{ij}\rangle_{\rm coll}\langle g_\text{n}^q\rangle_{\rm coll}},
\qquad
g_\text{n}=\widehat{\boldsymbol\sigma}\cdot\mathbf g.
\label{eq:C5}
\end{equation}
The factorization~\eqref{1.24} used in the kinetic closure requires $R_q=1$. As shown in Fig.~\ref{fig:R4} for $q=4$, this condition is well satisfied far from the transition, but breaks down progressively on approaching the low-temperature edge of the synchronized active branch. Moreover, the deviation increases strongly with the order of the velocity moment, $R_4-1>R_2-1>R_1-1$. The direct relevance of this effect to $a_2$ can be seen from the pair fourth-moment increment. Let
\begin{equation}
S_4\equiv v_1^4+v_2^4.
\label{eq:C6}
\end{equation}
For the binary rule $\Delta_{ij}=I_{ij}\Delta$, the exact collisional increment can be easily obtained from the collision rules~\eqref{1.1}, with the result
\begin{equation}
\delta S_4=\delta S_4^{(\Delta=0)}+I_{ij}Q_4,
\label{eq:C7}
\end{equation}
where the identity $I_{ij}^n=I_{ij}$, for all $n>1$, has been used.

In Eq.~\eqref{eq:C7}, the quantity $\delta S_4^{(\Delta=0)}$ refers to the corresponding IHS contribution, whereas $Q_4(\mathbf g,\mathbf G,\widehat{\boldsymbol\sigma};\alpha,\Delta)$ is a velocity-dependent function which accounts for the contributions coming from $\Delta$. We will not display its exact expression here, which can be directly derived from the collision rules~\eqref{1.1} and \eqref{1.1b}. Now, multiplication of the homogeneous Enskog equation~\eqref{2.1} by $v^4$ and its integration in velocities yields the evolution equation for $\langle v^4\rangle$. In the steady state, it can be shown that $a_{2,\rm ss}+1\propto\langle\delta S_4\rangle_{\rm coll}$, so that the factorization approximation~\eqref{1.24} amounts to replacing
\begin{equation}
\langle I_{ij}Q_4\rangle_{\rm coll}
\longrightarrow
\langle I_{ij}\rangle_{\rm coll}\langle Q_4\rangle_{\rm coll}.
\label{eq:C8}
\end{equation}
Consequently, the corresponding neglected contribution is the covariance
\begin{equation}
\delta_4^{\rm corr}=\langle(I_{ij}-P)Q_4\rangle_{\rm coll}.
\label{eq:C9}
\end{equation}

\begin{figure*}[t]
\centering
\fullwidthgraphic{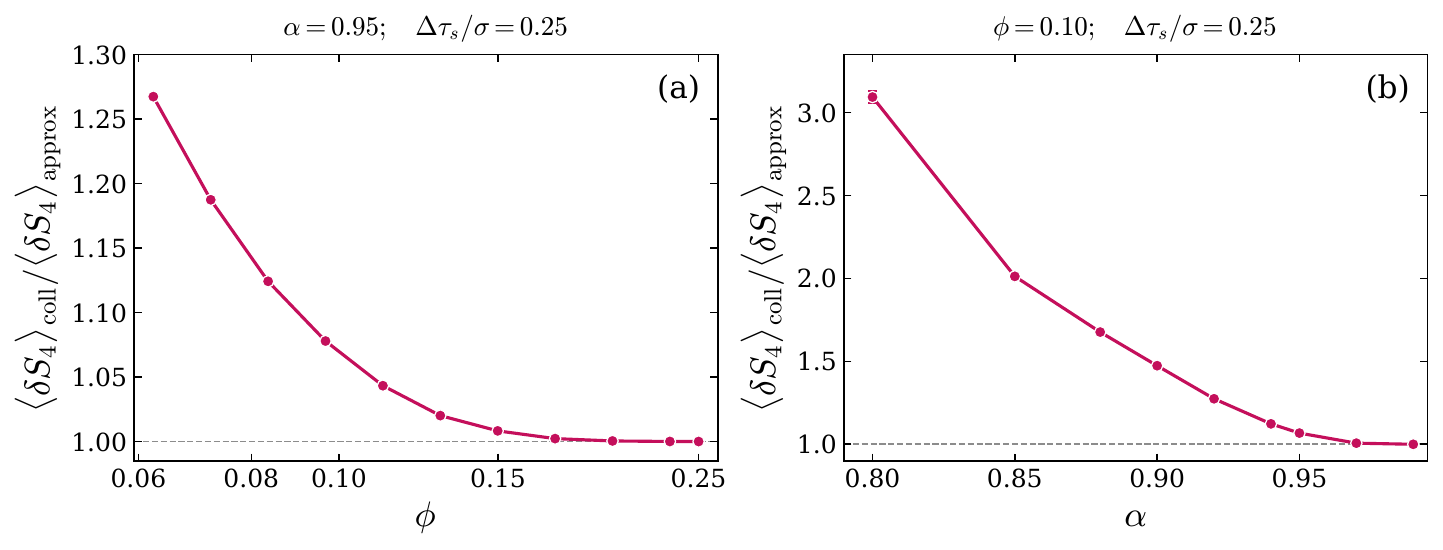}
\caption{Ratio between the actual collisional production of the pair fourth moment, $\langle\delta S_4\rangle_{\rm coll}$, and the corresponding value obtained after factorizing the synchronization state from the pre-collisional velocities, $\langle\delta S_4\rangle_{\rm approx}$, as a function of (a) the packing fraction $\phi$ for fixed $\alpha=0.95$ and (b) the coefficient of restitution $\alpha$ for fixed $\phi=0.10$. In both panels, $\Delta\tau_{\rm s}/\sigma=0.25$. Symbols are microscopic-clock EDMD results. The dashed horizontal line denotes the factorized limit, $\langle\delta S_4\rangle_{\rm coll}/\langle\delta S_4\rangle_{\rm approx}=1$. In all cases, $\gamma\sigma/\Delta = 0.04$.}
\label{fig:S4ratio}
\end{figure*}

For the sake of quantification, Figure~\ref{fig:S4ratio} compares the actual collisional fourth-moment production with the counterfactual value obtained by imposing the factorization approximation on the same ensemble of pre-collisional velocities. More specifically, we plot the ratio $\langle\delta S_4\rangle_{\rm coll}/\langle\delta S_4\rangle_{\rm approx}$, where $\langle\delta S_4\rangle_{\rm approx}$ is obtained by assuming $\delta_4^{\rm corr}=0$ while leaving the pre-collisional velocity ensemble unchanged. Hence, $\langle\delta S_4\rangle_{\rm coll}/\langle\delta S_4\rangle_{\rm approx}=1$ corresponds to an exact factorization. The results show that the approximation is essentially exact far from the transition, whereas pronounced deviations develop close to the low-temperature edge of the synchronized active branch. In particular, for $\alpha=0.80$, we have $\langle\delta S_4\rangle_{\rm coll}/\langle\delta S_4\rangle_{\rm approx}\simeq3.1$, showing that the neglected velocity--synchronization correlations strongly modify the collisional production of the fourth velocity moment. For the close-to-transition value of $\alpha=0.80$, additional numerical calculations reveal that neglecting velocity--synchronization correlations accounts for approximately $55\%$ of the discrepancy between the EDMD and theoretical predictions for the steady-state value of $a_2$.

The physical origin of these correlations is also natural. The synchronization variable records the time elapsed since the previous collision. Because the drag continuously reduces particle velocities during free flight, this time is itself correlated with the pre-collisional velocity. Collisions for which $I_{ij}=1$ are therefore not a random subset of all collisions. This correlation is absent from both the deterministic mean-field rule and the independent Poisson--Bernoulli rule.

\subsection{Asynchronous limit}

The discrepancy observed in the asynchronous limit has a different origin. When $\tau_{\rm s}\to\infty$, one has $\Delta_{ij}\equiv\Delta$, so neither the binary injection fluctuations nor the velocity--synchronization correlations discussed above are present. In this regime the relevant absorbing transition is the continuous activation--inactivation transition~\cite{MPSF25}, for which a finite fraction of particles can become dynamically inactive. The present kinetic theory explicitly restricts itself to the active branch and sets $P_{\rm a}=1$, and therefore does not retain this population dynamics. EDMD averages performed over all particles consequently mix active and inactive populations close to the continuous transition. When the same observables are conditioned on the active population, the discrepancy in $a_2$ is substantially reduced.

\section{Details of the EDMD simulations}
\label{app:EDMD}
\subsection{Event-driven molecular dynamics algorithm}

We simulate the hard-disk system using event-driven molecular dynamics~\cite{smallenburg2022efficient} of $N$ identical hard disks of diameter $\sigma$ and mass $m$ in a square box of side $L$ with periodic boundary conditions. Due to the hard-core nature of the particles, the time evolution of the system can be simulated exactly, up to numerical errors, by advancing time collision by collision.

Knowing the positions $\mathbf r_i$ and velocities $\mathbf v_i$ of all particles allows the next collision to be determined exactly. For a pair $(i,j)$, let $\mathbf r_{ij}=\mathbf r_i-\mathbf r_j$ and $\mathbf v_{ij}=\mathbf v_i-\mathbf v_j$ denote their relative position and velocity at the current time. Under viscous damping, their relative separation evolves as
\begin{equation}
\mathbf r_{ij}(t)=\mathbf r_{ij}+\frac{1-e^{-\gamma t}}{\gamma}\mathbf v_{ij}.
\end{equation}
A collision occurs when $|\mathbf r_{ij}(t)|=\sigma$. Solving this condition gives the earliest collision time
\begin{align}\label{eq:EDMD_drag_time}
t_{ij}^{\rm col}={}&-\frac{1}{\gamma}\ln\left[
1-\gamma\frac{-b-\sqrt{b^2-v_{ij}^2(r_{ij}^2-\sigma^2)}}{v_{ij}^2}
\right],
\end{align}
with
\begin{align}
b=\mathbf r_{ij}\cdot\mathbf v_{ij}.
\end{align}

A real positive $t_{ij}^{\rm col}$ exists only if the particles actually collide; otherwise the pair is discarded. After evaluating $t_{ij}^{\rm col}$ for all relevant pairs, the smallest value is selected and the system is advanced to that collision. The velocities of the colliding particles are then updated according to Eq.~\eqref{1.1}, with $\Delta_{ij}$ given by Eq.~\eqref{1.23}, and the procedure is repeated to determine the next collision.

\subsection{Measurement of observables}
The instantaneous granular temperature is measured directly from the particle velocities as
\begin{equation}
    T(t)=\frac{m}{2N} \sum_{i=1}^{N}   \mathbf v_i^2(t),
\end{equation}
and its steady-state value is obtained by averaging after the initial transient has been discarded. 

The shear viscosity is measured in a separate set of simulations in which a stationary shear is imposed through Lees--Edwards boundary conditions along $\hat{\mathbf u}_y$. Momentum injected by the boundaries is progressively dissipated in the bulk by the damping, and the stationary velocity profile is therefore not linear. The resulting hydrodynamic profile is instead:

\begin{equation}
\begin{aligned}
u_x(y)={}&-\frac{{{a}}L}{2}
\frac{\sinh[(y-L/2)/\lambda]}{\sinh[L/(2\lambda)]},\\
\eta={}&mn\gamma\lambda^2,\qquad 0\leq y\leq L.
\end{aligned}
\label{eq:shear_profile}
\end{equation}
where ${{a}}$ is the imposed shear rate and $\lambda$ is the characteristic penetration length of the velocity field, which is fitted to obtain the shear viscosity $\eta$. In our simulations, we use ${{a}}=\min\left[0.6\sqrt{T/m}/L,\sqrt{0.02\gamma nT/\eta}\right]$.

We measure the bulk viscosity by imposing a small periodic change in the box size, $L(t)=L_0[1+\epsilon\sin(\omega t)]$. The compression rate in two dimensions is $\vartheta(t)=\bm{\nabla}\cdot\mathbf u=2\dot L/L$. For small deformations, the collisional pressure responds as
\begin{equation}
P_{\rm c}(t)=P_{{\rm c},0}+c_n\,\delta n(t)+c_T\,\delta T(t)-\eta_{\rm b}\vartheta(t),
\label{eq:bulk_edmd_pressure}
\end{equation}
where $c_n$ and $c_T$ describe the pressure changes caused by density and temperature variations. Multiplying by $\vartheta$ and averaging over complete cycles removes the constant and density terms, giving
\begin{equation}
-\underbrace{\frac{\langle P_{\rm c}\vartheta\rangle}{\langle\vartheta^2\rangle}}_{B}
=\eta_{\rm b}-c_T\underbrace{\frac{\langle\delta T\,\vartheta\rangle}{\langle\vartheta^2\rangle}}_{C}.
\label{eq:bulk_edmd_fit}
\end{equation}
We measure $B$ and $C$ at two driving frequencies and fit $B=\eta_{\rm b}-c_TC$. The intercept gives the bulk viscosity, while the slope accounts for the pressure response to temperature oscillations. In our simulations, we use $\epsilon=0.04$ and verified that there was no change in $\eta_{\rm b}$ at slightly larger values.

%
\end{document}